\documentclass[12pt,a4paper,twoside]{report}
\usepackage[utf8]{inputenc}
\usepackage[english,french]{babel}
\usepackage[T1]{fontenc}
\usepackage{amsmath}
\usepackage{mathtools}
\usepackage{amsfonts}
\usepackage{amssymb}
\usepackage[scaled]{helvet}
\usepackage[pdftex, 
            pdfauthor={Robin Piron},
            pdftitle={Atoms in Dense Plasmas: Models, Applications, and Current Challenges},
            pdfsubject={Habilitation manuscript},
            pdfkeywords={dense plasmas, nonideal plasmas, strongly coupled plasmas, pressure ionization, atoms, atomic physics, equation of state, radiative properties, opacity, classical fluids, ion correlations, non-LTE plasmas, collisional-radiative model},
            pdfproducer={pdflatex},
            pdfcreator={pdflatex}]{hyperref}
\usepackage{doi}
\usepackage[left=1.5cm,right=1.5cm,top=2cm,bottom=2cm]{geometry}
\usepackage{fancyhdr}
\usepackage{lastpage}
\usepackage{graphicx}
\usepackage{titlesec}
\usepackage{color}
\usepackage{xcolor}
\definecolor{ceared}{HTML}{e50019}
\usepackage{cite}
\usepackage{comment}
\usepackage[nottoc,numbib]{tocbibind}
\usepackage{tikz}
\usetikzlibrary{arrows,shapes,fit,positioning,decorations.pathreplacing}
\usepackage{anyfontsize}
\titleformat{\chapter}[display]{\huge}{\vspace{-3cm}}{12pt}{\huge\bf \thechapter.~}

\renewcommand{\rho}{\varrho}
\renewcommand{\phi}{\varphi}
\renewcommand{\vec}{\mathbf}

\newcommand{\bra}[1]{\ensuremath{{\langle #1 |}}}
\newcommand{\ket}[1]{\ensuremath{{| #1 \rangle}}}
\newcommand{\moy}[1]{\ensuremath{{\langle #1 \rangle}}}

\newcommand{\ee}{\mathtt{e}}

\title{}

\author{Robin Piron}

\date{}

\begin{document}
%\maketitle
\newgeometry{left=0cm,right=0cm,top=0cm,bottom=0cm}
{

\thispagestyle{empty}
\noindent
\begin{tikzpicture}
%\draw[step=1.0,black,thin,xshift=0.5cm,yshift=0.5cm] (0.,0.) grid (210mm,290mm);

\node[inner sep=0,anchor=north east] at (210mm,290mm){\includegraphics[trim=0 0 2.5cm 0,height=289mm]{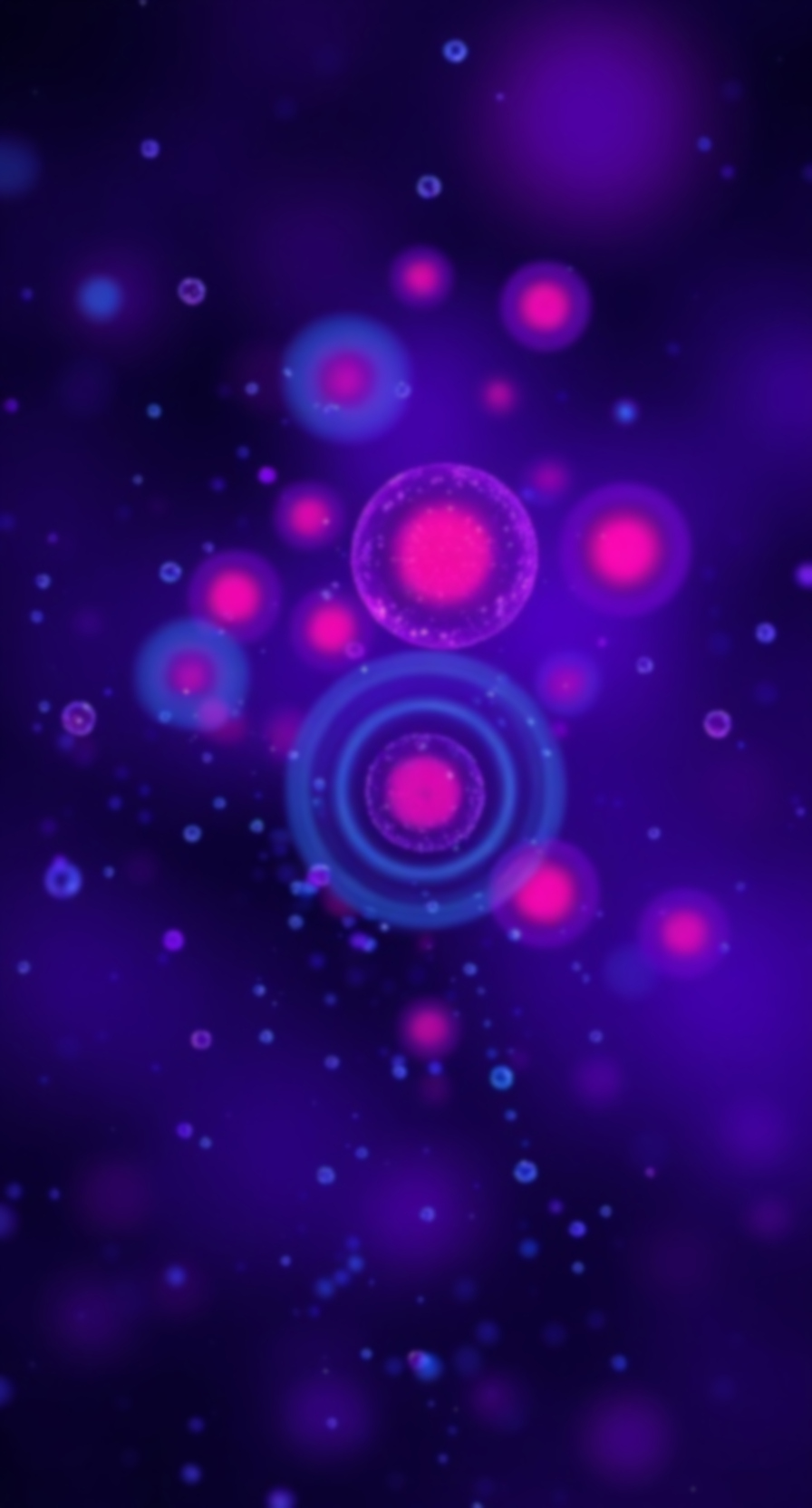}};
\fill[color=white] (0,290mm) -- (140mm,290mm) -- (180mm,0mm) -- (0,0);
\fill[color=white] (160mm,0) -- (160mm,15mm) -- (210mm,20mm) -- (210mm,0);
\fill[color=ceared] (132mm,290mm) -- (135mm,290mm) -- (170mm,60mm);
%\node[inner sep=0,anchor=north west] at (20mm,270mm){\includegraphics[height=25mm]{CEA_logo_perso.pdf}};
\node[inner sep=0,anchor=north west] at (20mm,210mm){
\begin{minipage}{12cm}
\begin{flushleft}\selectlanguage{english}
\fontsize{37pt}{40pt}\selectfont\color{ceared}\bf
ATOMS IN DENSE PLASMAS: MODELS, APPLICATIONS, AND CURRENT CHALLENGES
\end{flushleft}
\end{minipage}
};
\node[inner sep=0,anchor=north west] at (20mm,120mm){
\begin{minipage}{12cm}
\begin{flushleft}\selectlanguage{english}
\fontsize{16pt}{16pt}\selectfont\color{black}\bf
Habilitation thesis
\end{flushleft}
\end{minipage}
};
\draw[line width=3pt,color=ceared] (40mm,30mm) -- (45mm,30mm);
\node[inner sep=0,anchor=north west] at (40mm,25mm){
\begin{minipage}{12cm}
\begin{flushleft}\selectlanguage{english}
\fontsize{20pt}{20pt}\selectfont\color{black}\bf
Robin PIRON
\end{flushleft}
\end{minipage}
};
\end{tikzpicture}
}
\restoregeometry
\newpage
%{\thispagestyle{empty}\vspace*{\fill}\noindent Cover image: Chapel of St. Ursula of the Sorbonne seen from the courtyard in the 5th arrondissement of Paris. Adapted from Moonik (CC BY-SA 3.0 <https://creativecommons.org/licenses/by-sa/3.0>)}
{\thispagestyle{empty}\vspace*{\fill}\noindent Cover image: Image generated by DeepAI Image Generator (Public Domain).}
\newpage
{
\thispagestyle{empty}
~\vspace{6cm}\\
\begin{center}
\selectlanguage{english}
{\Large Habilitation thesis}
\end{center}

\begin{center}
\rule{17cm}{0.5pt}

\selectlanguage{english}
{\Huge Atoms in Dense Plasmas{}:\vspace{0.5cm}\\ Models, Applications, and Current Challenges}

\rule{17cm}{0.5pt}
\end{center}

\vspace{1cm}
\begin{center}
{{\Large Robin \textsc{Piron}}

\medskip

Docteur de l'\'Ecole Polytechnique

\medskip

Chercheur au CEA, DAM, DIF\\
F-91297 Arpajon, FRANCE

\medskip

Professeur associé à l'INSTN}
\end{center}
\vfill
%\begin{comment}
{
\selectlanguage{french}
Manuscrit présenté et soutenu publiquement pour obtenir l'Habilitation à Diriger des Recherches de Sorbonne Université, le 22 novembre 2024, devant le jury composé de:}\\
%Manuscrit soumis en vue d'obtenir l'Habilitation à Diriger des Recherches de Sorbonne Université,\break après soutenance, proposée le 22 novembre 2024, devant le jury composé de:}\\
\begin{center}
\begin{tabular}{@{}l l l l}
\textbf{Titre, Prénom, Nom} & \textbf{Statut} & \textbf{Affiliation} & \textbf{Qualité}\\
Dr HDR Annette \textsc{Calisti} & 
DR CNRS & 
PIIM & 
Rapporteur 
\\
Dr Charles Edward \textsc{Starrett} & 
Affiliate Pr. & 
LANL, AU & 
Rapporteur
\\
Dr HDR Claude \textsc{Deutsch} & 
DR Em. CNRS & 
LPGP &  
Examinateur
\\
Dr Christopher John \textsc{Fontes} & 
Affiliate Pr. & 
LANL, AU & 
Examinateur
\\
Dr HDR Frank \textsc{Rosmej} & 
PU & 
SU, LULI & 
%Examinateur
Président du Jury
\\
Dr HDR Gérard \textsc{Massacrier} & 
CR CNRS & 
CRAL & 
Rapporteur
\\
Dr HDR Stéphane \textsc{Bernard} & 
DR CEA & 
CEA & 
Examinateur
\\
Dr Yuri \textsc{Ralchenko} & 
Adjunct Pr. & 
NIST, CU &
Examinateur
\end{tabular}
\end{center}
}
%\end{comment}
%\newpage
\thispagestyle{empty}~
\newpage
%\begin{comment}
{
\thispagestyle{empty}
\textbf{Sigles des affiliations:}\medskip\\
\footnotesize
\begin{center}
\begin{tabular}{@{}l l}
CEA &
\begin{tabular}[t]{@{}l@{}}
Commissariat à l'\'Energie Atomique et aux \'Energies Alternatives\\
Bât. Le ponant D - 25 rue Leblanc\\
F-75015 Paris\\
FRANCE
\end{tabular}
\medskip\\
INSTN &
\begin{tabular}[t]{@{}l@{}}
Institut National des Sciences et Techniques Nucléaires\\
CEA Saclay - Bât. 399 - PC n°35\\
F-91191 Gif-sur-Yvette Cedex\\
FRANCE
\end{tabular}
\medskip\\
CNRS &
\begin{tabular}[t]{@{}l@{}}
Centre National de la Recherche Scientifique\\
3 rue Michel-Ange\\
F-75016 Paris\\
FRANCE
\end{tabular}
\medskip\\
PIIM &
\begin{tabular}[t]{@{}l@{}}
Laboratoire de Physique des Interactions Ioniques et Moléculaires\\
UMR 7345, CNRS - Aix-Marseille Université\\
Centre de Saint Jérôme, service 322\\
F-13397 Marseille Cedex 20\\
FRANCE
\end{tabular}
\medskip\\
LANL &
\begin{tabular}[t]{@{}l@{}}
Los Alamos National Laboratory\\
Los Alamos, NM 87545\\
USA
\end{tabular}
\medskip\\
AU &
\begin{tabular}[t]{@{}l@{}}
Auburn University\\
Auburn, AL 36849\\
USA
\end{tabular}
\medskip\\
LPGP &
\begin{tabular}[t]{@{}l@{}}
Laboratoire de Physique des Gaz et Plasmas\\
UMR 8578, CNRS - Université Paris-Saclay\\
Bât. 210 - rue Henri Becquerel\\
Université Paris-Saclay\\
F-91405 Orsay Cedex\\
FRANCE
\end{tabular}
\medskip\\
SU &
\begin{tabular}[t]{@{}l@{}}
Sorbonne Université\\
4 place Jussieu\\
F-75006 Paris\\
FRANCE
\end{tabular}
\medskip\\
LULI &
\begin{tabular}[t]{@{}l@{}}
Laboratoire pour l'Utilisation des Lasers Intenses\\
UMR 7605, \'Ecole Polytechnique - CEA - CNRS - Sorbonne Université\\
F-91128 Palaiseau\\
FRANCE
\end{tabular}
\medskip\\
CRAL &
\begin{tabular}[t]{@{}l@{}}
Centre de Recherche Astrophysique de Lyon\\
UMR 5574, Université de Lyon - \'Ecole normale supérieure de Lyon - CNRS\\
F-69364 Lyon Cedex 07\\
FRANCE
\end{tabular}
\medskip\\
NIST &
\begin{tabular}[t]{@{}l@{}}
National Institute of Standards and Technology\\
Gaithersburg, MD 20899-8422\\
USA
\end{tabular}
\medskip\\
CU &
\begin{tabular}[t]{@{}l@{}}
Clemson University\\
Clemson, SC 29631\\
USA
\end{tabular}
\end{tabular}
\end{center}
\bigskip
%$^*$ \textbf{Membres assimilés à un Dr HDR}
}
%\end{comment}
\selectlanguage{english}
\newpage
\setcounter{page}{1}
\selectlanguage{english}
\renewcommand{\chaptername}{}

{
\titleformat{\chapter}[display]{\huge}{\vspace{-3cm}}{12pt}{\huge\bf}
%-------------------------
\chapter*{\vskip-1.8cm{}Acknowledgements\vskip-1.8cm{}}
%%-------------------------

I would like to thank Thomas Blenski, with whom I worked on dense-plasmas atomic modeling. Thomas was my Ph.D. advisor and has been my closest colleague for the last 18 years. I can say that \emph{he} taught me the craft of plasma microscopic modeling. I would also like to thank Franck Gilleron, with whom I have been working on non-LTE plasmas modeling since I was hired at CEA/DAM, 15 years ago. I think we both share the care of constantly going from the fundamental theory to the very details of its implementation, which is the only way to ensure that a model is relevant. My current understanding of the atomic physics of dense-plasmas and non-LTE plasmas owes much to joint work and discussions with them.

I would also like to thank Bogdan Cichocki, who was involved in many studies reviewed here, including the early discussions about the development of the VAMPIRES model, and Clément Caizergues, who worked on the linear response of the VAAQP model, as a Ph.D. student co-supervised by Thomas Blenski and myself. 

As regards non-LTE plasma modeling, I would like to thank Olivier Peyrusse who is the author of the AVERROES code. Even though there was no daily collaboration with him, the knowledge base he left to our group and the experience of the field he shared with us are invaluable. Without him and an access to his numerical code, our progress on non-LTE plasma modeling would have been much slower. I am also grateful to Christopher Bowen, for encouraging me to work on NLTE plasma physics, and to the participants of the NLTE code-comparison workshops, in particular Christopher Fontes, Stephanie Hansen, Yuri Ralchenko, Howard Scott, and Evgeny Stambulchik for many fruitful exchanges and their constant input.

In parallel with my research activities, I was given the opportunity to teach, first in the Master in Fusion Sciences, and now in the Master in Physics of Plasma and Fusion, and in the Master in Large Facilities. I would like to thank Frank Rosmej for the confidence he has shown by entrusting me successively with two teaching modules. This gave me a foothold in teaching, which I believe is an essential complement to a researcher's life. I am also grateful for his constant support during the preparation of this habilitation. I would like to especially thank Andrea Ciardi and Serena Bastiani-Ceccotti with whom I have the pleasure to share lectures for several years now. I am also grateful to Guy Bonnaud, Catherine Krafft, Sophie Kazamias, Olivier Guilbaud, Benoit Canaud, Pierre Morel, and Roch Smets, who also supported my teaching activity over the years. I would also like to thank people of the LERMA, notably Franck Delahaye and Arno Vanthiegem for hosting me on many occasions.

Atomic physics of plasma has many outcomes, which result in collaborations on application studies. I would like to thank Jean-François Danel and Luc Kazandjian for a fruitful collaboration on supplementing Thomas-Fermi molecular dynamics simulations with quantum average-atom results for dense-plasmas equations-of-state calculations. I would also like to thank Laurent Jacquet for a productive collaboration on the design of X-ray sources on the OMEGA facility, Serena Bastiani-Ceccotti and Ambra Morana for a collaboration on NLTE spectroscopy experiments, and to Gilles Kluth for a collaboration on the application of machine learning to the modeling of NLTE plasmas. I would also like to thank Oleksandr Marchuk and Edward Marley for allowing me to reproduce their experimental spectra in the present manuscript.

Finally, I would like to express my sincere gratitude to Annette Calisti, Charles E. Starrett and Gérard Massacrier for accepting to review the present manuscript, for the time they spent and the care they took in preparing their reports. I would also like to thank very much Claude Deutsch, Christopher Fontes, Frank Rosmej, Stéphane Bernard and Yuri Ralchenko for agreeing to serve on the jury and attend the defense that I set up at the end of a conference week. I would also like to thank Stephanie Hansen, although she ultimately had to decline, for sincerely considering serving on this jury and for her encouragement.

%Many people cited in this manuscript passed away within the last 15 years. People of my generation only had a few occasions to meet them, and most often not timely enough to benefit at best from these exchanges. I would like to honor the memory of Michael E. Fisher, Avi Bar Shalom, Walter R. Johnson, Jacques Bauche, Charlotte Froese-Fischer, Vladimir Fortov, Starostin, Vladimir Novikov. It is really frustrating to remember that you once met these people and, a few years later, when it would have been really helpful to discuss with them, to just see that they are gone.

%I would like to honor (the memory of) Richard W. Lee, who used to run the RPHDM conference series as a very enthusiastic, very informal meeting point, so open and welcoming to young physicists. This is precious, and even the opportunity of organizing this defense somehow owes something to him.
%I would probably never had met half of the above-mentioned people without his way of running the conference.

%Finally, I would like to honor all the participants of the RPHDM conferences, and especially  Richard W. Lee who used to run this conference series as a very enthusiastic, very informal meeting occasion, so open and welcoming to young physicists. Even the opportunity of organizing the habilitation defense somehow owes something to him.
}
{ 
\newpage
\thispagestyle{empty}~
}

{
%\titleformat{\chapter}[display]{\huge}{\vspace{-3cm}}{12pt}{\huge\bf}
\titleformat{\chapter}[display]{\huge}{\vspace{-3cm}}{12pt}{\vskip-12pt\huge\bf}
\tableofcontents
}

%-------------------------
\chapter{Introduction}
%-------------------------
\section{Motivations for the Atomic Modeling of Plasmas}
Plasma physics is the physics of partially or fully ionized fluids (liquids or gases). Fundamentally, a plasma is just a collection of a large number of electrons and nuclei interacting through Coulomb potentials, both repulsive and attractive. Depending on the plasma conditions, part of the electrons may be bound to the nuclei. Atoms, ions, and free-electrons are concepts used in the interpretation of the microscopic structure of this physical system.
%, rather than elementary objects whose existence can be assumed as a premise.

Modeling the plasma microscopic structure is essential to the understanding of many astrophysical objects. The description of stellar interiors \cite{KippenhahnWeigertbook} and atmospheres \cite{MihalasStellarAtmospheres}, as well as of compact objects such as white dwarfs \cite{ShapiroTeukolsky}, relies on properties that are directly related to the microscopic structure of the plasma. These are the thermodynamic properties (equation of state), the radiative properties (opacity and emissivity) and other transport properties (direct-current conductivity, thermal conductivity, bulk viscosity...). The same properties are also useful for technological applications of high-energy-density plasmas such as controlled fusion (thermonuclear fusion using inertial or magnetic confinement) or X-ray sources (hohlraums, magnetic pinches, lithography sources).

Among the various models that may be used to describe plasmas, models that define a notion of atom or ion are particularly appealing. We mean here an idealized system that describes the plasma using effective 1-electron states (orbitals) stemming from a spherically symmetric 1-electron Hamiltonian. With such a model, one can greatly simplify the description of the microstates of the plasma, using the notion of atomic states.

A description of the atomic excited states in the plasma is essential to the modeling of spectral values, such as radiative properties. These can reveal the fluctuations around the average atomic state. The mathematical apparatus for building the atomic states is well established \cite{CondonShortley,Cowan,FroeseFischerbook,Grantbook}, and includes sophisticated methods of angular-momentum coupling \cite{Racah42,Racah42b,Racah43}.

Achieving an even somewhat complete description of molecular excited states is usually beyond reach. Even using a notion of ion or atom, a statistical treatment of energy levels is usually required for hot-plasma modeling, due to the combinatorial explosion of the number of levels when some shells are open. Analytic results from atomic physics give access to statistical properties of configurations~\cite{Moszkowski62} and transition arrays between configurations~\cite{Bauche79,Bauche82,Bauche85} (UTA, SOSA). This enables powerful detailed configuration accounting (DCA) approaches. In order to handle the calculation of spectra for ions with a highly complex electronic structure, coarser statistical approaches were developed, such as the Gaussian approximation~\cite{Perrot88b} and the super transition arrays (STA) formalism~\cite{BarShalom89,Blenski00}. On the other hand, in order to refine the UTA approach, finer approaches were also developed, such as the mixed-UTA~\cite{Mazevet06} and the partially resolved transition arrays (PRTA)~\cite{Iglesias12}. 

More generally, many phenomena are described using approaches based on the notion of atomic processes. One may cite, for instance, the collisional-radiative modeling of plasmas out of local thermodynamic equilibrium~\cite{MihalasStellarAtmospheres,Ralchenkobook}, the approaches to line-broadening mechanisms (see~\cite{Baranger58, SobelmanVainshteinYukov, Boercker87, Talin95,Stambulchik08}, and~\cite{Gomez22} for a review), or the notion of collision in the classical Boltzmann equation used to infer some transport properties (see, for instance, \cite{Ziman61}).

\section{Scientific Journey and Outline of the Manuscript}
The subject of my Ph.D. thesis \cite{PironPhD} was a variational model of atoms in a plasma (called VAAQP in its average-atom version). Its motivation was to address the thermodynamic consistency issue of atomic models of pressure-ionized plasmas. In the VAAQP approach, the average atomic state is obtained from the minimization of an approximate free-energy functional. In this model, all electrons, bound and continuum are treated on an equal footing, while the ion's surroundings is modeled using the notion of a statistical cavity around each ion.
 
I was given the opportunity to continue working on the atomic modeling of dense plasmas when I was hired at CEA, in a group in charge of the atomic physics of plasmas. In addition to continuing this research activity, I also started to work on the modeling of plasmas out of local thermodynamic equilibrium (non-LTE plasmas). My commitment to this topic was initially motivated by applications to the modeling of hohlraums and other X-ray sources, as well as by the use of emission spectroscopy for diagnostics in laser-plasma experiments. However, the modeling of non-LTE plasma also has a key role in the modeling of many astrophysical objects, for instance the stellar atmospheres and the solar corona.

In this manuscript, I mostly focus on the studies to which I have contributed in the field of dense-plasma atomic modeling. On this topic, I have been collaborating with Thomas Blenski from CEA/DRF, Bogdan Cichocki from the Institute of Theoretical Physics of Warsaw University, and Clément Caizergues, who was at the time a Ph.D. student co-supervised by Thomas Blenski and myself. At the end of the manuscript, I also comment briefly on the works to which I have contributed in the field of non-LTE plasma modeling. On the latter subject I have been collaborating with Franck Gilleron, who works in the same group as me. Our joint work also benefits much from the NLTE code-comparison workshops, in which we participate every two years.

Chapter~\ref{sec_state_of_art} gives a selected review of atomic models of plasma in order to state the particular issues of dense-plasma atomic modeling. This chapter notably includes results and comparisons with the VAAQP model as regards mean ionization and equation of state.

Following my Ph.D. thesis, several research prospects seemed promising. A first possible extension was the application of the VAAQP model to the calculation of radiative properties. Chapter \ref{ch_rad_prop} is devoted to this topic. I started working on the application of the VAAQP model to radiative properties in the independent-particle approximation, adapting the existing methods to the case of the VAAQP model (Sections~\ref{sec_fluctuations_DCA}, \ref{sec_reg_low_freq}, and references~\cite{Piron13,Piron18}). A more ambitious trail was the application of the self-consistent linear response to the VAAQP model. This subject was seemingly a good one for a Ph.D. thesis and, with Thomas Blenski, we proposed it for the Ph.D. thesis of Clément Caizergues \cite{CaizerguesPhD}. Section~\ref{sec_sc_linear_response}  deals with this topic, as well as references \cite{Caizergues14,Caizergues16}.

At the same period, I had the opportunity to collaborate with Jean-François Danel and Luc Kazandjian, from the same group as me, who were involved in molecular-dynamics simulations of plasmas. We worked on an approach allowing to supplement Thomas-Fermi molecular dynamics simulations with quantum average-atom results in order to better account for ion contribution to dense-plasmas equations of state \cite{Danel14,Danel16,Danel18}. Despite their useful applications, I will not comment these studies in the manuscript, for they do not properly pertain to atomic modeling of plasmas.

A highly ambitious subject emerging from the VAAQP study was the search for a new variational model of atoms in plasmas, this time including also a proper modeling of ion correlations in the plasma. Chapter~\ref{ch_ion_correlations} is devoted to this topic, which was and still constitutes a long-term research task. Thomas Blenski and I decided to address this subject without involving any student, because of the wide variety of physical notions it requires, and the highly uncertain outcome it had. Our colleague Bogdan Cichocki was involved in discussions at the early stages of thinking about this topic.

For the sake of this research, I was led to cope with the statistical mechanics of classical fluids. Because Thomas Blenski and I were not finding all theoretical tools we needed, we proposed new generalized free-energy functionals for the Debye-Hückel model with arbitrary interaction potential. In addition to providing us with useful free-energy functionals, these studies helped us to better understand the role of such functionals. This is the subject of Section~\ref{sec_classical_fluids} and references \cite{Piron16,Blenski17,Piron19a}.

Recently, we have crossed a significant milestone, proposing a variational atomic model of plasma accounting for both the electronic structure including continuum electrons, and ion-ion correlations in the plasma (VAMPIRES model). In this model, the average atomic state proceeds from the minimization of a generalized free energy and includes the average structure of the ion fluid, under the form of an equilibrium ion-ion correlation function. This is the subject of Section~\ref{sec_VAMPIRES} and of reference \cite{Piron19b,BlenskiPiron23}.

Finally, Chapter~\ref{ch_NLTE_plasmas} briefly outlines the studies to which I have contributed in the field of non-LTE plasma modeling. Part of this work is described in \cite{Gilleron15}. Section~\ref{sec_CR_modeling} is an introduction to the problem of collisional-radiative modeling of plasmas. Section~\ref{sec_NLTE_Xray_spectro} is devoted to the applications to X-ray-emission spectroscopy and to comparisons of models with various level of detail. Section~\ref{sec_NLTE_rad_hydro} regards the application to \textit{in-situ} calculation of radiative properties for radiation-hydrodynamics simulations. Finally, Section~\ref{sec_collisional_processes} presents a preliminary effort to apply dense-plasma atomic models to the calculation of collisional cross-sections of interest in non-LTE plasma modeling.

Throughout this manuscript, on each topic, I have tried to develop the problem statement, in order to situate the studies I contributed to. The main results and conclusions are recalled, without giving much detail on the derivations, which can be found in the articles cited above. My intent was more to present a synthetic and consistent view of my work and its common thread.  
Within each chapter, I attempted to sketch some research prospects regarding its topic. A significant part of the material covered in this manuscript may be found in the recent review article \cite{Piron24}.

{ 
\newpage
\thispagestyle{empty}~
}

%-------------------------
\chapter{Atomic Physics of Plasmas\label{sec_state_of_art}}
%-------------------------  

%-------------------------
\section{Atomic Modeling of Ideal Plasmas\label{sec_ideal_plasmas}}
%-------------------------  
In plasma physics, the ideal-gas approximation is often applied partially. One first defines a quasi-particle composed of a nucleus and a set of bound electrons that interact with the nucleus and among themselves. This is the definition of an \emph{ion} in this model. All electrons of the plasma that are not bound to a nucleus are viewed as \emph{free electrons}. One then makes the ideal-gas hypothesis on the system of ions and free electrons. This is the picture of an \emph{ideal plasma}. 
%Ions usually have a lower apparent charge than the bare nucleus, and the zero-coupling approximation thus has broader validity range in this context. 

Long-range attractive interactions in classical mechanics lead to the so-called \emph{classical Coulomb catastrophe}. One usually resorts to a quantum model for the electronic structure of the ion, which does not result in the Coulomb catastrophe. On the other hand, the ideal-gas approximation allows one to circumvent the catastrophe for the whole classical plasma of ions and free electrons.

In this context, the ion is seen as a charged, isolated system having a finite spatial extension. It is an \emph{isolated ion} since its electronic structure is obtained disregarding all the other particles of the plasma: other ions and free electrons. To describe this electronic structure requires one to address an $N$-body problem with $N \sim Z+1$ at most, where $Z$ is the atomic number of the nucleus.
%, which can be considered separately from the modeling of the whole plasma.

\subsection{Isolated Ion at Zero Temperature}
The problem of the isolated-ion electronic structure can be treated independently of the modeling of the whole plasma. A typical quantum Hamiltonian of the isolated ion having $Q$ bound electrons is the many-electron operator:
%\begin{align}
%\hat{H}_\text{isol.ion}^Q=\sum_{j=1}^{Q}\frac{\tilde{\vec{P}}_j^2}{2m_\text{e}}-\sum_{j=1}^{Q}\frac{Z \ee^2}{|\tilde{\vec{R}}_j|}
%+\sum_{j=1}^{Q}\sum_{\substack{k=1\\k\neq j}}^{Q}\frac{\ee^2}{|\tilde{\vec{R}}_j-\tilde{\vec{R}}_k|}
%\end{align}
\begin{align}
&\hat{H}_\text{isol.ion}=\hat{K}+\hat{V}_\text{nuc.}+\hat{W}
\end{align}
where $\hat{K}$ is the kinetic energy operator, $\hat{V}_\text{nuc.}$ is the external potential generated by the nucleus, and $\hat{W}$ is the electron-electron interaction operator.
\begin{align}
&\hat{K}=\sum_{\xi,\zeta}\bra{\phi_\xi}\tilde{H}_0\ket{\phi_\zeta}\hat{a}^\dagger_\xi\hat{a}_\zeta\ ;\ \tilde{H}_0\equiv\frac{\tilde{\vec{P}}^2}{2m_\text{e}}
\\
&\hat{V}_\text{nuc.}=\sum_{\xi,\zeta}\bra{\phi_\xi}\frac{-Z \ee^2}{|\tilde{\vec{R}}|}\ket{\phi_\zeta}\hat{a}^\dagger_\xi\hat{a}_\zeta
\\
&\hat{W}=\frac{1}{2}\sum_{\xi,\zeta,\xi',\zeta'}\bra{1:\phi_\xi\, 2:\phi_{\xi'}}\frac{\ee^2}{|\tilde{\vec{R}}_1-\tilde{\vec{R}}_2|}\ket{1:\phi_\zeta\, 2:\phi_{\zeta'}}\hat{a}^\dagger_\xi\hat{a}^\dagger_{\xi'}\hat{a}_\zeta\hat{a}_{\zeta'}
\label{eq_op_W}
\end{align}
where $\{\ket{\phi_\xi}\}$ is a basis of the 1-electron state space $\mathcal{E}$. $\tilde{\vec{P}}_j$ and $\tilde{\vec{R}}_j$ are the 1-electron momentum and position operators, respectively, acting on the $j$ electron. $\tilde{H}_0$ is the 1-electron free-particle Hamiltonian. For the sake of shortening the notation, we set $\ee^2=q_\text{e}^2/(4\pi\epsilon_0)$, with $q_\text{e}$ being the elementary charge. $\hat{a}^\dagger_\xi$ and $\hat{a}_{\xi}$ are the creation and annihilation operator of the 1-electron state $\xi$, respectively. Throughout the manuscript, we will denote $\hat{O}$ many-electron operators and $\tilde{O}$ 1-electron operators.

The problem of finding the stationary $Q$-electron states of the isolated-ion Hamiltonian $\hat{H}_\text{isol.ion}$ has a variational formulation (Ritz theorem). However, to address this problem, one resorts to approximate methods. Such methods frequently start with a model based on effective 1-electron states $\ket{\varphi_\xi}$, i.e.\ orbitals, which are solutions of a 1-electron Schrödinger equation associated with an effective spherically symmetric potential $v_\text{eff}(r)$.
\begin{align}
&\tilde{H}_\text{eff}\ket{\varphi_\xi}=\left(\tilde{H}_0+v_\text{eff}(|\tilde{\vec{R}}|)\right)\ket{\varphi_\xi}=\varepsilon_\xi\ket{\varphi_\xi}\\
&-\frac{\hbar^2}{2m_\text{e}}\nabla_{\vec{r}}^2\varphi_\xi(\vec{r})+v_\text{eff}(r)\varphi_\xi(\vec{r})=\varepsilon_\xi \varphi_\xi(\vec{r})
\label{eq_general_eff_pot}
\end{align}

Various models exist, which mostly differ in their way of obtaining the effective potential $v_\text{eff}(r)$. One may cite, for instance, the  Hartree-Fock-Slater~\cite{Slater51}, the optimized effective potential~\cite{Talman76}, or the parametric potential~\cite{Klapisch71} models. 
In the Hartree-Fock model~\cite{Slater28}, a nonlocal exchange term is added to Equation~\eqref{eq_general_eff_pot}. However, the problem is often restricted to an effective 1-electron problem having spherical symmetry. This is the restricted Hartree-Fock approach (see, for instance,~\cite{Cowan}).

Being spherically symmetric, the effective 1-electron Hamiltonian commutes with 1-electron angular momentum operators, which enables the separation of the angular part of the 1-electron states (spherical harmonics or spinors). Many-electron eigenstates of the many-electron Hamiltonian can be built subsequently from the orbitals, using a well-established mathematical apparatus. 

Starting from a basis of determinantal many-electron states, one evaluates the matrices of the squared total angular momentum $\hat{J}^2$, parity $\hat{\Pi}$ and many-electron Hamiltonian, treating the difference $\hat{U}-\hat{v}_\text{eff}$ as a perturbation. One may also add further perturbation operators such as spin-orbit coupling (fine-structure correction), or hyperfine-structure corrections.

The Wigner-Eckart theorem (see, for instance, \cite{Schiff}) allows to greatly reduce the number of matrix element to evaluate. Efficient methods of angular-momentum coupling, which allow one to circumvent the direct diagonalization of $J^2$, were thoroughly studied, both as regards the formalism~\cite{Racah42,Racah42b,Racah43,Judd,Yutsis} and the numerical methods~\cite{BarShalom88,Gaigalas01}.

Simultaneous diagonalization of the complete set of commuting operators $\hat{J}^2$, $\hat{J}_z$, $\hat{\Pi}$ and $\hat{H}_\text{isol.ion}$ yields the many-electron stationary states, grouped into degenerate energy levels. These are fine-structure levels if spin-orbit coupling is accounted for, hyperfine structure levels if hyperfine corrections are accounted for. Such a diagonalization may be performed within various approximation schemes (see \cite{Cowan,FroeseFischerbook,Grantbook}).

Without performing such a diagonalization procedure, it is also possible to calculate statistical properties of configurations \cite{Slater29,Moszkowski62} or of broader statistical objects:  super-configurations \cite{BarShalom89,Blenski00}.

\subsection{Saha-Boltzmann Model: A Variational Detailed Model of Ideal Plasma}
The approach in which one accounts for the various electronic states of the ions as distinct species is known as \emph{detailed modeling}. For an ideal plasma in thermal equilibrium, this approach yields the Saha-Boltzmann model of plasma~\cite{Saha20,Saha21}.

Accounting for each fine-structure level as a species is called a detailed level accounting (DLA) approach. One may also perform a detailed configuration accounting (DCA), grouping the levels according to their parent configuration. One may also group the configurations into super-configurations. The degree of statistical grouping defines the set of ion species $\Psi$ to consider. In this context, $\Psi$ will denote an energy level in a broad sense, rather than a single atomic state. Such an energy level may be a fine-structure level or the mean energy of a statistical object gathering several fine-structure levels, as for instance a configuration or a superconfiguration. Each energy level $\Psi$ has an energy $E_\Psi$ and a degeneracy $g_\Psi$ as main properties. Usually the reference of energies is taken such as the energy is zero for the bare nucleus and also for the free electrons.

The classical Hamiltonian for the ion-free-electron ideal plasma is the following:
\begin{align}
\mathcal{H}_\text{id}(\{\vec{p}_{\Psi,j}\},\{\vec{p}_{\text{e},j}\})=\sum_{\Psi=1}^{M}\sum_{j=1}^{N_\Psi}\left(\frac{p_{\Psi,j}^2}{2 m_\Psi}+E_\Psi\right)
+\sum_{j=1}^{N_\text{e}}\left(\frac{p_{\text{e},j}^2}{2 m_\text{e}}\right)
\label{eq_class_ideal_gas}
\end{align}
where the first sum runs over the $M$ ion species labelled $\Psi$ plus the free-electron species labelled by ``e''. In this Hamiltonian, interactions among the particles of the plasma are neglected. The ion-free-electron system is considered in the canonical ensemble and the free energy per ion of the classical ideal-gas mixture is: 
\begin{align}
\dot{F}_\text{id}\left(\{n_\Psi\},n_\text{e},T\right)
&=-\sum_{\Psi=1}^{M} \frac{n_\Psi}{n_\text{i}\beta} 
\left( \ln\left(\frac{g_\Psi e^{-\beta E_\Psi}}{n_\Psi\Lambda_\Psi^3}\right) + 1 \right)
-\frac{n_\text{e}}{n_\text{i}\beta} 
\left( \ln\left(\frac{2 }{n_\text{e}\Lambda_\text{e}^3}\right) + 1 \right)
\label{eq_Saha_free_energy}
\end{align}
where the $n_\Psi$'s are the numbers of particle of species $\Psi$ per unit volume, $n_\text{e}$ is the free-electron density, $n_\text{i}=\sum_{\Psi=1}^{M}{n_\Psi}$. $\Lambda_\Psi=h/\sqrt{2\pi m_\Psi k_\text{B} T}$ is the classical thermal length for species $\Psi$, $\Lambda_\text{e}=h/\sqrt{2\pi m_\text{e} k_\text{B} T}$ is the electron classical thermal length. We will denote by a dot the quantities per ion, in order to avoid confusion with total quantities or quantities per unit volume.

The equations of the Saha equilibrium model are obtained through a minimization of the free energy of the system with respect to the species populations, while also requiring the neutrality of the plasma and a fixed number of ions. Thus, transfers of population among the various ion species are allowed, and the populations are ultimately set by the condition of thermodynamic equilibrium. Only the populations, and not the quantities related to the shell structure of the ions, stem from the model.
\begin{align}
\dot{F}_\text{eq}(n_\text{i},T)
=\underset{\{n_\Psi\},n_\text{e}}{\text{Min}}\,\dot{F}\left(\{n_\Psi\},n_\text{e},T\right)
~&\text{ s. t. } \sum_\Psi n_\Psi =n_\text{i}
,\ \text{ s. t. } \sum_\Psi n_\Psi Z_\Psi^* =n_\text{e}
\end{align}

The latter minimization yields the following condition on the chemical potentials:
\begin{align}
&\mu_{\text{id},\Psi}(n_\Psi,T)+\mu_{\text{id,e}}(n_\text{e},T) Z_\Psi^* =\lambda_\text{i}\,\text{,~independent of $\Psi$}
\label{eq_Saha_chemical_equilibrium}
\end{align}
with the classical-ideal-gas chemical potentials being:
\begin{align}
&\mu_{\text{id},\Psi}(n_\Psi,T)
=\frac{\partial}{\partial n_\Psi}\left( n_\text{i}\dot{F}\left(\{n_\Psi\},T\right)\right)
=-\frac{1}{\beta} 
\ln\left(\frac{g_\Psi e^{-\beta E_\Psi} }{n_\Psi\Lambda_\Psi^3}\right)
\ ;\ 
\mu_{\text{id,e}}(n_\text{e},T)=-\frac{1}{\beta} 
\ln\left(\frac{2}{n_\text{e}\Lambda_\text{e}^3}\right)
\label{eq_id_chem_pot}
\end{align}

For a plasma of a pure substance, one usually assumes the thermal lengths $\Lambda_\Psi$ of all ion species to be equal: $\Lambda_\Psi\approx\Lambda_\text{i}$. From Equation~\eqref{eq_Saha_chemical_equilibrium} and the two constraints, one obtains the populations:
\begin{align}
&n_\Psi=n_\text{i}\frac{g_\Psi e^{-\beta(E_\Psi-\mu_{\text{id,e}}Z_\Psi^*)}}
{\sum_{\Psi'=1}^M g_{\Psi'} e^{-\beta(E_{\Psi'}-\mu_{\text{id,e}}Z_{\Psi'}^*)}}
\label{eq_id_Saha_pops}\\
&n_\text{e}=\sum_\Psi n_\Psi Z_\Psi^*
\label{eq_id_Saha_neutr}
\end{align}
The denominator in the right-hand side of Equation~\eqref{eq_id_Saha_pops} is a partition function.

From the populations follows notably the mean ionization of the plasma $Z^*=n_\text{e}/n_\text{i}$, as a value \emph{set by the thermodynamic equilibrium condition}. Obtaining the ionization state of the plasma as a result of its equilibrium state is among the purposes of modeling the plasma microscopic structure. 

%The plasma being treated in the ideal-gas approximation, the pair distribution functions among the particle of the plasma are identically $g_{\alpha,\gamma}=1$, and the mean inter-particle distance is $\Gamma(4/3) R_{\alpha,\gamma}$ with $R_{\alpha,\gamma}=(3/(4\pi\sqrt{n_\alpha n_\gamma}))^{1/3}$ (see~\cite{Chandrasekhar43}).

The thermodynamic quantities stemming from this model (internal energy, pressure...) can be obtained from the free energy, using the appropriate derivatives. They simply correspond to those of the ideal-gas mixture, taken with the equilibrium values of the species populations. Moreover, they obviously fulfill the virial theorem (in its non-relativistic version):
\begin{align}
P_\text{thermo}=n_\text{i}^2\frac{\partial \dot{F}_\text{eq}(n_\text{i},T)}{\partial n_\text{i}}=
P_\text{virial}=\frac{n_\text{i}}{3}\left( 2 \dot{U}_\text{eq}(n_\text{i},T)-\dot{U}_\text{inter,eq}(n_\text{i},T) \right)
\label{eq_virial_theorem}
\end{align}
where $\dot{U}_\text{eq}$ denotes the internal energy per ion, and $\dot{U}_\text{inter,eq}$ denotes the interaction energy per ion, which is zero for the ideal-gas mixture. This is an important feature as regards the consistency of the equation of state. 

An issue with the present Saha model is that, in principle, when accounting for the excited states in a complete manner, the partition function of Equation~\eqref{eq_id_Saha_pops} diverges because of the infinite number of bound states. Let us consider, for instance, the hydrogen-like atomic states. We have:
\begin{align}
\sum_{\Psi\text{ H-like}} g_\Psi e^{-\beta(E_\Psi-\mu_{\text{id,e}}Z_\Psi^*)}
=e^{\beta\mu_{\text{id,e}}(Z-1)}\sum_{n=1}^\infty 2n^2 e^{\frac{\beta Z^2}{2n^2}\alpha^2 m_\text{e}c^2}
\\
\lim_{n\rightarrow\infty}\left(2n^2 e^{\frac{\beta Z^2}{2n^2}\alpha^2 m_\text{e}c^2}\right)=2n^2+\beta Z^2\alpha^2 m_\text{e}c^2+O\left(\frac{1}{n^2}\right)
\label{eq_divergence_Saha}
\end{align}
The two first terms of the right-hand side of Equation~\eqref{eq_divergence_Saha} lead to the divergence of the sum.

The solution to this puzzle is to be found in the distortion of the continuum wave functions and in the non-ideal corrections to the Saha equilibrium. These somehow restrict the set of states to account for in the calculation. We postpone the discussion of this point to Section~\ref{sec_bound_state_suppr} and consider that the sum is in practice truncated at some value of the principal quantum number.

Finally, a straightforward extension of the Saha model consists of replacing the classical-ideal-gas free energy for the free electrons with that of the Fermi ideal gas.
\begin{align}
&\dot{F}\left(\{n_\Psi\},T\right)
=-\sum_{\Psi=1}^{M} \frac{n_\Psi}{n_\text{i}\beta} 
\left( \ln\left(\frac{g_\Psi e^{-\beta E_\Psi}}{n_\Psi\Lambda_\Psi^3}\right) + 1 \right)
+\frac{f_\text{e}^\text{F}(n_\text{e},T)}{n_\text{i}}
\label{eq_Saha_Fermi_gas}\\
&f_\text{e}^\text{F}(n_\text{e},T)=n_\text{e}\mu^\text{F}_\text{id,e}(n_\text{e},T)-\frac{2}{3}u^\text{F}_\text{e}(n_\text{e},T)\\
&u^\text{F}_\text{e}(n_\text{e},T)=\frac{4}{\beta\sqrt{\pi}\Lambda_\text{e}^3}I_{3/2}\left(\beta\mu^\text{F}_\text{id,e}(n_\text{e})
\right)\\
&n_\text{e}=\frac{4}{\sqrt{\pi}\Lambda_\text{e}^3}I_{1/2}\left(\beta\mu^\text{F}_\text{id,e}(n_\text{e},T)\right)
\end{align}
where the sum in Equation~\eqref{eq_Saha_Fermi_gas} only runs over the $M$ ion species. $f_\text{e}^\text{F}(n_\text{e},T)$ and $u^\text{F}_\text{e}(n_\text{e},T)$ are the free and internal energies per unit volume of a Fermi gas of density $n_\text{e}$ and temperature $T$, respectively. $\mu^\text{F}_\text{id,e}(n_\text{e},T)$ is the corresponding canonical chemical potential. This approach leads to Equations~\eqref{eq_id_Saha_pops} and \eqref{eq_id_Saha_neutr}, substituting $\mu^\text{F}_\text{id,e}$ for $\mu_\text{id,e}$.

\subsection{Average-Atom Model of Isolated Ion from a Variational Perspective\label{sec_AAII}}
The case of an ideal plasma of isolated ions can also be addressed through an \emph{average-atom} approach. In this kind of approach, instead of accounting for the many-electron states in a detailed fashion, one only aims to describe the average many-electron state of the plasma, associating fractional occupation numbers with the orbitals. The finite-temperature density-functional theory~(DFT; see \cite{Hohenberg64, KohnSham65a, Mermin65}) offers a sound theoretical basis for such models. In order to model an average isolated ion, we just have to restrict interactions to the ion nucleus and bound electrons and to consider that any continuum electron participate in a uniform, non-interacting electron density $n_\text{e}$.

The free energy per ion of such a system can be written as follows:
\begin{align}
\dot{F}&\left\{\left\{p_\xi\right\},\underline{v}_\text{trial},n_\text{e};n_\text{i},T\right\}
=\dot{F}_\text{id,i}(n_\text{i},T)+\dot{F}_\text{id,e}(n_\text{e};n_\text{i},T)+\Delta F_1\left\{\left\{p_\xi\right\},\underline{v}_\text{trial}\right\}
\label{eq_AAII_free_energy}
\end{align}
where the functional dependencies are underlined. Here, the $\dot{F}_\text{id,i}$ term corresponds to the contribution of the nuclei ideal-gas and $\dot{F}_\text{id,e}$ corresponds to the contribution of the free-electron ideal gas:

\begin{align}
\dot{F}_\text{id,i}=\frac{1}{\beta}\left(\ln\left(n_\text{i}\Lambda_\text{i}^3\right)-1\right)
\ ; \ \dot{F}_\text{id,e}=\frac{n_\text{e}}{n_\text{i}\beta}\left(\ln\left(n_\text{e}\Lambda_\text{e}^3\right)-1\right)
\end{align}
where $\Lambda_\text{i}$ and $\Lambda_\text{e}$ are the nucleus and electron thermal lengths, respectively. As in the Saha model, the classical ideal-gas free energy of the electrons may also be replaced by the Fermi ideal-gas free energy:
\begin{align}
\dot{F}_\text{id,e}=\frac{f_\text{e}^\text{F}(n_\text{e},T)}{n_\text{i}}
\end{align}

The $\Delta F_1$ term corresponds to the free-energy of the average-ion electronic structure, that is: the interacting system of bound electrons and the nucleus. We treat this system using the Kohn-Sham method~\cite{KohnSham65a}, that is, we split $\Delta F_1$ into the three contributions:
\begin{align}
\Delta F_1\left\{\left\{p_\xi\right\}, \underline{v}_\text{trial};T\right\}
=\Delta F_1^0 + \Delta F_1^\text{el} + \Delta F_1^\text{xc}
\end{align}
where $\Delta F_1^0$ is the kinetic-entropic contribution to the free energy of a system of independent bound electrons, feeling an external potential $\underline{v}_\text{trial}(r)$ and having 1-electron-orbital occupation numbers $\{p_\xi\}$ that together yield the density $n(r)$ of the interacting bound electrons. $\Delta F_1^\text{el}$ is the direct electrostatic contribution, and $\Delta F_1^\text{xc}$ is the contribution of exchange and correlation to the free energy of the electronic structure. 

According to its definition, the expression of $\Delta F_1^0$ is:
\begin{align}
\Delta F_1^0\left\{\left\{p_\xi\right\}, \underline{v}_\text{trial};T\right\}
&=\sum_{\xi\text{ bound}} \left(
p_\xi \bra{\varphi_\xi} \tilde{H}_0 \ket{\varphi_\xi}
-T s_\xi\right)\\
&=\sum_{\xi\text{ bound}} \left(
p_\xi\varepsilon_\xi-\int d^3r \left\{n(r)v_\text{trial}(r)\right\}
-T s_\xi\right)\label{eq_AAII_dF10}
\end{align}
where the sum over the $\xi$-indices only runs over the bound 1-electron states (bound orbitals). $p_\xi$ is the mean occupation number of the 1-electron state $\xi$, and the corresponding contribution to the entropy of the effective non-interacting system is:
\begin{align}
s_\xi=s(p_\xi)=-k_\text{B}\left(p_\xi\ln\left(p_\xi\right)
+(1-p_\xi)\ln\left(1-p_\xi\right)\right)
\label{eq_dft_entropy}
\end{align}
$\varepsilon_\xi$ and $\varphi_\xi(\vec{r})$ are shorthand notations for $\varepsilon_\xi\left\{\underline{v}_\text{trial}\right\}$ and $\varphi_\xi\left\{\underline{v}_\text{trial};\vec{r}\right\}$, respectively. These are the eigenvalues and wave functions of the 1-electron states obtained in the trial potential $v_\text{trial}(r)$. In the non-relativistic case, they are obtained by solving the 1-electron Schrödinger equation:
\begin{align}
-\frac{\hbar^2}{2m_\text{e}}\nabla_{\vec{r}}^2\varphi_\xi(\vec{r})+v_\text{trial}(r)\varphi_\xi(\vec{r})=\varepsilon_\xi \varphi_\xi(\vec{r})
\label{eq_Schrod_1electron}
\end{align}
We take the convention of normalizing the $\varphi_\xi$ to unity.
The trial potential $v_\text{trial}(r)$ and occupation numbers $\{p_\xi\}$ are such that the density of the system of independent bound electrons is $n(r)$. In this context, $n(r)$ is a shorthand notation for $n\left\{\left\{p_\xi\right\},\underline{v}_\text{trial};r\right\}$:
\begin{align}
n\left\{\underline{v}_\text{trial},\{p_\xi\};r\right\}=\sum_{\xi\text{ bound}} p_\xi|\varphi_\xi(\vec{r})|^2\label{eq_AAII_n}
\end{align}

The direct electrostatic contribution $\Delta F_1^\text{el}$ can be written as a functional of $n(r)$:
\begin{align}
\Delta F_1^\text{el}&\left\{\left\{p_\xi\right\}, \underline{v}_\text{trial}\right\}
=\tilde{\Delta F}_1^\text{el}\left\{\underline{n}\right\}
=\ee^2\int d^3r\left\{\frac{-Z n(r)}{r}\right\}
+\frac{\ee^2}{2}\int d^3r d^3r'\left\{\frac{n(r)n(r')}{|\vec{r}-\vec{r}'|}\right\}
\label{eq_AAII_dF1el}
\end{align}
The exchange-correlation contribution can be approximated by:
\begin{align}
\Delta F_1^\text{xc}\left\{\left\{p_\xi\right\}, \underline{v}_\text{trial};T\right\}
&=\tilde{\Delta F}_1^\text{xc}\left\{\underline{n};T\right\}
=\int d^3r\left\{f_\text{xc}\left(n(r),T\right)\right\}\label{eq_AAII_dF1xc}
\end{align}
where $f_\text{xc}$ is the exchange-correlation free energy per unit volume of a homogeneous electron gas (local density approximation). 

We stress that in this model, as in the Saha model, there is a strong distinction between bound and free electrons since any electron that belongs to the continuum is considered as non-interacting, whereas bound electrons participate in $n(r)$ and interact both with the other bound electrons of the same ion and with its nucleus, as is seen from the expressions of $\Delta F_1^\text{el}$ and $\Delta F_1^\text{xc}$.

In order to obtain the equations of the model, we minimize the free energy per ion, requiring the additional constraint of overall neutrality:
\begin{align}
\dot{F}_\text{eq}(n_\text{i},T)
=\underset{p_\xi,v_\text{trial},n_\text{e}}{\text{Min}}\,&\dot{F}\left\{\{p_\xi\},\underline{v}_\text{trial},n_\text{e};n_\text{i},T\right\}
~\text{ s. t. } Z - \sum_{\xi\text{ bound}} p_\xi =\frac{n_\text{e}}{n_\text{i}}
\label{eq_minimization_AAII}
\end{align}
Performing this constrained minimization, we obtain the equations of the average-atom model of isolated-ion (AAII):
\begin{align}
&v_\text{trial}(r)=v_\text{el}(r)+\mu_\text{xc}\left(n(r),T\right)\label{eq_AAII_vtrial}\\
&p_\xi =p_\text{F}(\mu,T,\varepsilon_\xi) = \frac{1}{e^{\beta(\varepsilon_\xi-\mu)}+1}\label{eq_AAII_FermiDirac}\\
&\mu=\begin{cases}
\mu_\text{id,e}(n_\text{e},T)&\text{ (classical ideal gas)}\\
\mu_\text{id,e}^\text{F}(n_\text{e},T)&\text{ (Fermi ideal gas)}
\end{cases}
\label{eq_AAII_mu}
\\
&Z - \sum_{\xi\text{ bound}} p_\xi =\frac{n_\text{e}}{n_\text{i}}
\end{align}
where $\mu_\text{xc}(n,T)=\partial f_\text{xc}(n,T)/\partial n$, and where $v_\text{el}(r)$ is a shorthand notation for $v_\text{el}\left\{\underline{n};r\right\}$, defined as follows:
\begin{align}
v_\text{el}\left\{\underline{n};r\right\}
=\frac{\delta \tilde{\Delta F}_1^\text{el}}{\delta n(r)}=-\frac{Z\ee^2}{r}+\ee^2\int d^3r'\left\{\frac{n(r')}{|\vec{r}-\vec{r}'|}\right\}
\label{eq_def_vel}
\end{align}

As a consequence of addressing the average electronic configuration of an ion, the electronic structure has to be determined self-consistently with the occupation numbers. The calculation of the electronic structure cannot be separated from the statistical modeling, as it is in the Saha model. This may appear as a major drawback of this model but, on the other hand, the average-atom approach is intrinsically complete. There is no issue of practical limitation in the number of excited multielectron states that may be accounted for.

However, the average atom model has the same issue of divergence of the partition function as the Saha model. If one considers, for instance, the sum of the occupation numbers, in the non-degenerate limit of the Fermi-Dirac distribution:
\begin{align}
\sum_{\xi\text{ bound}} p_\xi^\text{F}=2\sum_{n=1}^\infty\sum_{\ell=0}^{n-1} (2\ell+1) e^{-\beta(\varepsilon_{n,\ell}-\mu)}
\end{align}
Assuming that the exchange-correlation potential $\mu_\text{xc}$ does not compensate the self interaction in $v_\text{el}$, we have:
\begin{align}
&\lim_{r\rightarrow\infty} v_\text{trial}(r)=-\frac{Z^*\ee^2}{r}
\ ;\ 
\lim_{n\rightarrow\infty}\varepsilon_{n,\ell} = -\frac{Z^{*\,2}}{2 n^2}\alpha^2 m_\text{e} c^2
\\
&\lim_{n\rightarrow\infty}
\sum_{\ell=0}^{n-1} 2(2\ell+1) e^{-\beta\varepsilon_{n,\ell}}
=2n^2+\beta Z^{*\,2} \alpha^2 m_\text{e}c^2+O\left(\frac{1}{n^2}\right)
\label{eq_divergence_AAII}
\end{align}

Again, we postpone a longer discussion of this point to Section~\ref{sec_bound_state_suppr} and will assume that the principal quantum number is limited somehow.

From the free energy at equilibrium, we can derive all the thermodynamic quantities of interest, in particular, the pressure:
\begin{align}
P=n_\text{i}^2\frac{\partial \dot{F}_\text{eq}}{\partial n_\text{i}}
=&
\begin{cases}
n_\text{i}k_\text{B} T+n_\text{e}k_\text{B} T &\text{(classical ideal gas)}
\\
n_\text{i}k_\text{B} T
-f_\text{e}^\text{F}(n_\text{e},T)
+n_\text{e}\mu^\text{F}_\text{id,e}(n_\text{e},T) &\text{(Fermi ideal gas)}
\end{cases}
\label{eq_AAII_pressure_classical}
\end{align}
This corresponds to the pressure of the ideal-gas mixture of ions and free-electrons. It may be shown easily (the method is described in~\cite{Slater33}) that the virial theorem is fulfilled in this model.

To conclude about this derivation, let us note that instead of considering as variables the arbitrary  occupation numbers ${p_\xi}$ and trial potential $v_\text{trial}(r)$, one can consider the electron density $n(r)$ as the variable, formally inverting the relation between $v_\text{trial}(r)$ and $n(r)$. In this case, one defines $v_\text{trial}\{\underline{n},n_\text{e};r\}$ as the external potential yielding the density $n(r)$ for a system of independent particle at equilibrium, that is, with $p_\xi=p_\text{F}(\mu,T,\varepsilon_\xi)$. This corresponds more closely to the usual standpoint of DFT, and we will use this one in the following derivations.

\begin{figure}[h]
\begin{center}
\includegraphics[width=8cm] {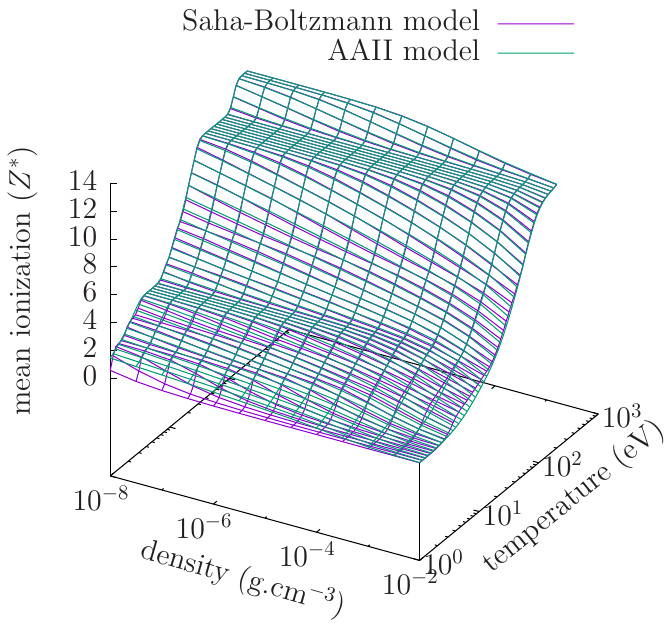} 
\end{center}
\caption{Mean ionization $Z^*$ of silicon stemming from the Saha equilibrium model, with detailed configuration accounting for the ion electron states, and from the average-atom model of isolated ion (AAII).
\label{fig_Saha_AAII}}
\end{figure}

Figure~\ref{fig_Saha_AAII} presents a comparison between the mean ionization: $Z^*\equiv n_\text{e}/n_\text{i}$ obtained from the Saha equilibrium model, using a detailed configuration accounting for the ion electron states, and the present average-atom model of isolated ion, for the case of silicon, with an arbitrary limitation of the principal quantum number to $n\leq 8$. Though the results differ slightly, they are rather close. The qualitative behavior of decreasing mean ionization when the density increases is similar, clearly exhibiting the lack of pressure ionization in these models.

%-------------------------
\section{Non-ideality Corrections to Isolated Ions\label{sec_nonideality_corrections}}
%-------------------------  

\subsection{Notion of Ionization-Potential Depression}
Non-ideal plasmas specifically correspond to those plasmas for which the interactions of the ions with the surrounding ions and free electrons cannot be disregarded.

In first approximation, one may assume that the internal structure of the ions remains the same and simply refines the ideal-gas approximation for the system of point-like ions and free electrons (ion-free-electron plasma) by accounting for the interaction energy.
Adding the interaction energy, one has the Hamiltonian:
\begin{align}
&\mathcal{H}(\{\vec{p}_{\Psi,j},\vec{r}_{\Psi,j}\},\{\vec{p}_{\text{e},j},\vec{r}_{\text{e},j}\})=
\mathcal{H}_\text{id}(\{\vec{p}_{\Psi,j}\},\{\vec{p}_{\text{e},j}\})
+\mathcal{W}_\text{IFE}(\{\vec{r}_{\Psi,j}\},\{\vec{r}_{\text{e},j}\})
\end{align}
\begin{align}
\mathcal{W}_\text{IFE}(\{\vec{r}_{\Psi,j}\},\{\vec{r}_{\text{e},j}\})
=&\frac{\ee^2}{2}
\sum_{\Psi=1}^{M}\sum_{j=1}^{N_\Psi}
\mathop{ \sum_{\Psi'=1}^{M}\sum_{k=1}^{N_{\Psi'}} }_{(\Psi',k)\neq (\Psi,j)}
\frac{Z^*_\Psi Z^*_{\Psi'}}{| \vec{r}_{\Psi,j} - \vec{r}_{\Psi',k} |}
-\ee^2\sum_{\Psi=1}^{M}\sum_{j=1}^{N_\Psi}\sum_{k=1}^{N_\text{e}}
\frac{Z^*_\Psi}{| \vec{r}_{\Psi,j} - \vec{r}_{\text{e},k} |}\nonumber\\
&+\frac{\ee^2}{2}\sum_{j=1}^{N_\text{e}}\mathop{\sum_{k=1}^{N_\text{e}}}_{k\neq j}\frac{1}{| \vec{r}_{\text{e},j} - \vec{r}_{\text{e},k} |}
\label{eq_hamiltonian_IFE}
\end{align}

Since the ions are assumed to be point-like in this approach, interactions can only be sensitive to the ion charge rather than to the detail of its electronic structure. All species $\Psi$ sharing a same ion charge $Z^*_\Psi$ behave the same. Consequently, changes only occur in energy differences among different charge states, yielding the notion of ionization-potential depression (IPD) or continuum lowering.

Two ways of accounting for this interaction energy are often described in the literature. They mostly lead to the same results and rather pertain to different standpoints on the problem than to strictly distinct approaches.

A first approach is to evaluate the average potentials $v_\Psi^*$ per unit charge, acting on the various point-like particles of the plasma, due to the interactions with all the other particles of the plasma. If we set $v_\Psi(\vec{r})$ to be the average potential around a particle of species $\Psi$, we have:
\begin{align}
v_\Psi^*=\lim_{r\rightarrow 0} \left( v_\Psi(\vec{r})-\frac{Z_\Psi \ee^2}{r}\right)
\end{align}

For each ion, one then adds the corresponding $v_\Psi^*$ potential to the binding energy $E_\Psi$ found from the Hamiltonian $\hat{H}_\text{isol.ion}^Q$ pertaining to the electronic structure. This may be interpreted as adding a constant perturbing potential to the Hamiltonian. This interpretation is, for instance, used in~\cite{StewartPyatt66}. The nucleus and each bound electron of the ion are subject to the same perturbing potential. This is consistent with the point-like ion hypothesis made in Equation~\eqref{eq_hamiltonian_IFE}. Of course, this hypothesis ceases to be relevant if $v_\Psi(\vec{r})$ varies appreciably over the scale of the electronic structure of the ion. In the context of detailed modeling, this approach results in substituting for the energies $E_\Psi$ in Equation~\eqref{eq_id_Saha_pops}:
\begin{align}
E_\Psi^*=E_\Psi+Z^*_\Psi v_\Psi^*
\end{align}
The correction to the ionization potential is thus:
\begin{align}
\Delta I_\Psi&=Z^*_{\Psi'} v_{\Psi'}^* - Z^*_\Psi v_\Psi^* \text{ with } Z^*_{\Psi'}=Z^*_\Psi+1
\\
&\approx v_{\Psi'}^*
\end{align}
In the average-atom model, each of the independent electrons (of charge $-1$) is subject to the same perturbing potential $v^*$. The energies of the orbitals, in Equation~\eqref{eq_AAII_FermiDirac}, then become:
\begin{align}
\varepsilon_\xi^*=\varepsilon_\xi-v^*
\end{align}
with $v^*$ calculated for the average ion charge $Z^*$.
%, and is often the first to come in mind when the expression ``continuum lowering'' is used.
 
However, this approach does not correspond to a unified treatment of both the atomic structure of the ions and the interactions among the particles of the plasma. Such a transposition of the interaction potentials in the plasma, stemming from a particular model, to the effective potentials pertaining to the ions' electronic structure is not formally justified. This procedure shifts the energy reference of each charge state without modifying the spectrum or the orbitals of the electronic structure \textit{per se}.

A slightly different standpoint, maybe less heuristic, on the accounting for the interaction energy is to add an approximate interaction contribution to the free energy of the ideal plasma. In statistical mechanics, such a contribution is called an \emph{excess free energy}.

In the detailed approach, Equation~\eqref{eq_Saha_free_energy} becomes:
\begin{align}
&\dot{F}\left(\{n_\Psi\},n_\text{e},T\right)
=\dot{F}_\text{id}\left(\{n_\Psi\},n_\text{e},T\right) + \dot{F}_\text{ex}\left(\{n_\Psi\},n_\text{e},T\right)
\label{eq_Saha_correction_Fex}
\end{align}
This results in adding excess chemical potentials $\mu_{\text{ex},\Psi}$, $\mu_{\text{ex,e}}$ in the chemical-equilibrium Equation~\eqref{eq_Saha_chemical_equilibrium}. This standpoint is, for instance, adopted in~\cite{Griem62,Hummer88}.
\begin{align}
&\mu_{\text{id},\Psi}(n_\Psi,T)+\mu_{\text{ex},\Psi}(\{n_\Psi\},n_\text{e},T)
+\left(\mu_{\text{id,e}}(n_\text{e},T)+\mu_{\text{ex,e}}(\{n_\Psi\},n_\text{e},T)\right) Z_\Psi^* =\lambda_\text{i}
\label{eq_Saha_chemical_equilibrium_excess}
\end{align}
In this framework, the effective correction to the ionization potential appears when one calculates, for instance, the ratio of populations for levels belonging to neighboring charge states. It results from the corrections to the chemical potentials of the neighboring charge states, and free electrons:
\begin{align}
\Delta I_\Psi
&= \mu_{\text{ex},\Psi'}-\mu_{\text{ex},\Psi}+\mu_{\text{ex},\text{e}} \text{ with $Z_{\Psi'}=Z_\Psi+1$}
\end{align}

In the average-atom case, one simply adds the excess free energy $\dot{F}_\text{ex}(n_\text{e},n_\text{i},Z^*,T)$ corresponding to the plasma of free electrons and a sole species of ions having the average charge $Z^*$, which depends on $n_\text{e}$. Equation~\eqref{eq_AAII_free_energy} becomes:
\begin{align}
\dot{F}&\left\{\left\{p_\xi\right\},\underline{v}_\text{trial},n_\text{e};n_\text{i},T\right\}
=\dot{F}_\text{id,i}(n_\text{i},T)+\dot{F}_\text{id,e}(n_\text{e};n_\text{i},T)+\dot{F}_\text{ex}(n_\text{e},n_\text{i},Z^*=n_\text{e}/n_\text{i},T)
+\Delta F_1\left\{\left\{p_\xi\right\},\underline{v}_\text{trial}\right\}
\label{eq_AAII_Fex_correction}
\end{align}
and Equation~\eqref{eq_AAII_mu} becomes
\begin{align}
\mu=\mu_\text{id,e}(n_\text{e},T)+\frac{\partial}{\partial n_\text{e}}\dot{F}_\text{ex}(n_\text{e},n_\text{i},Z^*=n_\text{e}/n_\text{i},T)
\end{align}
where one must account for the dependency of $Z^*$ on $n_\text{e}$ when calculating the derivative.

The underlying hypothesis common to these approaches is that the ion is point-like when compared to the typical inter-particle distance $d$ in the ion-free-electron plasma. 
\begin{align}
r_\text{outer} << d
\end{align}
In a detailed model, $r_\text{outer}$ is the largest mean radius of any populated orbital of any significantly populated multielectron state. In an average-atom model, $r_\text{outer}$ is the mean radius of the outer significantly populated orbital. The relevant inter-particle distance $d$ depends on the model used for the ion-free-electron plasma.

The point-like-ion hypothesis is the key argument in separating the modeling of interactions in the plasma from the modeling of the ion electronic structure. However, it excludes the case where some populated orbitals of an ion are perturbed by the effect of its surrounding particles, that is, the case of \emph{pressure-ionized} plasmas.

\subsection{Models for the Ionization-Potential Depression}
In order to avoid the classical Coulomb catastrophe, well-chosen hypotheses have to be made in order to approach the $\mathcal{W}_\text{IFE}$ term using classical mechanics. We summarize two broadly used models. However, other ways of circumventing the Coulomb collapse were also explored (see, e.g. \cite{Deutsch78}).

\subsubsection{Debye-Hückel Model}
The Debye-Hückel model~\cite{DebyeHuckel23} (DH) is the linearized mean-field approach to the ion-free-electron plasma. It accounts for the attractive interactions between ions and electrons and is valid in the limit of weakly coupled plasmas. The linearization with respect to the mean-field potential is strongly unjustified at short distances and leads to an unphysical behavior of correlation function at the origin. However, it allows one to circumvent the classical Coulomb catastrophe. 

In the DH model, the average potentials $v_\Psi^\text{DH}$, $v_\text{e}^\text{DH}$, excess free energy per ion $\dot{F}_{\text{ex}}^\text{DH}$, and chemical potentials $\mu_{\text{ex},\Psi}^\text{DH}$, $\mu_{\text{ex,e}}^\text{DH}$ can be written as:
\begin{align}
v_{\Psi}^\text{DH}(r)=Z^*_\Psi \ee^2 \frac{e^{-r/\lambda_\text{D}}}{r}\ ;\ 
v_{\text{e}}^\text{DH}(r)=-\ee^2 \frac{e^{-r/\lambda_\text{D}}}{r}\ ;\ 
\dot{F}_{\text{ex}}^\text{DH}=-\frac{1}{12\pi\beta n_\text{i}\lambda_\text{D}^3}\ ;\ \mu_{\text{ex},\Psi}^\text{DH}=-\frac{1}{2}\frac{Z^{*\,2}_\Psi \ee^2}{\lambda_\text{D}}
\ ;\ \mu_{\text{ex,e}}^\text{DH}=-\frac{1}{2}\frac{\ee^2}{\lambda_\text{D}}
\label{eq_DH_free_energy}
\end{align}
where $\lambda_\text{D}=\left( 4\pi \beta \ee^2 \left(n_\text{e}+\sum_{\Psi=1}^{M}n_\Psi Z_\Psi^2\right) \right)^{-1/2}$ is the Debye length.

It can be shown that the DH model fulfills the virial theorem of Equation~\eqref{eq_virial_theorem}. Thus, adding the excess free-energy $\dot{F}_{\text{ex}}^\text{DH}$ to the Saha model, as in Equation~\eqref{eq_Saha_correction_Fex}, preserves the virial theorem.
The correction to the ionization potentials is (see, for instance, \cite{Griem62}):
\begin{align}
&\Delta I_\Psi^\text{DH} = -\frac{(Z^*_\Psi + 1)\ee^2}{\lambda_\text{D}}
\end{align}
The latter correction is equivalent to applying the correction $v_{\Psi}^{*\text{DH}}$ using the upper charge state, as in~\cite{StewartPyatt66}.

In the average-atom context, the excess free energy of Equation~\eqref{eq_AAII_Fex_correction} is:
\begin{align}
\dot{F}_{\text{ex}}^\text{DH}=-\frac{1}{12\pi\beta n_\text{i}}\left(4\pi \ee^2 \beta \left(n_\text{e}+n_\text{i}\frac{n_\text{e}^2}{n_\text{i}^2}\right)\right)^{3/2}
\label{eq_DH_free_energy_AA}
\end{align}
Accounting for this interaction correction leads to shifting the 1-electron eigenvalues as follows:
\begin{align}
&\varepsilon_\xi^* =\varepsilon_\xi + \frac{(Z^*+1/2)\ee^2}{\lambda_\text{D}}
\label{eq_DH_AA_correction}
\end{align}

\subsubsection{Ion-Sphere Model}
A model often used to address the case of strongly coupled plasmas is the ion-sphere model (IS; see, for instance, \cite{StewartPyatt66}). In this model, one considers a point-like ion placed at the center of a sphere filled only with a frozen, uniform background of free electrons, with which it interacts. The uniform density of free electrons corresponds to the mean free-electron density of the plasma $n_\text{e}=\sum_\Psi n_\Psi Z_\Psi^*$, while the radius of the sphere is such that the ion sphere is neutral: 
\begin{align}
R_{Z^*_\Psi}=\left(\frac{3 Z_\Psi^*}{4\pi n_\text{e}}\right)^{1/3}
\end{align}
In the case of an average ion, $n_\text{e}=n_\text{i} Z^*$, the ion charge $Z_\Psi^*$ is replaced by $Z^*$ and the sphere radius is just the Wigner-Seitz radius $R_\text{WS}=(3 /(4\pi n_\text{i}))^{1/3}$.

In the IS model, the interaction energy of the central ion with the surrounding electrons in the sphere yields the correction $v_\Psi^{*\,\text{IS}}$ to the energy of the $\Psi$ species. The total interaction energy of the ion sphere, which corresponds to the energy added if one adds an ion to the system, gives the excess chemical potential.
\begin{align}
v_\Psi^{\text{IS}}(r)=
\frac{Z^{*}_\Psi \ee^2}{r}
+\frac{1}{2}\frac{Z^{*}_\Psi \ee^2}{R_{Z^*_\Psi}}\frac{r^2}{R_{Z^*_\Psi}^2}
-\frac{3}{2}\frac{Z^{*}_\Psi \ee^2}{R_{Z^*_\Psi}}
\ ;\ 
v_\Psi^{*\,\text{IS}}
=-\frac{3}{2}\frac{Z^{*}_\Psi \ee^2}{R_{Z^*_\Psi}}
\ ;\ 
\mu_{\text{ex},\Psi}^\text{IS} =-\frac{9}{10}\frac{Z^{*\,2}_\Psi \ee^2}{R_{Z^*_\Psi}}
\end{align}
This leads to the following correction to the ionization potentials:
\begin{align}
\Delta I_\Psi^\text{IS}&=-\frac{9}{10}\left(\frac{(Z_\Psi^*+1)^{2} \ee^2}{R_{Z^*_\Psi+1}}-\frac{Z^{*\,2}_\Psi \ee^2}{R_{Z^*_\Psi}}\right)\approx -\frac{3}{2}\frac{Z^{*}_\Psi \ee^2}{R_{Z^*_\Psi}}\text{ to first order in $Z^*$}
\end{align}
which, to first order, is equivalent  to applying the correction $v_{\Psi}^{*\text{IS}}$, as in~\cite{StewartPyatt66}.

On the interpretation of the IS model, two different physical pictures may be put forward. In the first picture, the plasma is seen as a highly structured set of neutral spheres, somehow resembling a solid-state situation (see, for instance, \cite{StewartPyatt66}). Another interpretation is that the medium surrounding the ion may be split in two regions: a spherical statistical cavity in which other ions do not enter, and a uniform neutral plasma beyond the cavity. This resembles more a liquid-state picture.

The hypothesis of the IS model which allows circumventing the Coulomb catastrophe is to neglect the polarization of the free electrons driven by the attractive long range potential. Free electrons then constitute a rigid uniform background neutralizing the ions. Rigorously speaking, it is applicable only when free electrons' degeneracy is so strong as to inhibit their polarization around the ions. 
This hypothesis is also the founding hypothesis of classical plasma models: one-component classical plasma (OCP) or multi-component classical plasma (MCP). However, in the IS model, the average ion density around the central ion is approached by a Heaviside function. This corresponds to the strong-coupling limit of the \emph{mean-field} model of the OCP/MCP (often called nonlinear Debye-Hückel model or Poisson-Boltzmann model).

However, in the case of strong coupling, the effect of correlations, beyond the reach of the mean-field approximation, is important. Models of simple fluid that account for these effects, such as the hyper-netted-chain model (HNC)~\cite{Morita58,Morita59}, exhibit a different behavior of the pair distribution functions for strongly coupled plasmas, as well as a different limit for the chemical potential. These models of fluid also better approach the Wigner crystal at conditions were the OCP/MCP system is crystallized. In this limit, the ion-sphere model of a lattice provides a relevant approximation. However, despite the similar name and the very close proximity of the equations, it is a different model from the IS model of fluid. Thus, whereas the DH model of ion-free-electron plasma can be seen as rigorously valid in the limit of weak coupling, the IS model of fluid only offers a qualitative description of strongly coupled classical plasmas.

\subsubsection{Stewart-Pyatt Model}
Despite the fact that ion sphere model is questionable for strongly coupled plasmas, the description of moderately coupled plasmas was often seen as a question of bridging between the DH and IS models, seen as two ``limits''. This question is notably addressed by Stewart and Pyatt in~\cite{StewartPyatt66}, using an approach inspired from the Thomas-Fermi model (see next section) and the mean-field model of the OCP. However, in \cite{StewartPyatt66}, the model is used to describe the electron cloud around an ion rather than the electronic structure of the ion itself. They obtain the following formula, which smoothly switches from the DH to the IS result, according to the ratio between the Debye length and the ion-sphere radius:
\begin{align}
\Delta I_\Psi^\text{SP}=
-\frac{1}{2\beta}\left(\frac{n_\text{e}}{n_\text{i}}+1\right)^{-1}
\left(\left[ \left(\frac{R_{Z_\Psi^*}}{\lambda^*_\text{D}}\right)^3+1\right]^{2/3}-1 \right)
\end{align}
with $\lambda^*_\text{D}=\left(4\pi\beta\ee^2 n_\text{e}\left(n_\text{e}/n_\text{i}+1\right)\right)^{-1/2}$.

\section{Divergence of Partition Functions, Suppression of Bound States, Screening and Limitations of the Non-ideality Corrections\label{sec_bound_state_suppr}}

The isolated-ion models of Section \ref{sec_ideal_plasmas} both exhibit divergences of their partition functions. One may relate this divergence issue to the Coulombic behavior of the atomic potential at large distances, typical of isolated-ion models.

First, let us remark that in the $n\rightarrow\infty$ limit, the 1-electron spectrum of the average-atom model of isolated ion forms a quasi-continuum having a density of states $\rho_\text{qc}$, which diverges for energies $\varepsilon\rightarrow 0^-$:
\begin{align}
\rho_\text{qc}(\varepsilon)=\frac{Z^{*\,2}}{\varepsilon^2}
\end{align}
The partition function of high-lying levels $\varepsilon>\varepsilon_\text{qc}$ may be seen as the diverging integral:
\begin{align}
\int_{\varepsilon_\text{qc}}^0 d\varepsilon \left\{ \rho_\text{qc}(\varepsilon)p^\text{F}(\mu,T,\varepsilon)\right\}
\end{align}
However, by limiting the partition function to the negative-energy part of spectrum, one disregards the distortion of the density of states induced by the atomic potential in the positive-energy part.

Let us consider two 1-electron Hamiltonians: the Hamiltonian of free electrons $\tilde{H}_0$ and the Hamiltonian $\tilde{H}_\text{eff}$ of electrons feeling an external effective potential. Eigenstates of each of those Hamiltonians constitute a complete orthonormal basis of the \emph{same} 1-electron-state space $\mathcal{E}$. The total number of states within each of these bases is thus the same. If those two bases share a common continuous label, for instance $\varepsilon$, then the density of states with respect to this label may change, but its integral over the label remains the same. This principle is the foundation of Levinson's theorem~\cite{Levinson49,Jauch57}. From this principle, one may expect the divergence of the density of states induced by a Coulomb-tail potential at $\varepsilon\rightarrow 0^-$ to be compensated elsewhere; more precisely, in the continuum.

A spherically symmetric potential does not couple the subspaces associated with the various orbital quantum number $\ell$. Consequently, it is possible to show that the total number of states is conserved for each value of $\ell$ (see, for instance,~\cite{Jauch57,Zhong-Qi06}). Moreover, for a spherically symmetric potential, the distortion of the density of states with respect to that of free particles is related to the \emph{scattering phase shift}.

Let us consider the radial wave functions of states $\xi=(\varepsilon,\ell,m)$ belonging to the continuum:
\begin{align}
&\varphi_{\varepsilon,\ell,m}(\vec{r})=\frac{P_{\varepsilon,\ell}(r)}{r}Y_{\ell,m_\ell}(\hat{r})\\
&\int_0^\infty dr\left\{P_{\varepsilon,\ell}(r)P_{\varepsilon',\ell}(r)\right\}=\delta(\varepsilon-\varepsilon')\text{ (normalization convention)}
\end{align}
In the case of a free particle (zero external potential, Hamiltonian $\tilde{H}_0$), the Schrödinger radial equation is the Bessel equation and the radial wave functions are:
\begin{align}
P_{\varepsilon,\ell}(r)=A_{\varepsilon}kr j_\ell(kr)\text{ with $k=\sqrt{2m_\text{e}\varepsilon}$}
\end{align}
where $j_\ell$ is the spherical Bessel function regular at 0, $y_\ell$ denoting the irregular one in the following~\cite{AbramowitzStegun}.
We define the local phase shift $\Delta_{\varepsilon,\ell}(R)$ of a continuum radial wave function $P_{\varepsilon,\ell}(r)$ with respect to the regular Bessel function as follows. Setting the potential to zero for $r\ge R$, $\Delta_{\varepsilon,\ell}(R)$ is such that
\begin{align}
&P_{\varepsilon,\ell}(r\ge R) = A_{\varepsilon}kr\left[ \cos(\Delta_{\varepsilon,\ell}(R)) j_\ell(kr)
-\sin(\Delta_{\varepsilon,\ell}(R)) y_\ell(kr) \right]
\label{eq_matching}
\end{align}

For a finite-range potential, the Schrödinger equation tends to the Bessel equation far from the origin, and the local phase shift has a finite asymptotic value, which is the scattering phase shift $\Delta_{\varepsilon,\ell}$.
\begin{align}
\lim_{r\rightarrow\infty}\Delta_{\varepsilon,\ell}(r)=\Delta_{\varepsilon,\ell}
\end{align}
The latter scattering phase shift is related to the change of the density of state as follows (see, for instance,~\cite{KittelQuantumTheoryofSolids}):
\begin{align}
&\Delta\rho_\ell(\varepsilon)=2 (2\ell+1)\frac{1}{\pi}\frac{\partial\Delta_{\varepsilon,\ell}}{\partial \varepsilon}
\label{eq_phaseshift_dos}
\end{align}
Applying the conservation of the total number of states for a given $\ell$, and remembering that there is no bound state for a free electron (no classically allowed region for $\varepsilon<0$), one immediately obtains:
\begin{align}
&\int_{-\infty}^0 d\varepsilon \left\{ \sum_{n=1}^{n_{\text{max},\ell}}2(2\ell+1)\delta(\varepsilon-\varepsilon_{n,\ell})\right\}
=-\int_{0}^\infty d\varepsilon \left\{\Delta\rho_\ell(\varepsilon) \right\}
\\
&n_{\text{max},\ell}=\lim_{\varepsilon\rightarrow 0}\frac{\Delta_{\varepsilon,\ell}}{\pi}
\end{align}
which corresponds to Levinson's theorem. For a finite-range potential, the number of discrete orbitals is thus finite~\cite{Bargmann52}.

For a Coulomb-tail potential, the solution of the Schrödinger equation tends to a combination of regular and irregular Coulomb wave functions:
\begin{align}
\lim_{r\rightarrow\infty}P_{\varepsilon,\ell}(r) = A_{k}\left[ \cos(\Delta_{\varepsilon,\ell}^\text{C}) F_\ell^\text{C}\left(-\frac{Z^*}{k};kr\right)
+\sin(\Delta_{\varepsilon,\ell}^\text{C}) G_\ell^\text{C}\left(-\frac{Z^*}{k};kr\right) \right]
\end{align}
where $F_\ell^\text{C}$, $G_\ell^\text{C}$ are the Coulomb wave functions regular and singular at 0, respectively~\cite{AbramowitzStegun}, and where $\Delta_{k,\ell}^\text{C}$ is a constant phase shift with respect to the Coulomb wave functions. This yields:
\begin{align}
\lim_{r\rightarrow\infty}\Delta_{\varepsilon,\ell}(r)\sim \frac{Z^*}{k}\ln(2kr)
\end{align}
This singularity of the scattering phase shift is fully consistent with the infinite number of bound states for a Coulomb-tail potential.

The detailed analysis of the compensation between bound and continuum parts of the partition function was studied thoroughly in the framework of the virial expansion by Beth and Uhlenbeck~\cite{BethUhlenbeck37}, Larkin~\cite{Larkin60}, Ebeling~\cite{Ebeling68,Ebeling74,Ebeling85} and Rogers~\cite{Rogers77,Rogers86}. Accounting for the density of state modification in the continuum yields the Planck-Larkin suppression of the diverging terms in Equations~\eqref{eq_divergence_Saha} and \eqref{eq_divergence_AAII}.
In itself, the divergence of the partition functions related to the infinite number of bound states is mostly due to a lack of proper accounting for the continuum.

However, non-ideal effects such as the screening by free electrons and the perturbation by surrounding ions may also limit \emph{physically} the range of the effective atomic potential and consequently the number of bound states. In fact, in the context of the virial expansion, the Debye-Hückel correction appears at the second order, together with the Planck-Larkin regularization (see, for instance,~\cite{Larkin60}).

For the most weakly bound states (either in the sense of many-electron states or in the sense of 1-electron orbitals), the non-ideality correction to the energy may be of the same order as the energy itself, or even greater. The question then is: how to treat these states that potentially end up in the continuum ?

In~\cite{Herzfeld16}, Herzfeld truncate heuristically the set of bound states in the Saha equilibrium for hydrogen by disregarding hydrogenic orbitals whose mean radii are larger than or of same order as the inter-particle distance. Qualitatively, this corresponds to limiting the spatial extension of existing states to the size of an ion sphere. In~\cite{Urey24,Fermi24}, Urey and Fermi independently elaborate on this idea, introducing, in the free energy, an excluded-volume term associated with the volume occupied by the various hydrogenic states. Minimizing the free energy, Fermi concludes that the population of a state drops when the total volume is smaller than the excluded volume of this state times the population of the most populated state.

In~\cite{Larkin60}, Larkin relates the truncation to an assumed Debye-Hückel decay of the 1-electron effective potential. Qualitatively, this corresponds to limiting the spatial extension of existing states to the Debye length. However, without a unified treatment for both the electronic structure and the screening in the whole plasma, the argument for using the Debye-Hückel potential in the ion electronic structure is heuristic.

In~\cite{StewartPyatt66}, Stewart and Pyatt recommend suppressing any level which has an occupied orbital whose energy is smaller than the ionization-potential depression.

One may put forward that, if one sees the non-ideality correction as a lowering of the continuum, then the intersection of this lowered continuum boundary with the isolated-ion potential sets a restriction on the range of the potential, yielding the truncation of the bound state spectrum. This qualitative standpoint is fully consistent with that of suppressing the states whose energy lies above the lowered continuum. 

\begin{figure}[t]
\includegraphics[width=8cm] {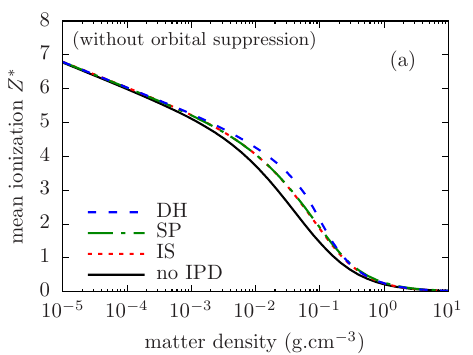} 
\hspace{1cm}
\includegraphics[width=8cm] {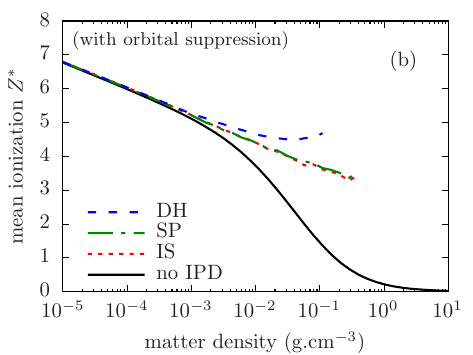} 
\caption{Mean ionization $Z^*$ of a silicon plasma at 20 eV temperature, as a function of matter density. Comparison between average-atom model of isolated ion without any continuum lowering (no IPD, principal quantum number limited to $n=8$), with Debye-Hückel continuum lowering (DH), with ion-sphere continuum lowering (IS), and with Stewart-Pyatt continuum lowering (SP). Comparisons are shown both without suppression of bound orbitals ({a}) and with suppression ({b}). In the latter case, the curves stop where suppression of a subshell having more than 10\% of the electrons occurs (regime of significant pressure ionization).
\label{fig_DH_IS_suppr}}
\end{figure}

Most of the effect of accounting for non-ideality corrections is in the modification of the partition function. Figure~\ref{fig_DH_IS_suppr} shows the effect of various ionization potential depression models, either disregarding or performing the suppression of bound orbitals. Looking at Figure~\ref{fig_DH_IS_suppr}\,a, one can see that without suppressing any orbital, the effect on the mean ionization remains moderate, the Debye-Hückel model yielding the largest effect. In the case considered, the Stewart-Pyatt formula leads to results that are close to those of the ion-sphere model. Looking at Figure~\ref{fig_DH_IS_suppr}\,b, one can see that performing the associated suppression of bound orbitals greatly increases the impact on the mean ionization.

\begin{figure}[p]
\begin{center}
\includegraphics[width=8cm] {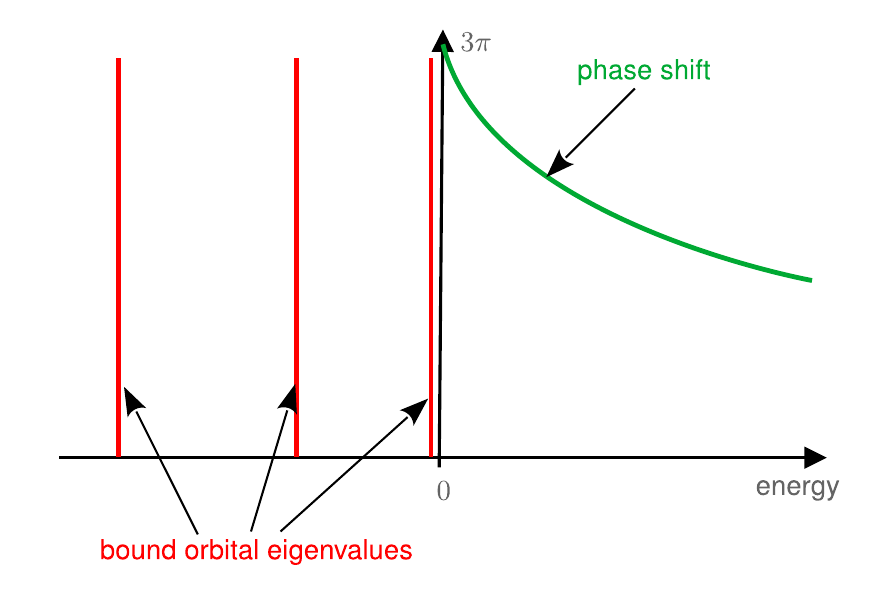} 
\includegraphics[width=8cm] {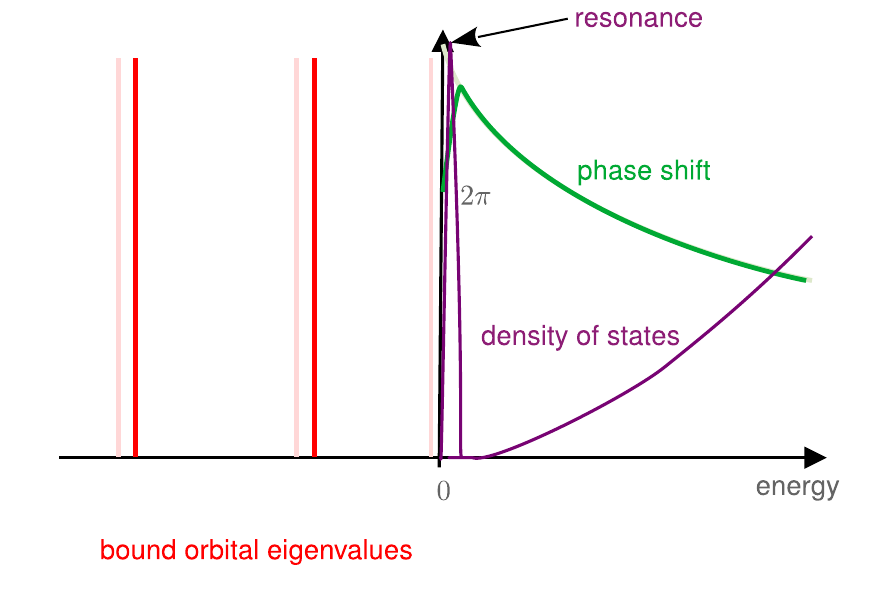} 
\end{center}
\caption{Schematic view of the delocalization of a bound orbital, with a resonance appearing in the continuum, as described by Levinson's theorem. 
\label{fig_schema_resonance}}
\end{figure}
\begin{figure}[p]
\begin{center}
\includegraphics[width=9cm] {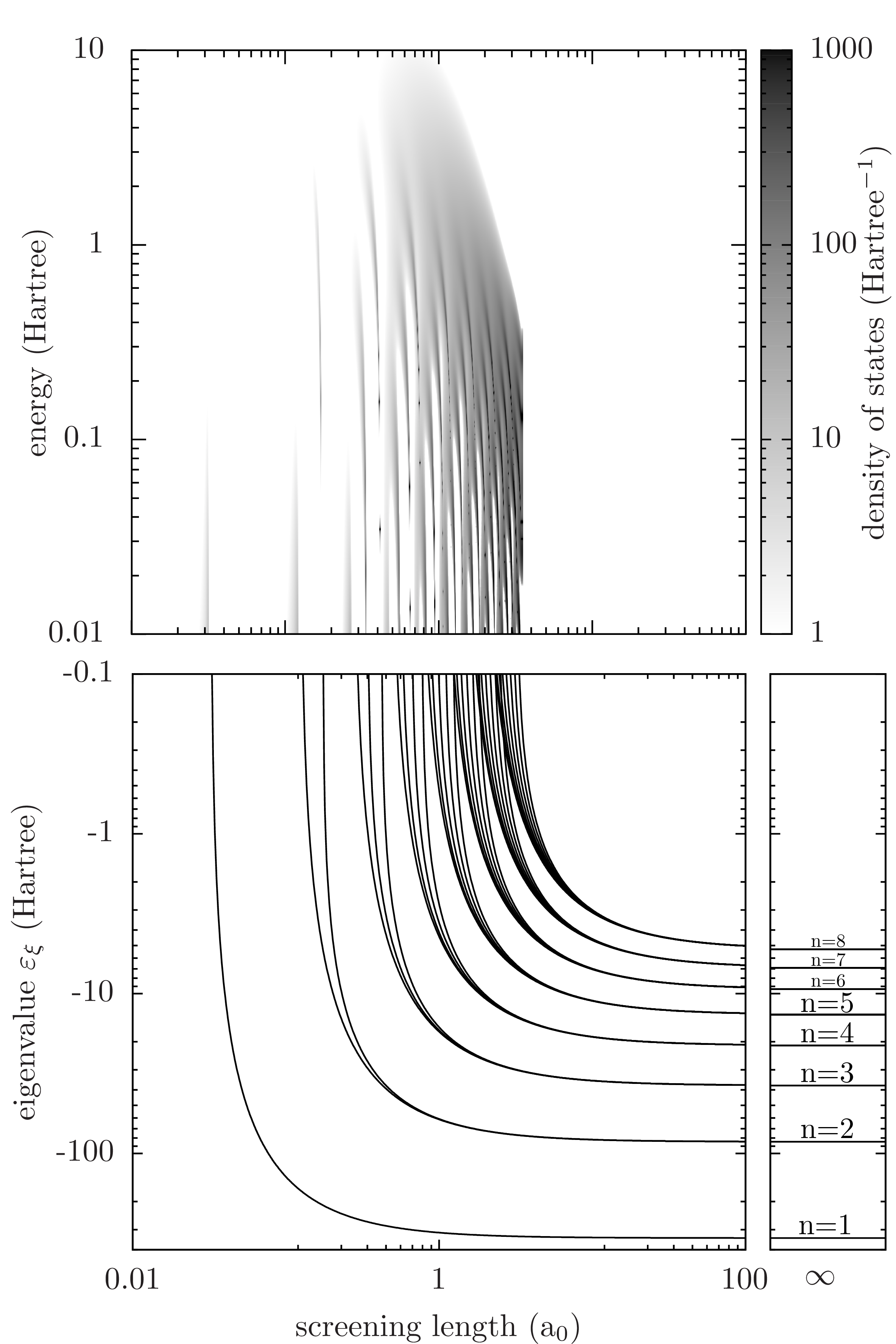} 
\end{center}
\caption{Electron in a screened Coulomb potential with charge $Z=26$. Eigenvalues as functions of the screening length, for principal quantum numbers up to $n=8$, and ion contribution to the density of states $\Delta\rho(\varepsilon)$ of Equation~\eqref{eq_phaseshift_dos}, showing the corresponding resonances. 
\label{fig_Yukawa_screening}}
\end{figure}

However, in the context of a finite-range potential, a correct accounting for the effect of the potential on the continuum remains a crucial issue. In this context, even if there is no singularity of the density of state to compensate, a compensation still occurs between \emph{finite} quantities, through the appearance of resonances.

For any small perturbation of the potential that leads to the passing of a bound state above the continuum limit, Levinson's theorem informs us that the phase shift at zero energy jumps accordingly. Since this small perturbation cannot lead to a modification of the phase-shift at energies that are large compared to the perturbation, a rapid variation of the phase shift necessarily occurs just above zero energy. This rapid variation corresponds to a sharp peak in the continuum density of states, called a resonance (see Figure~\ref{fig_schema_resonance}). In the end, the density of state, defined over the whole spectrum with Dirac $\delta$ in the negative energy part, just evolves continuously when one of the Dirac $\delta$'s crosses the zero energy. In this way, any observable remains continuous.

As an elementary illustration of the effect of a finite-range potential, Figure~\ref{fig_Yukawa_screening} presents the 1-electron eigenvalues and the total $\Delta\rho(\varepsilon)=\sum_\ell \Delta\rho_\ell(\varepsilon)$ contribution to the density of states for a screened Coulomb potential, with charge $Z=26$, as a function of the screening length. For infinite screening length, one recovers the usual hydrogen-like eigenvalues, with their accidental $\ell$-degeneracy. As the screening length is decreased, one can see how the accidental degeneracy is removed and how the eigenvalues are gradually shifted towards the continuum until the removal of the orbital from the discrete spectrum. Once the orbital is removed from the discrete spectrum, one can see how it is compensated for by a resonance in the continuum, which gradually spreads out as screening length is further decreased.

In conclusion, even assuming that a continuum-lowering model is yielding the correct energy shift, a sharp suppression of the bound states does not correspond to what stems from a screened potential with a proper accounting for the continuum.

In practice, the continuum-lowering argument or the introduction of Planck-Larkin partition functions suffices to justify the suppression of weakly bound states, which have negligible populations and do not yield significant contribution of the corresponding resonances. This suppression enables the convergence of the Saha partition function. However, when it comes to suppressing populated many-electron states or orbitals, this method is no longer valid, and in fact the whole point-like-ion hypothesis used in the treatment of interactions breaks down.

Whenever the interactions of the ions with the surrounding ions and free electrons have an impact on the electronic structure, we will speak of \emph{pressure-ionized plasma}. In this case, giving a relevant answer to the problem requires one to account for the interactions among particles of the plasma directly in the calculation of the ion electronic structure, while properly accounting for the continuum. As far as possible, such a description should account for both the polarization of free electrons around the ions, and the interactions of ions with their neighbors.

%-------------------------
\section{Atomic Models of Pressure-Ionized Plasmas\label{sec_pressure_ionized_plasmas}}
%-------------------------
In the following, we will focus on the average-atom description of the plasma since it offers a simpler framework for such modeling. However, most of the presented models can be extended to a detailed description of the plasma.

The modeling of dense, pressure-ionized plasmas has historically been addressed using self-consistent-field models of the ion electronic structure, including all electrons (bound and continuum) and accounting for the surrounding ions through the notion of a Wigner-Seitz cavity. These models focus on the electronic structure around a bare nucleus and depart from the formalism of correlations in the plasma. 
Being models of the ion electronic structure, all these models necessarily rely, to some extent, on quantum mechanics for the electrons and thus avoid the Coulomb catastrophe.

Depending on the model, the WS cavity is seen either as a neutral spherical cell in which the ion is enclosed (ion-in-cell picture) or as a statistical cavity within which surrounding ions do not enter and beyond which they are uniformly distributed (ion-in-jellium picture). 

A common feature of these models of pressure-ionized plasmas is that the resulting atomic potential has finite range, and thus naturally leads to a finite number of bound states.

\subsection{Thomas-Fermi Ion-in-Cell Model}

\begin{figure}[t]
\begin{center}
\includegraphics[width=8cm] {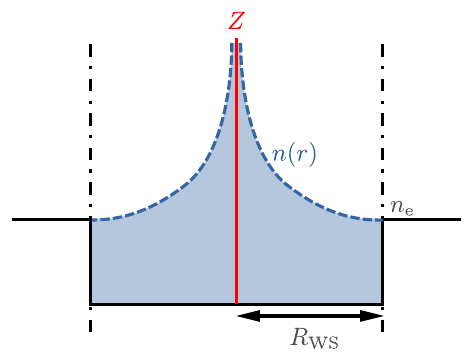} 
\end{center}
\caption{Schematic picture of the Thomas-Fermi model. 
\label{fig_schema_TF}}
\end{figure}

The Thomas-Fermi (TF) model is a semiclassical mean-field model of the ion electronic structure, accounting for all electrons and for the surrounding ions through the notion of an ion cell (see Figure~\ref{fig_schema_TF}). In the TF model, the electrons inside the ion cell are seen as a negatively charged fluid, behaving locally as an ideal gas. The atomic structure stems from the hydrostatic equilibrium of this charged ideal gas around the positively charged nucleus. Application of the TF model at finite temperature to dense plasmas was first proposed in~\cite{Feynman49}. A numerical method for solving the TF set of equations was given in~\cite{Latter55}. The equations of the TF model are:

\begin{align}
&\nabla^2 v_\text{el}(r)=-4\pi \ee^2 n(r) \label{eq_TF_SCF}\\
&\lim_{r\rightarrow 0}v_\text{el}(r)=-\frac{Z\ee^2}{r} \label{eq_TF_boundary_0}\\
&v_\text{el}(R_\text{WS})=0 \label{eq_TF_boundary_WS}\\
&\int_\text{WS} d^3r\left\{n(r)\right\}=Z \label{eq_TF_neutrality_WS}\\
&n(r)=\frac{4}{\sqrt{\pi}\Lambda_\text{e}^3}I_{1/2}\left(\beta\left(\mu^\text{F}_\text{id,e}(n_\text{e})-v_\text{el}(r)\right)\right) \label{eq_TF_density}
\end{align}
where the WS denotes that the integral is performed only within the WS sphere. $I_{1/2}$ is the Fermi integral of order $1/2$.

Equation~\eqref{eq_TF_SCF} is the Poisson equation. Equations~\eqref{eq_TF_boundary_0} and  \eqref{eq_TF_boundary_WS} are the boundary conditions at the origin and at the WS radius, respectively. The latter sets the reference of the energies.
Equation~\eqref{eq_TF_neutrality_WS} is the condition of neutrality of the ion sphere. Equation~\eqref{eq_TF_density} corresponds to the local-ideal-Fermi-gas hypothesis.
Equation~\eqref{eq_TF_SCF}, together with Equation~\eqref{eq_TF_density}, may be seen as a mean-field approximation.

An oft-used extension of the TF model consists of adding a local exchange or an exchange-correlation contribution to the electrostatic potential. This is called the Thomas-Fermi-Dirac model, referring to~\cite{Dirac30} in which a local exchange term was derived. In this case, Equation~\eqref{eq_TF_density} is replaced by:
\begin{align}
&n(r)
=\frac{4}{\sqrt{\pi}\Lambda_\text{e}^3}I_{1/2}\left(\beta\left(\mu^\text{F}_\text{id,e}(n_\text{e})-v_\text{trial}(r)\right)\right)
\label{eq_TFD_density}\\
&v_\text{trial}(r)=v_\text{el}(r)+\mu_\text{xc}(n(r))-\mu_\text{xc}(n_\text{e})
\label{eq_TFD_vtrial}
\end{align}
with $\mu_\text{xc}={\partial f_\text{xc}(n)}/{\partial n}$ being the chemical potential associated with $f_\text{xc}$, an approximate exchange-correlation contribution to the free-energy per unit volume of a uniform electron gas.

Besides its heuristic setup, one may also derive the Thomas-Fermi-Dirac model from a variational principle \cite{Mermin65}. One approximates the free energy per ion as the free energy of an ion cell, filled with an electron gas, locally considered as an ideal Fermi gas of density $n(r)$.
\begin{align}
\dot{F}&\{\underline{n};n_\text{i},T\}
=\int_\text{WS} d^3r\left\{ f_\text{e}^\text{F}(n(r),T)+f_\text{xc}(n(r),T) \right\}
+\int_\text{WS} d^3r\left\{ -\frac{Z n(r) \ee^2}{r} +\frac{\ee^2}{2}\int_\text{WS} d^3r' \left\{ \frac{n(r)n(r)}{|\vec{r}-\vec{r}'|}\right\} \right\}
\end{align}
It is worth noting that this free energy does not include terms related to the ion motion or interactions. This is among the shortfalls of such kind of model, which only focuses on the electronic structure of a central ion. In a first approximation, an ion ideal-gas free energy contribution can be trivially added. However, a proper accounting for the ion-ion interactions in such a model is a far more difficult subject, on which we will elaborate later.

One performs the minimization of the free energy per ion while requiring the neutrality of the ion cell
\begin{align}
\dot{F}_\text{eq}(n_\text{i},T)
=\underset{\underline{n}}{\text{Min}}\,&\dot{F}\{\underline{n};n_\text{i},T\}
~\text{ s. t. } Z=\int_\text{WS} d^3r\left\{ n(r) \right\}
\label{eq_minimization_TF}
\end{align}
The result of this constrained minimization is equivalent to Equations~\eqref{eq_TF_SCF}-\eqref{eq_TF_neutrality_WS} and \eqref{eq_TFD_density}.

Starting from the equilibrium free energy, one can rigorously write the thermodynamic
quantities using the appropriate derivatives. One notably finds, for the pressure:
\begin{align}
P_\text{thermo}(n_\text{i},T)
=-f_\text{e}^\text{F}(n_\text{e},T)-f_\text{xc}(n_\text{e},T)
+n_\text{e}\mu^\text{F}_\text{id,e}(n_\text{e},T)+n_\text{e}\mu_\text{xc}(n_\text{e},T)
\end{align}
It can also be shown that the model fulfills the virial theorem~\cite{Feynman49}.

In the TF model, the electron density is a local function of the potential. In fact, the hypothesis of locally having an ideal Fermi gas may be recovered from a local-density approximation of the quantum kinetic free energy of independent particles~\cite{Kirzhnits57}. Since the electrostatic potential is zero at the WS radius, the density at the WS radius is equal to that obtained from the chemical potential:
\begin{align}
Z^*=\frac{n_\text{e}(R_\text{WS})}{n_\text{i}}=\frac{n_\text{e}}{n_\text{i}}
\label{eq_TF_mean_ioniz}
\end{align}

\begin{figure}[h]
\includegraphics[width=8cm] {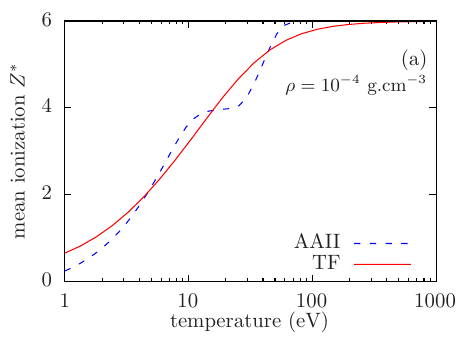} 
\hspace{1cm}
\includegraphics[width=8cm] {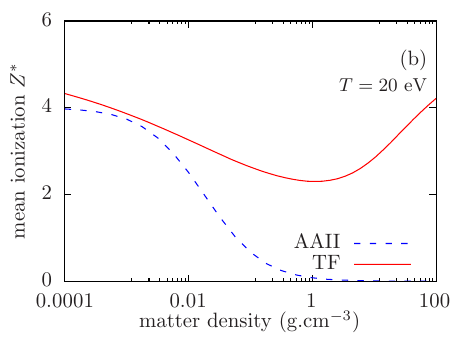} 
\caption{Mean ionization $Z^*$ of a C plasma along the $10^{-4}$-g.cm$^{-3}$-isochore ({a}), and along the 20-eV-isotherm ({b}). Comparison between the Thomas-Fermi model (TF) and the average-atom model of isolated ion (AAII).
\label{fig_TF_AAII}}
\end{figure}

Due to its semiclassical character, the TF model does not yield a shell structure in the sense of quantum mechanics. Consequently, there are no ionization plateaus, as in the mean ionization of the quantum isolated ion. As an illustration, Figure~\ref{fig_TF_AAII}\,a displays the TF mean ionization as a function of temperature for carbon at $10^{-4}$ g.cm$^{-3}$, compared to that of a quantum isolated ion.

On the other hand, in the TF model, pressure ionization is obtained through a squeezing of the ion cell when density is increased. As an example, Figure~\ref{fig_TF_AAII}\,b shows the TF mean ionization as a function of matter density for carbon at a temperature of 20 eV.

\break

In addition to being used in equation-of-state calculations, the TF model was also used for the calculation of radiative properties. Such calculations were performed either resorting to a heuristic use of the TF potential in orbital calculations~\cite{Latter55b} or from rigorous approaches to the dynamic semiclassical model~\cite{Ball73,Ishikawa98b,Caizergues14}. In the latter case, the unphysical behavior of the TF electron density in the vicinity of the nucleus has strong consequences on the photoabsorption cross-section at high frequencies~\cite{Ishikawa98b}. Moreover, the lack of shell structure implies the absence of line emission and absorption in the spectra.

Finally, let us remark that in the Thomas Fermi model, the equations are restricted to the WS cell. This renders the model versatile in the sense that it is rather insensitive to the modeling of the medium outside the WS sphere. On the one hand, one may interpret the ion sphere as an element of a highly ordered pile of neutral spheres. In~\cite{Slater35}, the TF ion cell is used in this way, as an approximation of the polyhedral WS cell of a metal lattice.  On the other hand, one may interpret the ion cell as a statistical cavity surrounded by a homogeneous neutral plasma (jellium), as suggested later by Liberman in the context of his INFERNO model~\cite{Liberman79}.
The coexistence of these two possible interpretations relates to the similar ambiguity of the physical picture underlying the ion-sphere model. This duality of interpretation left an imprint on the models proposed later for a quantum extension of the TF model.

\subsection{Quantum Ion-in-Cell Models\label{sec_quantum_ion_cell}}

Among the first quantum extensions to the TF model was the model of Rozsnyai~\cite{Rozsnyai72}. This model is based on the solid-state picture of the ion cell. The bound electrons are described by resorting to energy bands, whose boundaries are obtained from the Wigner-Seitz cellular method (zeros of the wave function and of its derivative, see~\cite{Wigner33,Wigner34} or the monograph~\cite{AshcroftMermin}). Positive-energy spectrum (with respect to the effective potential at infinity) is approximated using the TF approach, with a restriction on the energy integration in order to only cover the classically allowed range. The treatment of continuum electrons is therefore not consistent with that of bound electrons. In particular, the contributions of resonances or energy bands in the continuum are disregarded. However, because of the treatment of bound electrons through energy bands, the pressure ionization of a bound state occurs gradually and does not result in a proper discontinuity of observables.

A variant of this model resorts to wave functions calculated with boundary conditions applied at infinity. In this case, the boundary condition is the exponential decay of the wave function, or, in practice, the matching onto localized zero-field solutions at the WS radius (third kind modified spherical Bessel function). This kind of model is, for instance, used in~\cite{Blenski00}. In this model, due to the semiclassical treatment of the continuum and the discrete nature of bound states, pressure ionization of a bound state results in a discontinuity of observables.

In practice, the equations of the latter model are the same as in the TF model, except that the electron density is partially calculated from quantum mechanics. Namely, one retains Equations~\eqref{eq_TF_SCF}-\eqref{eq_TF_neutrality_WS} and \eqref{eq_TFD_vtrial}, whereas the electron density is given by:
\begin{align}
n(r)=&\sum_{\xi\text{ bound}}p_\text{F}(\mu,T,\varepsilon_\xi)|\varphi_\xi(\vec{r})|^2
+\frac{4}{\sqrt{\pi}\Lambda_\text{e}^3}I_{1/2}^\text{inc.}\left(\beta\left(\mu^\text{F}_\text{id,e}(n_\text{e})-v_\text{trial}(r)\right);
-\beta v_\text{trial}(r)
\right)
\label{eq_Rozsnyai_density}
\end{align}
instead of Equation~\eqref{eq_TFD_density}. The orbitals $\{\ket{\varphi_\xi}\}$ are obtained solving Equation~\eqref{eq_Schrod_1electron} only for the bound states, since the sum only runs over the discrete part of the spectrum. $p_\text{F}$ is the Fermi-Dirac distribution and $I_{1/2}^\text{inc.}$ is the incomplete Fermi integral defined as follows:
\begin{align}
I_{1/2}^\text{inc.}\left(y;z\right)
=\int_z^\infty dx\left\{ \frac{x^{1/2}}{e^{x-y}+1} \right\}
\end{align}

Another slightly different variant of this model approximates the positive-energy spectrum using the non-degenerate limit of the 1-electron distribution instead of the Fermi-Dirac distribution~\cite{Rosmej11}\footnote{In~\cite{Rosmej11}, this model is called ``finite-temperature ion-sphere model'', whereas what we call in the present document ``ion-sphere model'' is called ``uniform electron gas model''.}. Let us also mention that some models also use bands for both the negative and positive parts of the energy spectrum~\cite{Massacrier21}, with applications to matter in which an ion lattice may subsist.

\begin{figure}[t]
\begin{center}
\includegraphics[width=8cm] {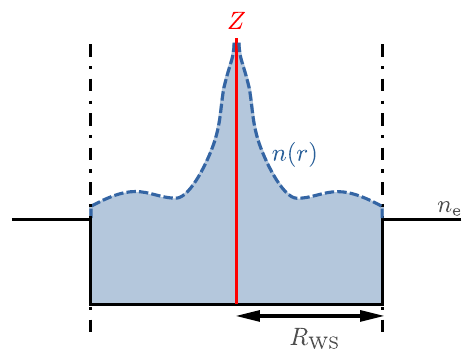} 
\end{center}
\caption{Schematic picture of Liberman's INFERNO model. The electron density is represented with a discontinuity at the WS radius, consistently with the interpretation proposed by Liberman of a homogeneous jellium surrounding the WS sphere.
\label{fig_schema_INFERNO}}
\end{figure}

The first fully quantum model of the ion cell in a plasma was Liberman's model named ``INFERNO''~\cite{Liberman79,Liberman82,INFERNO}. Contrary to Rozsnyai, Liberman proposes the physical picture of an ion cell surrounded by a finite-temperature jellium, as sketched in Figure~\ref{fig_schema_INFERNO}. A jellium is a homogeneous electron gas, neutralized by a homogeneous ion background. 

The equations of the INFERNO model are the same as in the TF model, with an electron density fully calculated from quantum mechanics. One thus keeps Equations~\eqref{eq_TF_SCF}-\eqref{eq_TF_neutrality_WS} and \eqref{eq_TFD_vtrial}, with an  electron density given by:
\begin{align}
&n(r)=\sum_{\xi}p_\text{F}(\mu,T,\varepsilon_\xi)|\varphi_\xi(\vec{r})|^2
\label{eq_inferno_density}
\end{align}
where the orbitals $\{\ket{\varphi_\xi}\}$ are obtained solving Equation~\eqref{eq_Schrod_1electron} for both the bound and continuum states, since the sum runs over both the discrete and the continuum part of the spectrum. For the latter part, the sum is to be understood as an integral over the momentum. Contrary to the Rozsnyai model and its variants, the INFERNO model accounts for the resonances in the continuum. 

\begin{figure}[t]
\begin{center}
\includegraphics[width=15cm] {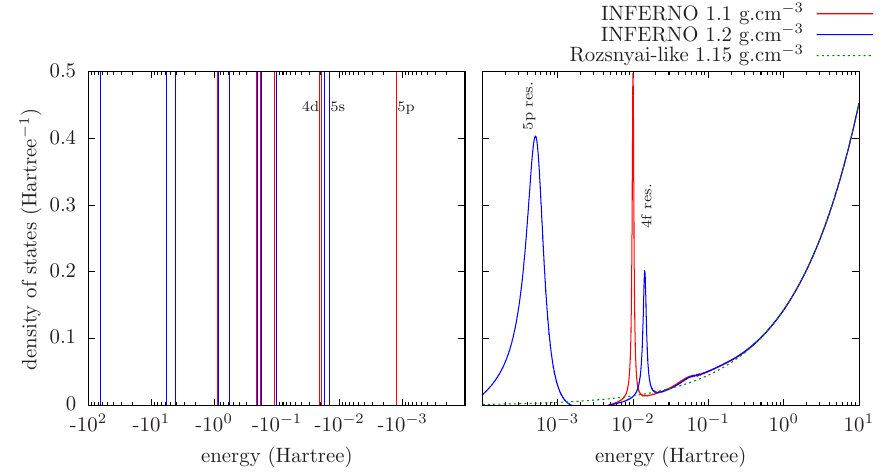} 
\end{center}
\caption{Density of states obtained from the INFERNO model in the case of silicon at 5 eV temperature and matter densities of 1.1 and 1.2 10$^{-2}$ g.cm$^{-3}$. In the negative energy ranges, the Dirac distributions are represented by vertical lines. For the sake of comparison, the density of states stemming from the Rozsnyai-like model of Equation~\eqref{eq_Rozsnyai_density}, at 5 eV temperature and 1.15 10$^{-2}$ g.cm$^{-3}$, is also shown, only in the positive energy range.
\label{fig_dos_INFERNO}
}
\end{figure}

Figure~\ref{fig_dos_INFERNO} shows the density of states obtained from the INFERNO model for silicon at 5 eV temperature and matter densities of 1.1 and 1.2 10$^{-2}$ g.cm$^{-3}$. At these conditions, one can observe resonances related to the delocalizations of the 5p and 4f subshells. In particular, between 1.1 and 1.2 10$^{-2}$ g.cm$^{-3}$, the 5p subshell is pressure-ionized, yielding a sharp resonance in the continuous spectrum. In order to illustrate the lack of resonances in Rozsnyai-like models, we also display the density of states obtained when using Equation~\eqref{eq_Rozsnyai_density} for the electron density, at similar plasma conditions. The discrete spectrum from the Rozsnyai-like model is not shown, to avoid obfuscation of the figure.

Consistently, with the picture of an ion cell surrounded by a neutral jellium, the boundary condition applied to the wave functions $\varphi_\xi$ at the WS radius is just the matching onto the zero-potential solution (a linear combination of Bessel functions, defining a phase shift). Like in the TF model, the medium surrounding the ion cell does not interact with the content of the ion cell. Consequently, the model can be formulated using equations restricted to the WS cell, with the jellium surrounding the WS sphere playing no direct role in the model. 

The quantum electron density of Equation~\eqref{eq_inferno_density} is a nonlocal functional of the self-consistent potential. As a consequence, even if the potential is zero at the WS radius, the electron density $n(R_\text{WS})$ in general differs from the electron density obtained from the chemical potential $n_\text{e}$. The latter corresponds to the asymptotic value $\lim_{r\rightarrow\infty}n(r)$.

This yields an ambiguity in the definition of the mean ionization $Z^*$, which is closely related to the ambiguity in the physical interpretation of the model. Either the electron density has a discontinuity at the WS radius, or it is continuous but electrons of the jellium have a chemical potential that is different from those of the ion cell.

Connected to this interpretation issue is the problem of defining the pressure in the model (electron pressure at the WS boundary versus electron pressure stemming from a jellium of density $n_\text{e}$). More generally, due to the lack of variational formulation for this model, any thermodynamic quantity is defined heuristically and may have more than one possible definition. Indeed, Liberman proposed two versions of his model (denoted A and T), differing in the region of integration for the free and internal energies~\cite{Liberman79,Liberman82,INFERNO}. Thermodynamic consistency among these quantities is in general not assured. 

The sharp cut-off of the equations at the WS radius also implies that the virial theorem is not fulfilled. When trying to derive the virial theorem for the system, surface terms appear at the WS radius, which results in the impossibility of fulfilling the theorem (see, for instance~\cite{PironPhD}).

Nevertheless, Rozsnyai's and Liberman's models are among the most often used when dealing with pressure-ionized plasma, both in their respective average-atom versions (see, for instance,~\cite{Wilson06,Penicaud09}) or in a modified version adapted to fixed configurations (see, for instance,~\cite{Blenski00}). To some degree of approximation, these models account for both the quantum shell structure of the ion and the pressure ionization phenomenon. Both also have relatively low computation costs, favored by the restriction of the equations to the WS cell. Of course, INFERNO involves a much higher computational cost than Rozsnyai's model, due to the quantum treatment of the continuum.

Moreover, a variant of Rozsnyai's model was used in~\cite{Rosmej11} as the starting point to obtain an approximate, closed formula fitting the atomic potential. Such a fit for the atomic potential can be used to infer corrections to the isolated-ion energies, from perturbation theory~\cite{Li12,Iglesias19}. Such an approach yields analytical formulae for the line shifts, which showed agreement with experimental measurements~\cite{Beiersdorfer19} of line shifts in He-like ions at electron densities of the order of $10^{23}$ cm$^{-3}$~\cite{Li20,Zeng22b}. In prior studies, the simpler model of the ion-sphere had also been used, outside the context of point-like ion hypothesis, to calculate an analytical perturbing potential and infer line shifts~\cite{Massacrier90}.

\subsection{Ion-in-Jellium Models}

\begin{figure}[b]
\includegraphics[width=8cm] {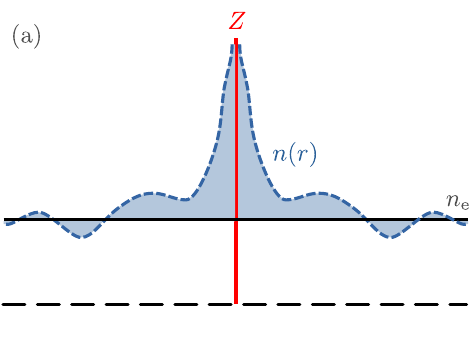} 
\hspace{1cm}
\includegraphics[width=8cm] {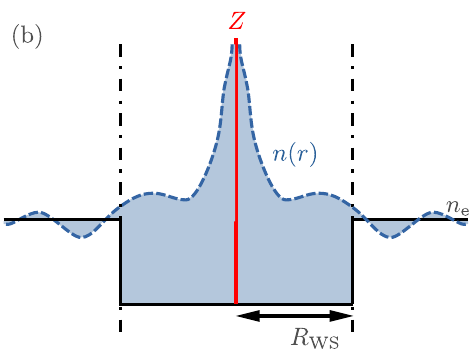} 
\caption{Schematic pictures of an impurity in a jellium ({a}), and of an ion-in-jellium model such as AJCI or VAAQP ({b}).
\label{fig_schema_ion_jellium}}
\end{figure}

Models of an impurity (or a defect) in a jellium were developed during the 1970s in the context of solid-state physics~\cite{Arponen73,Manninen75,Jena78}. In these models, the perturbation generated by the impurity may extend spatially far from its origin. There is no restriction to a particular cell (see Figure~\ref{fig_schema_ion_jellium}\,a).

A first extension of the treatment of an impurity in a jellium to the modeling of an ion in a plasma was suggested by Perrot in the 1990s, in his ``Atome dans le Jellium de Charge Impos\'ee'' model (AJCI, atom in a jellium with fixed charge). In his model, Perrot introduces a WS statistical cavity in the jellium, much like the picture proposed by Liberman. However, he also considers an ion extending in the whole space, rather than enclosed within a cell (see Figure~\ref{fig_schema_ion_jellium}\,b). Consistently, the neutrality is assumed to hold in the whole space rather than in the ion cell. In this model, the uniform ion background of the jellium surrounding the cavity interacts with the electron density, which asymptotically tends to the jellium density. This leads to the charge density:
\begin{align}
q_\text{e} \left( n(r)-n_\text{i}Z^*\theta(r-R_\text{WS}) \right)
=q_\text{e} \left( n(r)-n_\text{e}\theta(r-R_\text{WS}) \right)
\end{align} 

The AJCI model, like models of impurity in metals, resorts to a fixed jellium density $n_\text{e}$, given as an input to the model. It is also lacking a variational derivation. However, the notion of an ion extending beyond the WS sphere, up to infinity, allows one in principle to solve the problem of surface terms in the virial theorem.

Starting from the founding ideas of the AJCI model, a model of a variational average-atom in a quantum plasma (VAAQP) was proposed and studied~\cite{Blenski07a,Blenski07b,Piron11,Piron11b}. This showed that building an atom-in-jellium model within a variational framework enables one to set the jellium density from the thermodynamic equilibrium condition and to fulfill the virial theorem.

Formally, to treat the nuclei-electron plasma as a set of ions, we resort to a reasoning called a ``cluster'' decomposition. Let $O$ be a quantity that may be calculated for any set of $K$ nuclei, with spatial configuration $(\vec{R}_1 ... \vec{R}_K)$, including the empty set. We may then write (see~\cite{Felderhof82} for more detail):
\begin{align}
O(\vec{R}_1 ... \vec{R}_K)
= O(\emptyset)
+\sum_{j=1}^{K} \Delta O_1(\vec{R}_j)
+\frac{1}{2}\sum_{j=1}^{K}\sum_{\substack{k=1\\k\neq j}}^{K} \Delta O_2 (\vec{R}_j,\vec{R}_k) +...
\label{eq_cluster_formal}
\end{align}
defining the $\Delta O_K$ terms recursively, so as to assure the equality for each value of $K$:
\begin{align}
&\Delta O_1(\vec{R})
=O(\vec{R})-O(\emptyset)
\\
&\Delta O_2(\vec{R}_1,\vec{R}_2)
=O(\vec{R}_1,\vec{R}_2)
-\Delta O_1(\vec{R}_1)
-\Delta O_1(\vec{R}_2)
+O(\emptyset)
\\
&\hspace{0.5cm}\vdots\nonumber
\end{align}
The quantity $O$ has a clustering property if the terms in Equation~\eqref{eq_cluster_formal} exhibit a decreasing ordering, which makes Equation~\eqref{eq_cluster_formal} a convergent expansion. 

In the VAAQP model, we first assume that the electron density $n(\vec{R}_1 ... \vec{R}_{N_\text{i}};\vec{r})$ for a system of $N_\text{i}$ ion is correctly described by limiting the cluster expansion to the zeroth and first order only:
\begin{align}
n(\vec{R}_1 ... \vec{R}_{N_\text{i}};\vec{r})
=n_0(\vec{r})+\sum_{j=1}^{N_\text{i}}\Delta n_1(\vec{R}_j;\vec{r})
=n_\text{e}+\sum_{j=1}^{N_\text{i}}q(|\vec{r}-\vec{R}_j|)
\label{eq_cluster_density}
\end{align}
where the zeroth-order term $n_0$ is identified as the homogeneous jellium density $n_\text{e}$, and the first-order term corresponds to the sum of spherically symmetric clouds of displaced electrons, corresponding each to an ion in a jellium.
We also assume the first-order cluster expansion to hold for the free energy of the system. This leads us to write the free energy per ion as follows:
\begin{align}
\dot{F}(n_\text{i},T)=\dot{F}_0(n_\text{e};n_\text{i},T)+\Delta F_1\{n_\text{e},\underline{q};T\}
\label{eq_VAAQP_cluster_F}
\end{align}
Here, $\dot{F}_0=(f_\text{e}^\text{F}(n_\text{e},T)+f_\text{xc}(n_\text{e},T))/n_\text{i}$ is the free energy per ion of the uniform electron gas. We choose to treat $\Delta F_1$ using a density-functional formalism~\cite{Hohenberg64,KohnSham65a,Mermin65} and decompose the $\Delta F_1$ as suggested by Kohn and Sham~\cite{KohnSham65a}:
\begin{align}
\Delta F_1\{n_\text{e},\underline{q};T\}
=\Delta F_1^0\{n_\text{e},\underline{q};T\}
+\Delta F_1^\text{el}\{n_\text{e},\underline{q};T\}
+\Delta F_1^\text{xc}\{n_\text{e},\underline{q};T\}
\label{eq_VAAQP_dF1}
\end{align}
$\Delta F_1^0$ corresponds to the kinetic and entropic contribution to the free energy of a system of independent electrons subject to an external potential $v_\text{trial}\left\{n_\text{e},\underline{q};r\right\}$ that yields the electron density $n_\text{e}+q(r)$, with the contribution from a homogeneous system of density $n_\text{e}$ subtracted.
\begin{align}
&\Delta F_1^0\left\{n_\text{e},\underline{q};T\right\}
=\sum_\xi
\int d^3r \Big\{
p_\text{F}(\varepsilon_\xi;n_\text{e},T)\Big[\vphantom{\frac{1}{1}}
\left(\varepsilon_\xi-v_\text{trial}(r)-T s_\text{F}(\varepsilon_\xi;n_\text{e},T)\right)|\varphi_\xi(\vec{r})|^2
\nonumber\\
&\hspace{8cm}
-\left(\varepsilon_\xi-T s_\text{F}(\varepsilon_\xi;n_\text{e},T)\right)|\varphi^0_\xi(\vec{r})|^2\Big]
\Big\}
\label{eq_VAAQP_dF10}\\
&s_\text{F}(\varepsilon_\xi;n_\text{e},T)=s\left(p_\text{F}(\varepsilon_\xi;n_\text{e},T)\right)
\end{align}
where the $\{\ket{\varphi_\xi}\}$ are obtained solving Equation~\eqref{eq_Schrod_1electron} for the bound and continuum states. As in INFERNO, the sum runs over both the discrete and continuum part of the spectrum. Here the $\{\ket{\varphi^0_\xi}\}$ correspond to the plane waves (eigenstates of $\tilde{H}_0$) and only contribute to the continuum part. 
$\Delta F_1^\text{xc}$ corresponds to the exchange and correlation contribution to the free energy, with the contribution from the homogeneous system subtracted, taken in the local density approximation:
\begin{align}
\Delta F_1^\text{xc}\{n_\text{e},\underline{q};T\}
=\int d^3r \left\{f_\text{xc}(n_\text{e}+q(r),T)-f_\text{xc}(n_\text{e},T)\right\}
\label{eq_VAAQP_dF1xc}
\end{align}

$\Delta F_1^\text{el}$ is the direct interaction term, in which we introduce the hypothesis of the WS cavity. We model the surrounding ions by a charge density $q_\text{e} n_\text{e}\theta(r-R_\text{WS})$, like in the AJCI model. This leads to:

\begin{align}
\Delta F_1^\text{el}=&
\int d^3r\left\{ -\frac{Z \left(n_\text{e}+q(r)-n_\text{e}\theta(r-R_\text{WS})\right) \ee^2}{r} 
\right.\nonumber\\&\left.
+\frac{\ee^2}{2}\int d^3r' \left\{ \frac{\left(n_\text{e}+q(r)-n_\text{e}\theta(r-R_\text{WS})\right)\left(n_\text{e}+q(r)-n_\text{e}\theta(r'-R_\text{WS})\right)}{|\vec{r}-\vec{r}'|}\right\} \right\}
\label{eq_VAAQP_dF1el}
\end{align}
Accordingly, the condition of neutrality in the whole space can be written as:
\begin{align}
Z=\int d^3r \left\{
n_\text{e}+q(r)-n_\text{e}\theta(r-R_\text{WS})
\right\}=\frac{n_\text{e}}{n_\text{i}}+\int d^3r \left\{q(r)\right\}
\end{align}

Finally, the VAAQP model is based on the minimization of the free energy with respect to the displaced-electron density $q(r)$ and jellium density $n_\text{e}$, while requiring the neutrality condition:
\begin{align}
\dot{F}_\text{eq}(N_\text{i},V,T)=&\underset{n_\text{e},\underline{q}}{\text{Min}}\,\dot{F}\left\{n_\text{e},\underline{q};n_\text{i},T\right\}
\text{ s.\,t. } \int d^3r \left\{
n_\text{e}+q(r)-n_\text{e}\theta(r-R_\text{WS})
\right\}=Z
\label{eq_minimiz_VAAQP}
\end{align}
This constrained minimization yields the following equations:
\begin{align}
&v_\text{trial}(r)=v_\text{el}(r)+\mu_\text{xc}(n_\text{e}+q(r))-\mu_\text{xc}(n_\text{e})\label{eq_VAAQP_vtrial}\\
&v_\text{el}(r)=-\frac{Z \ee^2}{r}+\ee^2\int d^3r' \left\{ \frac{n_\text{e}+q(r)-n_\text{e}\theta(r'-R_\text{WS})}{|\vec{r}-\vec{r}'|}\right\}\\
&\int d^3r\left\{ v_\text{el}(r)\theta(r-R_\text{WS}) \right\}=0\label{eq_VAAQP_n0_condition}
\end{align}
Equation~\eqref{eq_VAAQP_vtrial} gives the self-consistent potential of the VAAQP model, which is in general nonzero outside the WS sphere.

Equation~\eqref{eq_VAAQP_n0_condition} stems from the minimization condition with respect to the jellium density $n_\text{e}$ and allows finding its value. Thus, in the VAAQP model, the density of the uniform background is uniquely defined; it corresponds to the asymptotic electron density of each ion and is \emph{set by the thermodynamic equilibrium condition}.

From the equilibrium free energy per ion, it is possible to rigorously obtain the other thermodynamic quantities, by calculating the appropriate derivatives. For the pressure, the following formula is obtained:
\begin{align}
P_\text{thermo}=-f_\text{e}^\text{F}(n_\text{e},T)-f_\text{xc}(n_\text{e},T)
+n_\text{e}\mu^\text{F}_\text{id,e}(n_\text{e},T)+n_\text{e}\mu_\text{xc}(n_\text{e},T)+n_\text{e}v_\text{el}(R_\text{WS})
\label{eq_VAAQP_pressure}
\end{align}
The first four terms correspond to the pressure of an ideal Fermi gas of density $n_\text{e}$. The last term is related to the WS cavity. Moreover, it can be shown that the virial pressure leads to the same formula as Equation~\eqref{eq_VAAQP_pressure}, meaning that the \emph{virial theorem is fulfilled} in the VAAQP model.

In the contributions to the free energy expression of Equation~\eqref{eq_VAAQP_cluster_F}, the ions are disregarded. For that reason, the thermodynamic quantities from the VAAQP model may be viewed as electron contributions, which may be supplemented by ion contributions. Adding an ion ideal-gas contribution to the model is straightforward. However, adding the results of a model of interacting ions is more problematic because part of the ion-ion interactions is necessarily included in the VAAQP model through the WS-cavity hypothesis.

Like INFERNO, the VAAQP model allows for the description of the ion shell structure, while the WS cavity assumed in the model enables the description of pressure ionization. Treating the perturbation of the density in the whole space, the model also accounts for the Friedel oscillations (see, for instance,~\cite{AshcroftMermin}) of the displaced-electron density. The physical relevance of these oscillations in the case of ions in a plasma is rather unclear. However, accounting for them is essential to ensure the fulfillment of the virial theorem.

In the VAAQP model, the potential range is not strictly limited to the WS radius but has a strong decay, due to the total screening of the nucleus in the whole space. In practice, the variational Equation~\eqref{eq_VAAQP_n0_condition} most often constrains the atomic potential to take small values at the WS radius, of the order of the amplitude of Friedel oscillations. For that reason, the VAAQP model yields results that mostly agree with those of the INFERNO model, except in the low-temperature/high-density regime.

\begin{figure}[p]
\includegraphics[width=7.7cm] {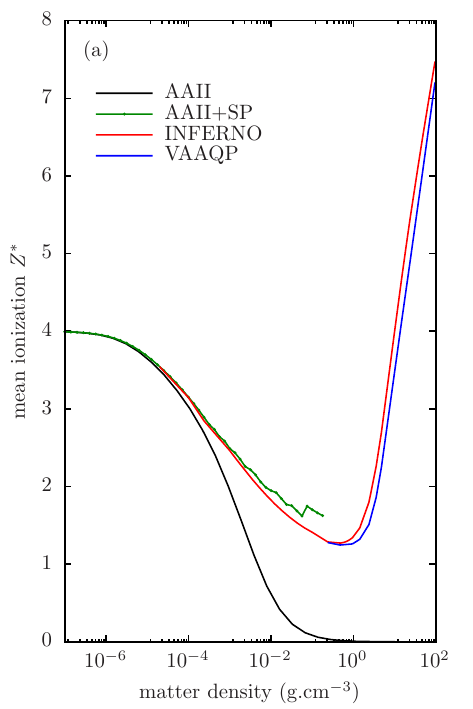} 
\hspace{1cm}
\includegraphics[width=7.7cm] {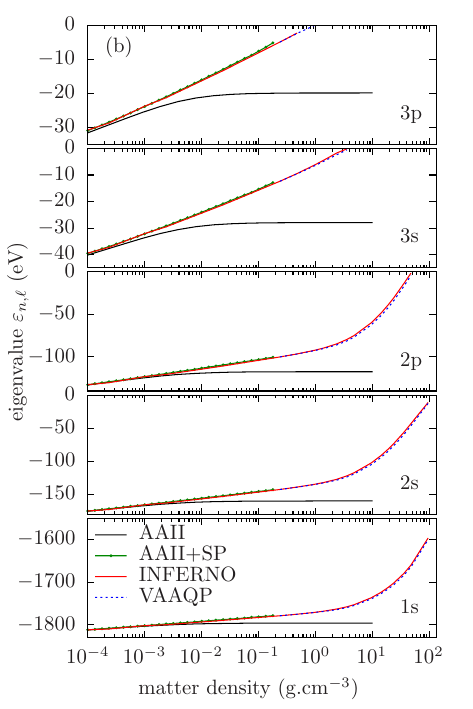} 
\caption{Mean ionization $Z^*$ ({a}), and 1-electron eigenvalues ({b}) as a function of matter density, for silicon at 5-eV temperature. Comparisons between INFERNO, VAAQP, average-atom model of isolated ion (AAII), and AAII with Stewart-Pyatt correction and suppression of orbitals (AAII+SP).
\label{fig_Si_ioniz_eigenvalues}}
\end{figure}

Figure~\ref{fig_Si_ioniz_eigenvalues} shows an example comparison of results from the isolated-ion, INFERNO, and VAAQP models in the case of silicon at 5-eV temperature. Both the mean ionization and the 1-electron energies are displayed.  As is seen from these figures, the energy correction, and consequently the mean ionization, are rather well estimated using the Stewart-Pyatt approach with the suppression of bound orbitals~\cite{StewartPyatt66}, up to cases of significant pressure ionization (here, around 0.1 g.cm$^{-3}$). The INFERNO and VAAQP models agree well in this regime. In the region of strong pressure ionization, the results from VAAQP depart significantly from those of INFERNO. Accordingly, the thermodynamic consistency of the INFERNO results is problematic in this region. However, differences in the 1-electron energies are less pronounced (see Figure~\ref{fig_Si_ioniz_eigenvalues}\,b).

The Thomas-Fermi, INFERNO and VAAQP models can be used to calculate electron contributions to the equation of state.  For the sake of supplementing the electron thermodynamic quantities with their corresponding ion contributions, it is perfectly justified to add an ion ideal-gas contribution. In order to obtain results that are somewhat more realistic, ion contributions are often estimated using models accounting for ion-ion interactions, like the OCP model, or semi-empirical approaches like that of Cowan, or the corrected rigid ion sphere model of Kerley \cite{Kerley80a,Kerley80b}. Theoretical justification for using such models is however less rigorous, and there is a need for a unified formalism to describe both the electron structure of ions and the ion fluid (see Section~\ref{ch_ion_correlations}). 

\begin{figure}[p]
\begin{center}
\includegraphics[width=8cm] {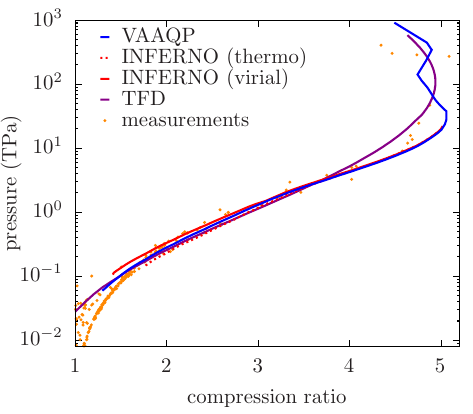} 
\end{center}
\caption{Principal Hugoniot of Aluminum. Comparison of the results from VAAQP, INFERNO and TFD models with experimental measurements. Results from the models were supplemented by a OCP ion contribution. For the INFERNO models, several definition of the thermodynamic quantities may be considered. Here, we present results from the thermodynamical and virial definitions of the pressure, with energy integrals restricted to the WS sphere (A-version in Liberman's notations). Measurments were taken from the database maintained by the Joint Institute for High Temperatures, Russian Academy of Sciences (\cite{rusbank} and the references therein).
%Data presented here are from \cite{Altshuler60,Altshuler60b,Altshuler61,Kormer62,Skidmore62,Isbell68,McQueen70,Altshuler74,Bakanova74,Altshuler77,vanThiel77,Marsh80,Altshuler81,Volkov81,Mitchell81,Ragan82,Vladimirov84,Ragan84}
\label{fig_EOS}}
\end{figure}

As an illustration of the application to equation-of-state calculations, Figure~\ref{fig_EOS} display the Aluminum principal Hugoniot as it is obtained from the TFD, INFERNO and VAAQP models, compared to measurements (see \cite{rusbank} and complete list of references therein). Due to the heuristic character of the thermodynamic quantities in the INFERNO model, several definitions can be chosen and may lead to different results. Here we have chosen to display only results from the thermodynamical and virial definitions of the pressure, with energy integrals restricted to the WS sphere (A-version in Liberman's notations).  One observes disagreement in the region where the Hugoniot is close to the cold-compression curve, which falls outside the validity domain of such plasma models.

As regards the VAAQP model, applications to the calculation of radiative properties were also studied. It is the subject of next chapter.

Let us point out that atom-in-jellium models seem a better starting point for improving the modeling of ion-ion correlations than ion-cell models since they account for the ion surroundings in the whole space. This is the subject of Chapter~\ref{ch_ion_correlations}

\vspace{-10pt}
\section{In brief}
Ideal plasma model is associated to isolated-ion model of the electronic structure. Non-ideality corrections are obtained by accouting for interactions among the particles of the ion-free-electron plasma, assuming point-like ions with fixed electronic structure. This leads to depressions of the ionization potentials.
Several approximate models are used to calculate these depressions, notably the Debye-Hückel, ion-sphere, and Stewart-Pyatt models.
%Two ways of introducing it co-exist: 1) adding to each internal component of the ion, the interaction energy stemming from the mean potential due to all other particles, and 2) adding an excess free-energy to the ideal-plasma free energy. It is shown that, in the Debye-Hückel model as well as in the ion-sphere model of fluid, both approaches lead to qualitatively similar results. 
These approaches are also used to set a restriction on the spatial range of the atomic structure, and truncate the partition function that diverges in the purely ideal case. 

Pressure ionized plasma are addressed through models resorting to the notion of Wigner-Seitz sphere. Two theoretical standpoints on the Wigner-Seitz sphere are co-existing: 1) plasma is seen as a pile of neutral cells, this picture finding its origin in solid-state physics, and 2) plasma is seen as a strongly-coupled fluid, with ion-ion correlations forming a statistical cavity.

The ``solid-state'' standpoint is that of band models such as that of Rozsnyai \cite{Rozsnyai72}, or that proposed by Massacrier \textit{et al.} \cite{Massacrier21}. Despite the physical picture of an ion surrounded by a jellium presented in the article, and even if it resorts to boundary conditions located at infinity for the wavefunctions, Liberman's INFERNO model \cite{Liberman79} also belongs to this category. 

The ``fluid-state'' standpoint was initally explored in Perrot's AJCI model, extending to ions the picture of an impurity in a jellium. The VAAQP model provided this approach with the variational derivation it was lacking, and notably enabled the determination of the plasma mean ionization from the thermodynamic equilibrium condition.

%-------------------------
\chapter[Application of the VAAQP Model to Photoabsorption...]{Application of the VAAQP Model to Photoabsorption in Dense Plasmas\label{ch_rad_prop}}
%-------------------------

\section{Absorption of Light by Matter}

\subsection{Absorption and Linear Response to a Perturbing Field}
From the electrodynamics of continuous media (see, for instance, \cite{LandauElectrodynamicsContinuousMedia}), one can relate the absorption coefficient $k_\text{abs}(\vec{k},\omega)$ of a medium (opacity per unit volume) to its complex electric susceptibility $\chi_{\vec{k},\omega}$.
\begin{align}
&k_\text{abs}(\vec{k},\omega) = 
\frac{\omega }{c\text{Re}(n_{\vec{k},\omega}^\text{ref})}
\text{Im}(\chi_{\vec{k},\omega})\label{eq_k_abs}
\\
&\text{Re}(n_{\vec{k},\omega}^\text{ref})
=\left(\frac{\text{Re}(1+\chi_{\vec{k},\omega})+|1+\chi_{\vec{k},\omega}|}{2}\right)^{1/2}\label{eq_real_nref}
\end{align}
where $n_{\vec{k},\omega}^\text{ref}$ corresponds to the complex refraction index of the medium.

The electric susceptibility is directly related to the dielectric function $\epsilon_{\vec{k},\omega}$ and dynamic conductivity $\sigma_{\vec{k},\omega}$:
\begin{align}
&\epsilon_{\vec{k},\omega}=\epsilon_0(1+\chi_{\vec{k},\omega})\ ;\ \sigma_{\vec{k},\omega}=i\omega \epsilon_{\vec{k},\omega}
\label{eq_diel_conductivity}
\end{align}
Also, one may express the absorption of the medium in terms of absorption cross-section per ion or atom $\sigma_\text{abs}$, or in terms of opacity per unit mass $\kappa$:
\begin{align}
\sigma_\text{abs}(\vec{k},\omega)=\frac{k_\text{abs}(\vec{k},\omega)}{n_\text{i}}\ ;\ \kappa(\vec{k},\omega)=\frac{\mathcal{N}}{M_\text{mol}}\sigma_\text{abs}(\vec{k},\omega)
\end{align}
where $\mathcal{N}$ is Avrogadro's number and $M_\text{mol}$ is the mean molar mass of the medium.

Let us now consider that each elementary volume $V$ of the continuous medium contains a macroscopic plasma composed of a large number $N_\text{i}$ of nuclei having atomic number $Z$ and $N_\text{i} Z$ electrons at a fixed temperature $T$. We will subsequently consider this system in the thermodynamic limit $N_\text{i}\rightarrow\infty$, $V\rightarrow\infty$, $N_\text{i}/V\rightarrow n_\text{i}$.

For simplicity, let us assume a fixed configuration of the nuclei in space. The static many-electron system has the Hamiltonian:
\begin{align}
\hat{H}_\text{static}=\hat{K}+\hat{V}_{\text{$N_\text{i}$ nuc.}}(\vec{R_1},...,\vec{R_{N_\text{i}}})+\hat{W}
\label{eq_H_static}
\end{align}
\begin{align}
\hat{V}_{\text{$N_\text{i}$ nuc.}}(\vec{R_1},...,\vec{R_{N_\text{i}}})
=\sum_{j=1}^{N_\text{i}}\int d^3r\left\{\frac{-Z \ee^2}{|\vec{r}-\vec{R}_j|}\hat{n}_\vec{r}\right\}
\end{align}
with $\hat{V}_{\text{$N_\text{i}$ nuc.}}$ being the sum of the external potentials generated by the fixed nuclei. $\hat{W}$ is defined in Equation~\eqref{eq_op_W}. $\hat{n}_\vec{r}=\hat{a}^\dagger_\vec{r}\hat{a}_\vec{r}$ is the operator giving the electron number density at $\vec{r}$.

We now consider that this system is perturbed by an external time-dependent monochromatic electromagnetic field. The nuclei being at fixed positions, we disregard their response to the field and focus on the response of the quantum many-electron system. 

%Choosing the Coulomb gauge for the potential, we may write the potential associated to the monochromatic plane wave as
%\begin{align}
%\vec{A}(\vec{r},t)=\vec{A}_0\sin(\omega t-\vec{k}.\vec{r})
%\end{align}
%with $\vec{k}.\vec{A}_0=0$ and no scalar potential.

In the minimal coupling approach (see, for instance, \cite{Schiff}, Chapter 11), the perturbing field is associated with the following time-dependent operator $\hat{V}_\text{pert}(t)$:
\begin{align}
\hat{V}_\text{pert}(t)=\sum_{\xi,\zeta}\int_V d^3r\left\{
\varphi_\xi^*(\vec{r})
\left(q_\text{e}\Phi(\vec{r},t)+\frac{q_\text{e} i\hbar}{2m_\text{e}}
\left(
\nabla_{\vec{r}}.\vec{A}(\vec{r},t)
+2\vec{A}(\vec{r},t).\nabla_{\vec{r}}
\right)
+\frac{q_\text{e}^2}{2m_\text{e}}\vec{A}^2(\vec{r},t)
\right)
\varphi_\zeta(\vec{r})
\right\}\hat{a}^\dagger_\xi\hat{a}_\zeta
\label{eq_minimal_coupling_operator}
\end{align}
where $\Phi(\vec{r},t)$ and $\vec{A}(\vec{r},t)$ are the scalar and vector potentials of the monochromatic perturbing field. These are such that the electric field has the form:
\begin{align}
\vec{E}(\vec{r},t)=-\vec{\nabla}\Phi(\vec{r},t)-\frac{\partial \vec{A}(\vec{r},t)}{\partial t}=\vec{E}_0\cos\left(\omega t-\vec{k}.\vec{r}\right)
\end{align}
In the following, we will neglect the non-linear term in the right-hand side of Equation \eqref{eq_minimal_coupling_operator}. The explicit expression of $\hat{V}_\text{pert}(t)$ depends on the choice of a gauge. However, without loss of generality, we may write it as:
\begin{align}
\hat{V}_\text{pert}(t)=
\delta\hat{v}_{\text{pert,S},\vec{k}}\sin(\omega t)
+\delta\hat{v}_{\text{pert,C},\vec{k}}\cos(\omega t)
\end{align}

From the quantum linear-response theory of the many-electron system (see, for instance, \cite{FetterWalecka}) one can evaluate the power absorbed by the system, due to these harmonic perturbations. 
One can thus relate the electric susceptibility of Equation~\eqref{eq_k_abs} to the susceptibility of the quantum system $\chi^{\delta\hat{v}_{\text{pert},\bullet,\vec{k}}}_{V,\omega}$ associated to the operators $\delta\hat{v}_{\text{pert},\bullet,\vec{k}}$, where $\bullet$ stands for the S or C label. The latter susceptibility is just the Fourier transform of the retarded time-response function to this operator, which expresses the linear response of the observable $\moy{\delta\hat{v}_{\text{pert},\bullet,\vec{k}}}$ to the perturbation $\delta\hat{v}_{\text{pert},\bullet,\vec{k}}$ in the system of volume $V$.

\begin{align}
&\chi_ {\vec{k},\omega}=\frac{\chi^{\delta\hat{v}_{\text{pert},\bullet,\vec{k}}}_{V,\omega}}{\epsilon_0\vec{E}_0^2 V}
\label{eq_LR_chi_chi}
\\
&\chi^{\delta\hat{v}_{\text{pert},\bullet,\vec{k}}}_{V,\omega}
=-\frac{i}{\hbar}\int_{-\infty}^\infty dt \Big\{ \text{Tr}\left\{\hat{\rho}\left[\delta\hat{v}_{\text{pert},\bullet,\vec{k}}^\text{Heisen.}(t),\delta\hat{v}_{\text{pert},\bullet,\vec{k}}^\text{Heisen.}(0)\right]\right\}\theta(t)e^{i\omega t} \Big\}
\label{eq_LR_chi_R}
\end{align}
where $\hat{\rho}$ is the density matrix of the static system, and where $\delta\hat{v}_{\text{pert},\bullet,\vec{k}}^\text{Heisen.}(t)$ corresponds to the operator $\delta\hat{v}_{\text{pert},\bullet,\vec{k}}$ in the Heisenberg picture related to $\hat{H}_\text{static}$. The Heaviside function $\theta(t)$ enforces causality in the time response. In Equation~\eqref{eq_LR_chi_R}, the square brackets denote a commutator.

Equations~\eqref{eq_LR_chi_chi} and \eqref{eq_LR_chi_R} may be seen as an expression of the fluctuation-dissipation theorem for our quantum many-electron system at finite temperature (see, for instance, \cite{LandauStatisticalPhysics}, \textsection 123 and 124)

\subsection{Atomic Response}

In atomic modeling of plasma, one decomposes the susceptibility of the macroscopic plasma $\chi^{\delta\hat{v}_{\text{pert},\bullet,\vec{k}}}_{V,\omega}$ into contributions from the various ions. In principle, such a decomposition stems from the considered atomic model. 

For the case of an ideal plasma of isolated ions, the susceptibility of the plasma is directly obtained summing the susceptibilities of each isolated ion, seen as an independent system, plus the susceptibility of the free-electron gas.
\begin{align}
\chi^{\delta\hat{v}_{\text{pert},\bullet,\vec{k}}}_{V,\omega}
&=\chi^{\delta\hat{v}_{\text{pert},\bullet,\vec{k}}}_{\text{e},V,\omega}
+\sum_{\Psi=1}^{M}N_\Psi\chi^{\delta\hat{v}_{\text{pert},\bullet,\vec{k}}}_{\Psi,V,\omega}
\end{align}
where $\chi^{\delta\hat{v}_{\text{pert},\bullet,\vec{k}}}_{\text{e},V,\omega}$ is the susceptibility of the ideal free-electron gas of density $n_\text{e}$, at temperature $T$, enclosed in the volume $V$. $\chi^{\delta\hat{v}_{\text{pert},\bullet,\vec{k}}}_{\Psi,V,\omega}$ is the susceptibility of an ion of species $\Psi$, calculated in the volume $V$. The latter corresponds to a system that is spatially localized, since continuum electrons are disregarded. For that reason, taking the isolated-ion system in the limit $V\rightarrow\infty$ does not lead to a divergence of $\chi^{\delta\hat{v}_{\text{pert},\bullet,\vec{k}}}_{\Psi,V,\omega}$. In the thermodynamic limit, we just get for the electric susceptibility:
\begin{align}
\chi_{\vec{k},\omega}&=\chi_{\text{e},\vec{k},\omega}
+\sum_{\Psi=1}^{M}n_\Psi \frac{\chi^{\delta\hat{v}_{\text{pert},\bullet,\vec{k}}}_{\Psi,\omega}}{\epsilon_0\vec{E}_0^2}
\end{align}
with $\chi^{\delta\hat{v}_{\text{pert},\bullet,\vec{k}}}_{\Psi,\omega}$ being calculated in an infinite volume.
In the case of the average-atom model of the isolated ion, all contributions $\chi_{\Psi,\vec{k},\omega}$ are considered equal to the average one, $\chi_{\text{AAII},\vec{k},\omega}$:
\begin{align}
\chi_{\vec{k},\omega}&=\chi_{\text{e},\vec{k},\omega}
+n_\text{i} \frac{\chi^{\delta\hat{v}_{\text{pert},\bullet,\vec{k}}}_{\text{AAII},\omega}}{\epsilon_0\vec{E}_0^2}
\equiv \chi_{\text{e},\vec{k},\omega} + \chi_{\text{AAII},\vec{k},\omega}
\label{eq_AAII_chi}
\end{align}

%In the case of an ion-cell model, one may see the response of the plasma as the sum of the responses of each ion cell. 

For the VAAQP model, the cluster expansion of Eq.~\eqref{eq_cluster_formal}, which is used for the electron density and free energy, can be extended to the plasma susceptibility~\cite{Blenski92,Blenski94,Felderhof95a}:
\begin{align}
\chi^{\delta\hat{v}_{\text{pert},\bullet,\vec{k}}}_{V,\omega}
&=\chi^{\delta\hat{v}_{\text{pert},\bullet,\vec{k}}}_{0,V,\omega}+
\sum_{j=1}^{N_\text{i}}\Delta\chi^{\delta\hat{v}_{\text{pert},\bullet,\vec{k}}}_{1,V,\omega}(\vec{R}_j)
\end{align}
The zeroth-order corresponds to the susceptibility of a homogeneous electron gas of density $n_\text{e}$, at temperature $T$, enclosed in the volume $V$. The first-order term is the susceptibility of a cloud of displaced electrons around a single nucleus, surrounded by a cavity beyond which other ions are seen as an homogeneous charge density, minus the susceptibility of a homogeneous electron gas. The latter subtraction yields a divergence-free quantity when $V\rightarrow\infty$, even if the continuum electrons are accounted for.
In the thermodynamic limit, we just get for the electric susceptibility:
\begin{align}
\chi_{\vec{k},\omega}
&=\chi_{\text{e},\vec{k},\omega}+
n_\text{i}\frac{\Delta\chi^{\delta\hat{v}_{\text{pert},\bullet,\vec{k}}}_{1,\omega}}{\epsilon_0\vec{E}_0^2}\equiv \chi_{\text{e},\vec{k},\omega}+\Delta\chi_{1,\vec{k},\omega}
\label{eq_cluster_chi}
\end{align}
where $\Delta\chi^{\delta\hat{v}_{\text{pert},\bullet,\vec{k}}}_{1,\omega}$ is evaluated in an infinite volume.

\subsection{Electric-Dipole Approximation}
The perturbation induced by an isolated ion, or by the VAAQP ion (nucleus and cavity) has a limited spatial range. For wavelengths large compared to the typical range of this perturbation, the dipole approximation is justified. Making the dipole approximation in the Babushkin gauge (see \cite{Grant74} for a discussion on the gauge choice), and setting the $z$-axis along $\vec{k}$, the operator $\hat{V}_\text{pert}(t)$ is expressed as:
\begin{align}
\hat{V}_\text{pert}(t)=\int d^3r\big\{ \underbrace{q_\text{e}E_0 z}_{\equiv\delta v_{\text{pert},\omega}(\vec{r})} \hat{n}_\vec{r} \big\}\cos(\omega t)
\end{align}
Choosing the $z$ axis along the direction of propagation, we get from Equation~\eqref{eq_LR_chi_R}:
\begin{align}
\chi^{\delta\hat{v}_{\text{pert,C},\vec{k}}}_{V,\omega}= q_\text{e}^2E_0^2 \int_V d^3r d^3r'\left\{z z'\mathcal{D}_\omega^\text{R}(\vec{r},\vec{r}')\right\}
\end{align}
where $\mathcal{D}_\omega^{R}$ is the density-susceptibility matrix, that is:
\begin{align}
\mathcal{D}_\omega^\text{R}(\vec{r},\vec{r}')
=-\frac{i}{\hbar}\int_{-\infty}^{+\infty} d\tau \left\{ 
\text{Tr}\left(\hat{\rho}\left[\hat{n}_{\vec{r}}^\text{Heisen.}(\tau),\hat{n}_{\vec{r}'}^\text{Heisen.}(0)\right]\right)\theta(\tau)e^{i\omega\tau}
\right\}
\end{align}

In the framework of the time-dependent density-functional theory (TD-DFT; see
\cite{Stott80b,Zangwill80,Runge84,Dhara87,Gosh88}), one can properly relate the density-susceptibility matrix to the response of the electron density to the frequency-dependent perturbing potential $\delta v_{\text{pert},\omega}(\vec{r})$:
\begin{align}
\delta n_\omega(\vec{r})
=
\int d^3r'\left\{\mathcal{D}_\omega^\text{R}(\vec{r},\vec{r}')
\delta v_{\text{pert},\omega}(\vec{r}')\right\}
\end{align}
where $\delta n_\omega(\vec{r})$ is the frequency-dependent perturbation of the density, resulting from $\delta v_{\text{pert},\omega}(\vec{r})$.

In the isolated-ion case, writing Equation~\eqref{eq_AAII_chi} in this approximation, we obtain:
\begin{align}
\chi_{\text{AAII},\vec{k},\omega}=4\pi\ee^2 n_i\int d^3rd^3r'\left\{z z'
\mathcal{D}^{\text{R}}_{\text{AAII},\omega}(\vec{r},\vec{r}')\right\}
\end{align}
In the VAAQP context, Equation~\eqref{eq_cluster_chi} leads to:
\begin{align}
\Delta\chi_{1,\vec{k},\omega}=4\pi\ee^2 n_i\int d^3rd^3r'\left\{z z' \left[
\mathcal{D}^{\text{R}}_{1,\omega}(\vec{r},\vec{r}')-\mathcal{D}^{\text{R}}_{0,\omega}(\vec{r},\vec{r}') \right]\right\}
\label{eq_cluster_Deltachi1_D}
\end{align}
These equations directly relate the electric susceptibility of the plasma to the atomic polarizability, and are thus the Clausius-Mossotti relations associated to these atomic models.

\section{Independent Particle Approximation and the Effect of Screening}
\subsection{Generalities}
The simplest approximate approach to the atomic retarded susceptibility is to use the retarded susceptibility of the effective system of independent particles. This amounts to replacing, for the 1-center system (isolated ion or VAAQP ion), the Hamiltonian $\hat{H}_\text{static}$ by $\tilde{H}_\text{eff}$. The corresponding susceptibility can be obtained directly from the dynamic perturbation theory of the effective 1-electron system.
\begin{align}
&\text{Im}\,\mathcal{D}_\omega^{\text{R}}(\vec{r},\vec{r}') 
\approx \text{Im}\,\mathcal{D}_\omega^{\text{R},\text{indep}}(\vec{r},\vec{r}')=\sum_{\xi,\zeta}\left(p_\text{F}(\mu,T,\varepsilon_\xi)-p_\text{F}(\mu,T,\varepsilon_\zeta)\right)
\frac{\varphi_\xi^*(\vec{r})\varphi_\xi(\vec{r}')\varphi_\zeta(\vec{r})\varphi_\zeta^*(\vec{r}')}
{\varepsilon_\zeta-\varepsilon_\xi-\hbar\omega}
\label{eq_indep_part_approx}
\end{align}
where $\xi$, $\zeta$ label the orbitals of the average-atom model, and where the sums run over both the discrete and the continuum part of the 1-electron spectrum in the case of the VAAQP ion.

%This approach disregards collective phenomena such as plasma oscillations, which stem from the feedback of electrons on the potential (potential induced by the density perturbation). It also disregards the mixing of channels that occurs because perturbed wavefunction can mix stationary wavefunctions from both the bound and continuum parts of the spectrum.

Averaging over the polarization and direction of propagation, the independent-particle approximation leads to the average-atom Kubo-Greenwood formula:
\begin{align}
\text{Im}(\chi_\omega) =4\pi n_\text{i}\ee^2\frac{\pi}{3}& \sum_{\xi,\zeta} \left(p_\text{F}(\mu,T,\varepsilon_\xi)-p_\text{F}(\mu,T,\varepsilon_\zeta)\right)
\left| \langle\varphi_\xi \right| \tilde{\vec{R}}\left|\varphi_\zeta \rangle \right|^2\delta(\hbar\omega-\hbar\omega_{\zeta,\xi})
\label{eq_AA_Kubo_Greenwood}
\end{align}
where $\hbar\omega_{\zeta,\xi}=\varepsilon_\zeta-\varepsilon_\xi$.
Because channel-mixing is disregarded, the photoabsorption can be decomposed into bound-bound, bound-continuum, and continuum-continuum contributions.
Although we use the average-atom model here as an example for the discussion, a similar treatment can be performed in the case of a more detailed model, yielding the contributions of the various excited states to the plasma electric susceptibility function. The overwhelming majority of models used for opacity calculations is based on the independent-particle approximation. 

The oscillator strengths are the numbers defined as follows (see, for instance, \cite{BetheSalpeterbook,Cowan}):
\begin{align}
f_{\xi,\zeta}=\frac{2}{3}\frac{m_\text{e}}{\hbar^2}\hbar\omega_{\zeta,\xi}
\left| \langle\varphi_\xi \right| \vec{R}\left|\varphi_\zeta \rangle \right|^2
\label{eq_def_osc_str}
\end{align}
when $\xi$, $\zeta$ belong to the discrete part of the spectrum. In the case where either $\xi$ or $\zeta$ belongs to the continuum, this expression is to be understood as a density of oscillator strength, also called a differential oscillator strength. 

When both $\xi$ and $\zeta$ belong to the continuum, the dipole matrix elements appearing in Equation~\eqref{eq_def_osc_str} are conditionally convergent integrals. In the case of isolated ions, continuum electrons are excluded from the ions' electronic structure, which amounts to set the populations $p_\text{F}$ of continuum orbitals to zero. The continuum-continuum matrix elements thus disappear. The free-electron contribution to the susceptibility may however be accounted for in the $\chi_{\text{e},\omega}$ term of equation~\eqref{eq_AAII_chi}.

In the case of the VAAQP model, continuum orbitals have nonzero populations. Subtraction of the contribution of the homogeneous plasma in principle ensures the convergence of the first-order term $\Delta\chi_{1,\omega}$. In practice, in the independent-particle approximation, there is no need to subtract the contribution of the homogeneous plasma. One can use the Ehrenfest theorem in order to recast the dipole matrix elements into their acceleration form (see, for instance, \cite{BetheSalpeterbook}), yielding convergent integrals:
\begin{align}
\left| \langle\varphi_\xi \right| \vec{R}\left|\varphi_\zeta \rangle \right|
=\frac{1}{m_\text{e}\omega_{\zeta,\xi}^2}
\left|\int d^3r \left\{
 \langle\varphi_\xi | \vec{r}\rangle
\vec{\nabla}_{\vec{r}}v_\text{trial}(r)
 \langle\vec{r} | \varphi_\zeta\rangle
\right\} \right|
\label{eq_accel_form}
\end{align}
where $v_\text{trial}(r)$ is the potential associated with the orbitals $\{\ket{\varphi_\xi} \}$.
%The role of dipole matrix elements involving two continuum orbitals is rather specific to plasma physics because it implies non-zero populations for the continuum orbitals.
\begin{figure}[t]
\begin{center}
\includegraphics[width=8cm] {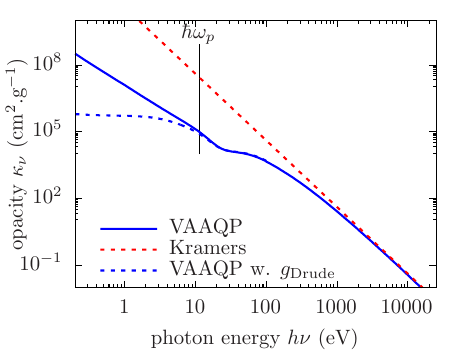} 
\end{center}
\caption{Continuum-continuum contribution to the opacity of a silicon plasma at 5 eV temperature and 2.36 g.cm$^{-3}$ matter denisty. Calculation using orbitals from the VAAQP model, and comparison with the Kramers formula and the opacity corrected using the $g_\text{Drude}$ function of Equation~\eqref{eq_g_Drude}.
\label{fig_free_free}}
\end{figure}

An example of the continuum-continuum contribution to the opacity of a plasma is given in Figure~\ref{fig_free_free}, corresponding to silicon at 2.36 g.cm$^{-3}$ matter density and 5 eV temperature. In the presented calculation, the double sum, or more precisely the double integral, of Equation~\eqref{eq_AA_Kubo_Greenwood} was performed using continuum wave functions obtained from the VAAQP model. Approximate methods allowing one to avoid the double summation exist (see, for instance,~\cite{Perrot96}). One can see in this figure that, at high frequency, one recovers the Kramers classical result~\cite{Kramers23,Zeldovich} involving the bare-nucleus charge, and not an effective ion charge. This is expected since the atomic response at high frequencies essentially involves the electrons located in the vicinity of the nucleus. At low frequencies, one recovers the elastic-scattering limit (see \cite{Somerville64,Ashkin66}):
\begin{align}
\lim_{\omega\rightarrow 0}\text{Im}(\chi_\omega) = \frac{32}{3}\frac{n_\text{i}\ee^2}{\hbar m_\text{e} \omega^3}\sum_{\ell}(\ell+1)\int_0^\infty d\varepsilon
\left\{
\varepsilon\frac{\partial p_\text{F}(\varepsilon)}{\partial \varepsilon} \sin^2
\left(\Delta_{\varepsilon,\ell+1}-\Delta_{\varepsilon,\ell}\right)
\right\}
\label{eq_Somerville}
\end{align}
with $\Delta_{\varepsilon,\ell}$ being the scattering phase shift. This behavior yields a $1/\omega^2$-divergence of the opacity at zero frequency, often called ``infrared divergence''.

\subsection{Effects of Screening}

When the effective potential defining the orbitals is screened, the oscillator strengths have a behavior that is qualitatively different from those of an isolated ion. The underlying reasons are closely related to the limitation of the number of bound states.

For a potential having a Coulomb-tail (pure Coulomb potential or potential stemming from an isolated-ion model), the continuity of the cross-section across the ionization threshold may easily be expressed through the matching of the two quantities:

\begin{align}
\left.f_{\xi,\zeta}\frac{1}{d\varepsilon_\xi/dn_\xi}\right|_{\varepsilon_\xi\rightarrow 0-}
=
\left.f_{\xi,\zeta}\frac{n_{\xi}^3}{Z^{*\,2}}\right|_{\varepsilon_\xi\rightarrow 0-}
=
\left.\frac{df_{\ell_\xi,\zeta}(\varepsilon)}{d\varepsilon}\right|_{\varepsilon\rightarrow 0+}
\label{eq_continuity_osc_str}
\end{align}
where $\xi\equiv(n_{\xi},\ell_{\xi},m)$ for a bound orbital and $\xi\equiv(\varepsilon,\ell_{\xi},m)$ for a continuum orbital, $n_\xi$ and $\ell_\xi$ being the principal and orbital quantum numbers, respectively. $(d\varepsilon_\xi/dn_\xi)^{-1}$ is related to the density of states of the quasi-continuum of infinitely close discrete states in the $n\rightarrow\infty$ limit, for a Coulomb-tail potential.

For Coulomb-tail potentials, we have a finite value of the differential oscillator strength at the threshold. With screening, the behavior of radial wave functions at infinity is changed: the radial wave functions tend to Bessel functions instead of Coulomb functions in the case of Coulomb potential. Due to the related change in the normalization coefficient, the differential oscillator strength smoothly goes  to zero~\cite{Shore75}. 
This change in the behavior of oscillator strengths becomes more pronounced as density is increased since the potential is screened over shorter distances.

\begin{figure}[p]
\begin{center}
\includegraphics[width=16cm] {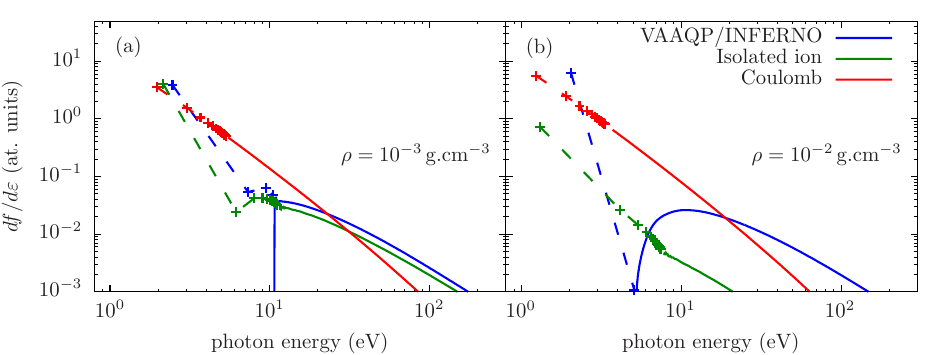} 
\end{center}
\caption{Differential oscillator strengths for the $4s-p$ (continuous solid lines) and oscillator strengths for the discrete $4s-np$ transitions (crosses connected by dashed lines, with values being multiplied by the $(d\varepsilon_\xi/dn_\xi)^{-1}$ term of Equation~\eqref{eq_continuity_osc_str}), for a silicon plasma at 5-eV temperature and matter densities of $10^{-3}$ ({a}) and $10^{-2}$ g.cm$^{-3}$ ({b}). Comparison between results from VAAQP/INFERNO (in blue, same results at these conditions), from an isolated ion having the average configuration taken from VAAQP/INFERNO (in green) and from a Coulomb potential with charge $Z^*$ taken from VAAQP/INFERNO (in red).
\label{fig_osc_strengths_drop}}
\end{figure}
\begin{figure}[p]
\begin{minipage}{8.5cm}
\begin{center}
\includegraphics[width=8cm] {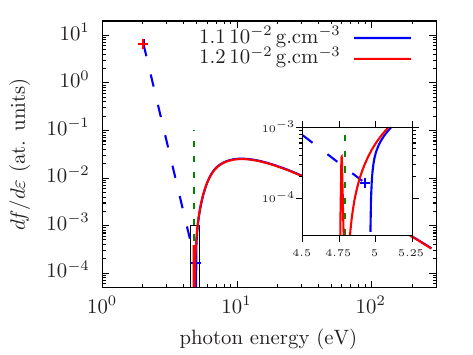} 
\end{center}
\caption{Differential oscillator strengths for the $4s-p$ (continuous solid lines) and oscillator strengths for the $4s-np$ transitions (crosses connected by dashed lines, values multiplied by the $(d\varepsilon_\xi/dn_\xi)^{-1}$ term of Equation~\eqref{eq_continuity_osc_str}), for a silicon plasma at 5-eV temperature and matter densities of $1.1\,10^{-2}$ and $1.2\,10^{-2}$ g.cm$^{-3}$. Results are taken from VAAQP/INFERNO (same results at these conditions). Between the two matter densities considered, the 5p subshell disappears and is replaced by the corresponding resonance in the $p$ continuous spectrum. The vertical dotted line in green indicates the position of the 4s photo-ionization threshold obtained from the Stewart-Pyatt formula~\cite{StewartPyatt66}. 
\label{fig_osc_strengths_press_ioniz}}
\end{minipage}
\hspace{1cm}
\begin{minipage}{8.5cm}
\begin{center}
\includegraphics[width=8cm] {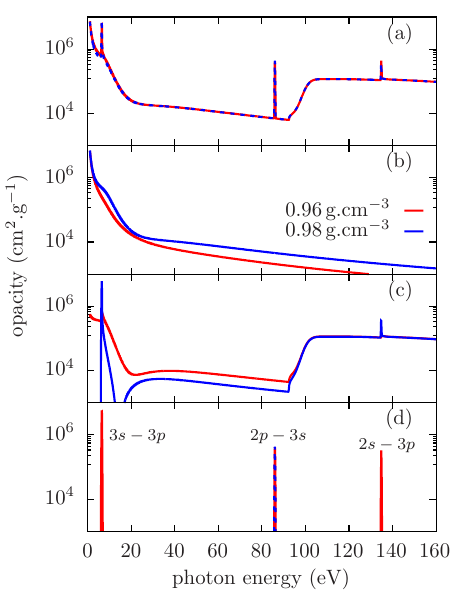} 
\end{center}
\caption{Opacity of a Si plasma at $5$ eV temperature, for matter densities of $0.96$ and $0.98$ g.cm$^{-3}$ (delocalization of the 3p subshell). Total opacity (a), continuum-continuum (b), bound-continuum (c) and bound-bound (d) contributions. For the sake of plotting, an arbitrary 1 eV line broadening was applied to all contributions.
\label{fig_continuity_spectrum}}
\end{minipage}
\end{figure}

Figures~\ref{fig_osc_strengths_drop}\,a and b show an illustration of this oscillator-strength drop near the photo-ionization threshold in the case of silicon at $5$ eV temperature and matter densities of $10^{-3}$ and $10^{-2}$ g.cm$^{-3}$, respectively. In these figures, oscillator strengths are multiplied by the $(d\varepsilon_\xi/dn_\xi)^{-1}$ term of Equation~\eqref{eq_continuity_osc_str}, in order to emphasize the continuity with differential oscillator strengths. The results from the VAAQP model (or INFERNO, with both models being in agreement) are compared to those of an isolated ion with an average configuration fixed to the VAAQP average configuration, as well as to results from a Coulomb potential with a charge fixed to the VAAQP mean ionization. For both the Coulomb potential and the isolated-ion, in principle, the set of bound states is infinite, as well as the series of oscillator strengths. At the low density of $10^{-3}$ g.cm$^{-3}$, one can see that despite the qualitatively different behavior of oscillator strengths near the photo-ionization threshold, a quantitative agreement is obtained between the VAAQP model and the isolated-ion. On the contrary, at the higher density of $10^{-2}$ g.cm$^{-3}$, the change of behavior has larger quantitative impact.

Using a model such as INFERNO or VAAQP, one accounts for both the decrease in the oscillator strength and the appearance of a resonance when a bound state disappears. Using an isolated ion with continuum lowering in order to suppress subshells does not account for either of these phenomena. Figure~\ref{fig_osc_strengths_press_ioniz} shows the oscillator strengths at two matter densities between which the 5p subshell gets pressure-ionized. One can easily see how the discrete oscillator strength is replaced by an equivalent contribution from a resonance in the differential oscillator strength. Thus, the corresponding bound-bound channel does not disappear but is replaced by a contribution to the bound-continuum channels. For the sake of comparison, the energy of the photoionization threshold obtained from the average-atom of an isolated ion with Stewart-Pyatt IPD is also shown. The location of the threshold is in good agreement with VAAQP in this case of relatively low density, but the cross-section of an isolated-ion model would be different. Moreover, obtaining a fully continuous variation of the opacity with density requires correct and consistent accounting for the resonances in the bound-continuum and continuum-continuum contributions to the opacity. Figure~\ref{fig_continuity_spectrum} illustrates the compensations occurring between the bound-bound, bound-continuum and continuum-continuum contributions to the opacity when a bound states disappear and is replaced by the related resonance in the continuum. In the present case, the 3p subshell becomes pressure-ionized and one sees the $2s-3p$ and $3s-3p$ features disappearing from the bound-bound spectrum (d) and being compensated in the bound-continuum spectrum (c). The $3p-s$ and $3p-d$ bound-continuum contributions disappear  also and are compensated in the continuum-continuum spectrum (b). As a result from these compensations, the total opacity remains continuous, as shown in (a).

\section{Fluctuations around the Average Atomic State and Detailed Modeling \label{sec_fluctuations_DCA}}

Whereas the average microstate of a whole macroscopic plasma may virtually sample many atomic excited states, the average-atom approach is based on the average atomic state of the plasma. 

In terms of detailed atomic modeling, spectral quantities such as the opacity or emissivity may reveal the contributions of the various atomic states that have significant populations, because different atomic states contribute at distinct frequencies. Even when the resulting spectral features are unresolved due to physical broadening, the statistical distribution among the various atomic states yields a statistical broadening of the unresolved feature.

For that reason, some degree of detailed modeling is required to obtain realistic estimates of spectral quantities. The physical pictures underlying average-atom models can often be extended to more detailed modeling. For instance, the models~\cite{Perrot87b,Blenski00} rely on various extensions of the quantum ion-cell model to detailed configuration or super-configuration accounting.  

The variational approach leading to the VAAQP model can be formally generalized to configurations or super-configurations~\cite{Blenski07b}.
In the case of configurations $C\equiv 1s^{Q_{1s}^C}\,2s^{Q_{2s}^C}\,2p^{Q_{2p}^C}...$, the free energy to minimize is:
\begin{align}
&\dot{F}\left\{ \left\{ P_C \right\},n_\text{e},\left\{q^C(\vec{r})\right\}, \left\{R_C\right\};n_i,Z,T \right\}\nonumber\\
&=\dot{F}_0(n_\text{e};n_\text{i},T)+\sum_C P_C \left( 
\Delta F_1^C\left\{ n_\text{e},q^C(\vec{r}),R_C;n_i,Z,T \right\}
+k_BT \ln P_C \right)\label{eq_cluster_config}
\end{align}
where $P_C$ is the probability of configuration $C$, and $R_C$ its cavity radius. First order contribution to the free energy $\Delta F_1^{C}$ is decomposed as in Equation~\eqref{eq_VAAQP_dF1}, with $\Delta F_1^{0,C}$ and $q^C(r)$ expressed as:
\begin{align}
\Delta F_1^{0,C}\left\{n_\text{e},\{\underline{q}^C\};T\right\}
=&\sum_{\xi\text{ bound}}
\int d^3r \Big\{
\frac{Q^C_\xi}{g_\xi}\Big[
\left(\varepsilon_\xi^C-v_\text{trial}^C(r)-T s\left(\frac{Q^C_\xi}{g_\xi}\right)\right)|\varphi_\xi^C(\vec{r})|^2\Big]\Big\}
\nonumber\\
&+
\sum_{\xi\text{ continuum}}
\int d^3r \Big\{
p_\text{F}(\varepsilon_\xi;n_\text{e},T)\Big[\vphantom{\frac{1}{1}}
\left(\varepsilon_\xi-v_\text{trial}^C(r)-T s_\text{F}(\varepsilon_\xi;n_\text{e},T)\right)|\varphi_\xi^C(\vec{r})|^2
\nonumber\\&
\hspace{6cm}-\left(\varepsilon_\xi-T s_\text{F}(\varepsilon_\xi;n_\text{e},T)\right)|\varphi^0_\xi(\vec{r})|^2\Big]
\Big\}
\label{eq_VDCA_dF10}\\
q^C(r)=&\sum_{\xi\text{ bound}}
\frac{Q^C_\xi}{g_\xi}|\varphi_\xi^C(\vec{r})|^2
+\sum_{\xi\text{ continuum}}
p_\text{F}(\varepsilon_\xi;n_\text{e},T)\left(|\varphi_\xi^C(\vec{r})|^2-|\varphi^0_\xi(\vec{r})|^2\right)
\label{eq_VDCA_q}
\end{align}
with the $\{\ket{\varphi_\xi^C}\}$ being obtained by solving the 1-electron Schrödinger Equation~\eqref{eq_Schrod_1electron} associated with the trial potential $v_\text{trial}^C(r)$.

$\Delta F_1^{\text{el},C}$ and $\Delta F_1^{\text{xc},C}$ have the same expressions as in Equations~\eqref{eq_VAAQP_dF1el} and \eqref{eq_VAAQP_dF1xc}, respectively, substituting $q^C(r)$ for $q(r)$ and $R_C$ for $R_\text{WS}$. In the above expressions, and in the following, $Q^C_\xi$ denotes the number of electrons in the subshell to which belong the orbital $\xi$, and $g_\xi$ the subshell degeneracy.

The neutrality condition is required to hold on average over the configurations, according to the 1st order cluster expansion of the density (Equation~\eqref{eq_cluster_formal}):
\begin{align}
Z=\frac{n_\text{e}}{n_\text{i}}+\sum_C P_C \int d^3r \left\{q^C(r)\right\}
\end{align}
The constrained minimization of $\dot{F}$ yields the following equations:
\begin{align}
&v_\text{trial}^{C}(r)=v_\text{el}^{C}(r)+\mu_\text{xc}\left(n^C(r)\right)-\mu_\text{xc}\left(n_\text{e}\right)\label{eq_var_SCF}\\
&v_\text{el}^{C}(R_{C})={v}^*_\text{el}\,\text{,~independent of $C$}\label{eq_var_config_consist}\\
&P_C=\frac{g_C}{\Xi}e^{-\beta\left(\Delta F_1^C - \left(\mu_\text{e}+\mu_\text{xc}(n_\text{e})+\bar{v}_\text{el}\right)Q_C^*\right)}\label{eq_var_config_proba}\\
&\sum_C P_C \int d^3r\left\{v_\text{el}^{C}(r) \theta(r-R_C)\right\}=0\label{eq_var_suppl_cond}
\end{align}

Equation~\eqref{eq_var_config_consist} is a self-consistent condition involving all configuration potentials. Thus, solving Equations \eqref{eq_var_SCF}--\eqref{eq_var_suppl_cond} is in practice beyond reach. However, we used this approach in \cite{Piron13} to build an approximate DCA model,noted hereafter VAAQP-DCA. The latter resorts only to results from the VAAQP model and is based on the following approximations:
\begin{align}
&v^C_\text{trial}(r)\approxeq v^\text{AA}_\text{trial}(r)
\ ;\ 
n_\text{e} \approxeq n_\text{e}^\text{AA}\label{approx_VAAQP_DCA}
\end{align}
where $v^\text{AA}_\text{trial}$ and $n_\text{e}^\text{AA}$ are the trial potential and asymptotic density stemming from the VAAQP approach, respectively.
Within this approximation, the one-electron orbitals $\{\ket{\varphi_\xi^C}\}$ and eigenvalues $\{\varepsilon_\xi^C\}$ for all configurations $C$ are fixed to those of the VAAQP model: $\{\ket{\varphi_\xi^\text{AA}}\}$, $\{\varepsilon_\xi^\text{AA}\}$. Orbital relaxation is thus neglected. This limits in particular the precision of the energies of spectral features. This kind of approach can nonetheless bring useful results when spectroscopy-grade precision is not required, and often gives reasonable estimates of the main spectral features and mean opacities. Moreover, it is also possible to account perturbatively for orbital relaxation using the static linear response~\cite{Liberman94,Hansen23}.

In Equations~\eqref{eq_VDCA_dF10} and \eqref{eq_VDCA_q}, only the bound states contribution to the electron density depends on the configuration $C$, through the occupation numbers $Q_\xi^C$.
We immediately get from the neutrality condition that
\begin{align}
&Z=\underbrace{\sum_{\xi\text{ bound}} \frac{Q_\xi^C}{g_\xi^C}}_{Q_C} 
- \underbrace{\sum_{\xi\text{ bound}}  p_F(\varepsilon_\xi)}_{Q_\text{AA}}  + \frac{4\pi}{3}R_C^3n_0^\text{AA}
+ \underbrace{\int d^3r\left\{q^\text{AA}(r)\right\}}_{Z-Z^*_\text{AA}}
\label{eq_neutr_simplified}
\end{align}
with $q^{AA}(r)$ being the average-atom displaced electron density.
Thus, in the VAAQP-DCA model, the deviation of the bound electron number $Q_C$ with respect to that of the average atom is simply balanced by an adjustment of the cavity radius $R_C$. Given that $R_C$ has to be positive, we should in principle limit ourselves to configurations such that $Q_C-Q_\text{AA}\leq Z^*_\text{AA}$. Other configurations cannot be described using continuum wavefunctions fixed to those of the average-atom.

\begin{figure}[p]
\begin{center}
\includegraphics[width=8cm] {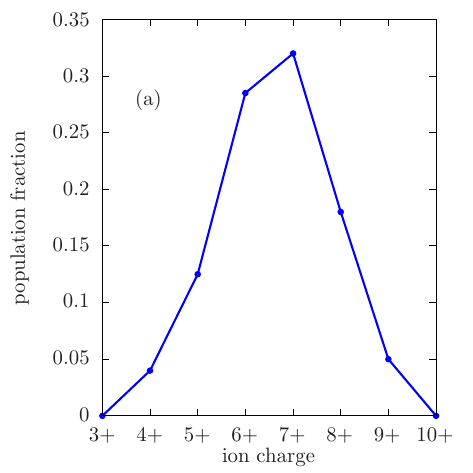} 
\hspace{1cm}
\includegraphics[width=8cm] {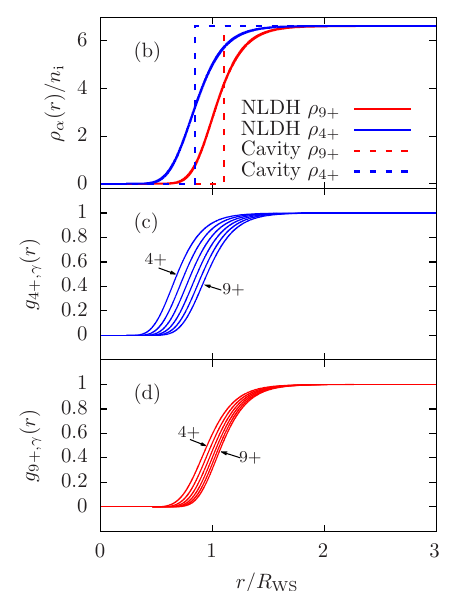} 
\end{center}
\caption{Example of a 6-component classical plasma, typical of the charge-state distribution of an iron plasma at solid density (7.8\,g.cm$^{-3}$) and 40-eV temperature, after~\cite{Piron13} ({a}). Ion-ion coupling of significantly populated ions spans from $\Gamma_{4+,4+}=1.4$ to $\Gamma_{9+,9+}=7.6$.
Plots of the average ion charge density $\rho_\Psi(r)$ around an ion of of species $\Psi$ ({b}) and of the related pair distribution functions $g_{\Psi,\Psi'}(r)$ ({c},{d}) obtained using the NLDH model of the MCP, for the two charge states $\Psi=4+$ and $9+$. Comparison with the Heaviside approximation of $\rho_\Psi(r)$ is shown in ({b}).
\label{fig_OCP_NLDH_IS_g_MCP}}
\end{figure}
\begin{figure}[p]
\includegraphics[width=17cm] {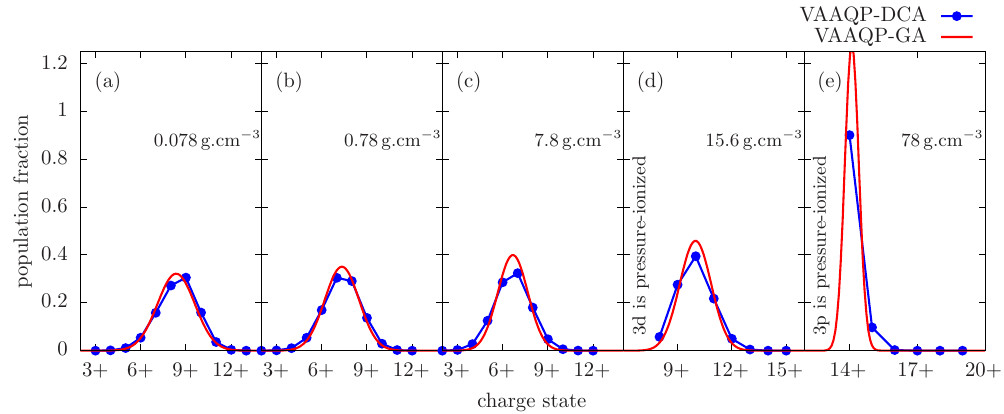} 
\caption{Charge state distributions of an iron plasma at 40-eV temperature, for various matter densities ranging from 1/100th ({a}) to 10 times ({e}) solid density. Comparison between a DCA approach reconstructed from the VAAQP model (VAAQP-DCA)~\cite{Piron13} and an approximate treatment of this model using the Gaussian approximation~\cite{Perrot88b,Piron13} (VAAQP-GA).
\label{fig_CSD_Fe}}
\end{figure}

The formation of a statistical cavity around an ion can be recovered from the mean-field model (nonlinear DH model) of a classical plasma of charged ions. Figure~\ref{fig_OCP_NLDH_IS_g_MCP} presents the result from the nonlinear DH model for a mixture of ions following the charge-state distribution obtained from the VAAQP-DCA model in the case of an Iron plasma at solid density and 40 eV temperature. A comparison with the cavity hypothesis is shown. As is easily understood, higher ion charges are associated with larger cavity radii, because repulsion of other ions is stronger.

The change of $R_C$ in order to ensure neutrality for each ion charge state (Equation~\eqref{eq_neutr_simplified}), which is a consequence of the approximation of Equations~\eqref{approx_VAAQP_DCA}, may be seen as a crude way of accounting for the ion charge in the ion-ion correlation functions.

However, fixing the trial potential, the continuum contribution to the density is fixed to that of the average atom. The model behaves as if continuum electron density was insensitive to the changes of bound-electron configuration and charge state, whereas the ion-ion correlations adapt.
In addition to the crude modeling of ion-ion correlation functions through Heaviside functions, we somehow assume that each ion environment is correctly described through the \emph{equilibrium} correlation functions corresponding to its charge state. These are strong hypotheses, which can obviously be questioned.

Even resorting to simple models, the detailed modeling of plasma still remains an implementation challenge. For elements of moderate or high atomic numbers, especially at high temperatures, the number of excited states that contribute to radiative properties may be enormous. Statistical approaches are available to reduce the number of species, leading to various levels of detail in the spectra. 

%However, one often has to make a tradeoff between the level of detail and the completeness of the approach. Detailed modeling is also used in the collisional-radiative modeling of non-equilibrium plasmas~\cite{Ralchenkobook}. In this context, the number of states or statistical objects that may be accounted for is limited by the rank of the subsequent collisional-radiative matrix. The issue of choosing a relevant tradeoff is even more important in this context (see Section~\ref{sec_completeness_precision_tradeoff}).

From an average-atom standpoint, the populations of the various levels may be obtained from the analysis of fluctuations around the average atomic state (see \cite{LandauStatisticalPhysics}, Chapter 12, and \cite{Green64}). Starting from an approximate detailed model that resorts to the average-atom energies and orbitals, models of fluctuations can be used to perform a statistical approach. An example of such an approach is the Gaussian approximation~\cite{Perrot88b}, which was applied to the VAAQP-DCA model in~\cite{Piron13}. In the context of opacity calculations, the intensity of radiative transitions depends on products of two occupation numbers (occupation of initial shell times vacancy of final shell). The approximation of independent fluctuations of occupation numbers~\cite{Shalitin84,Stein85} leads to an overestimation of the statistical broadening. Correlated fluctuations~\cite{Perrot88b,Blenski90} are required to obtain realistic estimates. 

Figure~\ref{fig_CSD_Fe} shows the charge state distributions obtained for iron at 40 eV temperature, at various matter densities, using either the VAAQP-DCA model or its approximate, statistical treatment through the Gaussian approximation (denoted VAAQP-GA). 

In the case of pressure ionization, the removal of some orbitals from the discrete 1-electron spectrum results in the removal of any configuration having non-zero population of these orbitals. This ultimately leads to a truncation of the charge state distribution, pushing it towards higher charge states.

In Figure~\ref{fig_CSD_Fe}, one can see how the charge state distribution of the DCA model is pushed towards higher ionization stages as available configurations for the lowest ionization stages are removed. The case of 15.6 g.cm$^{-3}$ (2-fold compression, Figure~\ref{fig_CSD_Fe}\,d) illustrates the pressure ionization of the 3d subshell, whereas the case of 78 g.cm$^{-3}$ (10-fold compression, Figure~\ref{fig_CSD_Fe}\,e) illustrates the pressure ionization of the 3p subshell. The Gaussian approximation to the fluctuations cannot account properly for this cut-off, but it still yields the same qualitative trend of a narrow peak on the average charge state. 

\begin{figure}[t]
\begin{center}
\includegraphics[width=8cm] {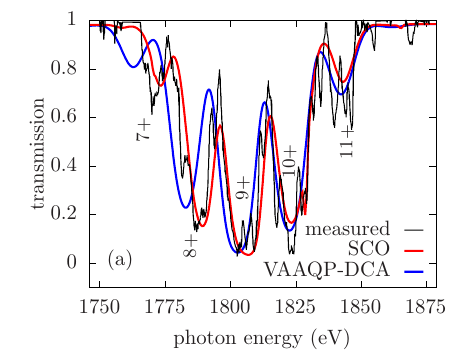} 
\hspace{0.5cm}
\includegraphics[width=8cm] {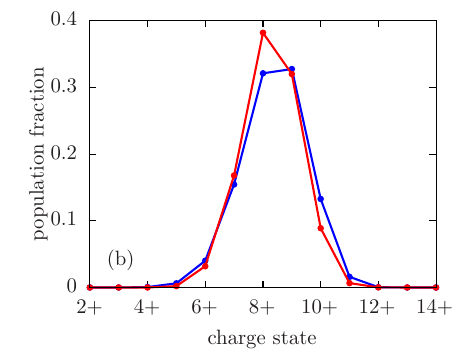} 
\end{center}
\caption{Transmission of a silicon plasma from \vphantom{~\cite{Perrot87b,Blenski00,Blenski07b,Piron13,Green64,Perrot88b,Shalitin84,Stein85,Blenski90}
}~\cite{Wei08}, at areal density of 80$\mu$g.cm$^{-2}$. Estimated plasma conditions are 60 eV temperature and 45 mg.cm$^{-3}$ matter density. Comparison between results from the VAAQP-DCA approach~\cite{Piron13}, from the SCO approach~\cite{Blenski00}, and measured transmission spectrum from~\cite{Wei08}. In the calculations, an arbitrary line width of 4 eV was added to the statistical width to mimic the physical broadening and instrumental resolution.
\label{fig_Si_SG2}}
\end{figure}

As an illustration of the need for detailed modeling to obtain realistic spectra, Figure~\ref{fig_Si_SG2}\,b displays the ion charge state distributions for silicon at 60 eV temperature and 45 mg.cm$^{-3}$ matter density resulting from two detailed models: the DCA model from~\cite{Piron13} and the STA model of~\cite{Blenski00}. In Figure~\ref{fig_Si_SG2}a the corresponding transmission spectra are shown and compared to the measured spectrum from~\cite{Wei08} (areal density of 80$\mu$g.cm$^{-2}$). The latter measure was in fact performed on a plasma of SiO$_2$. However, the ionization competition between oxygen and silicon in the mixture was studied using an isolated-ion detailed model with Stewart-Pyatt continuum lowering in~\cite{Yongjun15}. Its effect on the populations of silicon was shown to be rather weak. 

In Figure~\ref{fig_Si_SG2}\,a, one can easily identify the contributions from the various charge states. One may note a slight shift in the positions of some spectral features obtained from the VAAQP-DCA model. This is due to the lack of orbital relaxation in this approach, whereas it is accounted for in the STA model of~\cite{Blenski00}.

\section{Various Implications of Collective Phenomena\label{sec_collective}}

What the linear response of the atom describe is really the response of the electron clouds around the atom, which, depending on the model, may include the response of continuum electrons. This picture is essentially relevant to frequencies at which small displacements of the electrons occur due to the perturbing field. Essentially, this is well adapted in the dielectric regime typical of frequencies higher than the plasma frequency $\omega_\text{P}$
\begin{align}
\omega_\text{P}=\sqrt{\frac{4\pi n_\text{e}\ee^2}{m_\text{e}}}\label{eq_plasma_freq}
\end{align}
%From the standpoint of classical mechanics, this corresponds to the regime of Bremstrahlung by electrons that are weakly deflected passing in the vicinity of an ions, at atomic scale. 

Close to the plasma frequency, the perturbing field is resonant with the natural frequency of the free-electron gas, collective excitations of free electrons (plasmons, a.k.a. Langmuir waves) are then expected to play a crucial role. In a model that rely on fictitious independent particles in an effective potential, the collective behavior of the many-particle system stems from the self-consistency of the potential. In the framework of the independent-particle approximation of previous section, the self-consistency is accounted for only in the static screening of the atom. There is no self-consistent accounting for the potential induced by the dynamical perturbation of the density. This is a shortfall, and its impact is not necessary limited to the vicinity of the plasma frequency. Indeed, close to the atom, polarization of electrons takes place and may generate a complex coupling between the exciting field and the plasmons at various frequencies (notion of local plasma frequency).

At frequencies lower than the plasma frequency, the quasi-static collective behavior of the plasma, i.e.\ the screening, prevents the electromagnetic field from propagating in the plasma. The plasma behaves as a conductor, and speaking in terms of dynamic conductivity is more relevant than in terms of opacity.

In this low-frequency regime, free electrons may travel at long distances, greater than inter-ionic distances. It becomes necessary to account for the presence of other ions since the collisions have an impact on the conductivity. Physical modeling of the plasma at the scale of multiple ions is well known in the context of classical kinetic theory of plasmas. More precisely, modeling of low-frequency inverse Bremsstrahlung absorption is possible through the Boltzmann equation in the relaxation-time approximation (see, for instance, \cite{Bekefi}). In this description, the average atom model may provide a relevant estimate of the collision frequency through the electron-ion scattering cross-section.

For a correct description of radiative properties in all regimes, it is desirable to address each of these topics. Moreover, as is clear from Equation~\eqref{eq_plasma_freq}, collective effects have an impact on a frequency range that extends farther as density is increased. They thus have a specific importance for dense plasmas. However, a fully consistent description of all these phenomena is still an open problem. The next section deals with a heuristic way of accounting for the electron-ion scattering in the dynamic conductivity at low frequency. Section~\ref{sec_sc_linear_response} outlines our effort to tackle the dynamic self-consistent linear response of the VAAQP model.

\section{Collisions and Regularization at Low Frequencies\label{sec_reg_low_freq}}

The continuum-continuum opacity obtained from the independent-particle approximation exhibits an unphysical divergence at zero frequency (see Figure~\ref{fig_free_free}). On the other hand, the Drude model, which accounts for collisions (drag force), yields a finite value of the direct-current conductivity, as well as of the corresponding opacity. 
%Somerville, Ashkin formula and limit.

In~\cite{Perrot96}, a very simple, heuristic approach is proposed in order to recover the Drude-like collective behavior at low frequency. Very similar approaches are also described in~\cite{Johnson06,Kuchiev08,Johnson09}.

From the Boltzmann equation in the relaxation-time approximation, one can derive Ziman's static conductivity~\cite{Ziman61}:
\begin{align}
\sigma_\text{Ziman} = -\frac{2}{3}\frac{n_\text{e} q_\text{e}^2\hbar^2}{m_\text{e}^2}\int \frac{d^3k}{(2\pi)^3}\left\{\frac{k^2}{\omega_\text{col}(k)}  \left.\frac{\partial f_0(\varepsilon)}{\partial \varepsilon}\right|_{\varepsilon_k}\right\}
\label{eq_Ziman}
\end{align}
where $\varepsilon_k=\hbar^2k^2/(2m_\text{e})$, $f_0$ is the free-electron energy distribution (normalized to unity) and $\omega_\text{col}$ is the collision frequency. In our case, $n_\text{e}f_0$ is just the Fermi-Dirac distribution $p_\text{F}$.

In the quantum-mechanical framework, the collision frequency $\omega_\text{col}(k)$ of electrons can be related to the net rate of elastic scattering out of the momentum $\hbar\vec{k}$. We may estimate the latter by summing the electron-ion elastic-scattering cross-section in the limit of weak scattering (see, for instance, \cite{SobelmanVainshteinYukov}). One obtains:
\begin{align}
\omega_\text{col}(k)=\frac{4\pi n_i\hbar}{m_\text{e} k}\sum_\ell (\ell+1) \sin^2
\left(\Delta_{\varepsilon_k,\ell+1}-\Delta_{\varepsilon_k,\ell}\right)
\label{eq_omega_coll}
\end{align}
with $\Delta_{\varepsilon,\ell}$ being the scattering phase shift. 

The method of correction of~\cite{Perrot96} for the conductivity or opacity is as follows. Starting from Equation~\eqref{eq_diel_conductivity} and from the elastic-scattering limit of Equation~\eqref{eq_Somerville}, one writes the low-frequency dynamic conductivity. Identifying the collision frequency of Equation~\eqref{eq_omega_coll} in this expression, we obtain:
\begin{align}
\lim_{\omega\rightarrow 0}\text{Re}\left( \sigma_{\omega}\right)
=-\frac{2}{3}\frac{q_\text{e}^2\hbar^2}{m_\text{e}^2}
\int \frac{d^3k}{(2\pi)^3}
\left\{
k^2 \left.\frac{\partial p_\text{F}(\varepsilon)}{\partial \varepsilon}\right|_{\varepsilon_k}
\frac{\omega_\text{col}(k)}{\omega^2}\right\}
\end{align}
Then, by analogy with Ziman's formula Equation~\eqref{eq_Ziman} one builds a correcting factor $g_\text{Drude}$ that allows removing the singularity and recovering Ziman's result at zero frequency, while having no effect at high frequency. 
\begin{align}
&g_\text{Drude}(k,\omega)=\frac{\omega^2}{\omega^2+\omega_\text{col}^2(k)}\rightarrow
\begin{cases}
\omega^2/\omega_\text{col}^2&\text{ for $\omega << \omega_\text{col}$}\\
1&\text{ for $\omega >> \omega_\text{col}$}
\end{cases}
\label{eq_g_Drude}
\end{align}
The resulting, regularized continuum-continuum contribution to the electric susceptibility then writes:
\begin{align}
&\text{Im}(\chi_\omega^\text{reg}) =4\pi n_\text{i}\ee^2\frac{\pi}{3} 2\sum_{\ell,m_\ell}\sum_{\ell',m_\ell'}\int_0^\infty d\varepsilon
\left\{g_\text{Drude}(k_\varepsilon,\omega) \left(p_\text{F}(\varepsilon)-p_\text{F}(\varepsilon+\hbar\omega)\right)
\left| \langle\varphi_{\varepsilon,\ell,m_\ell} \right| \tilde{\vec{R}}\left|\varphi_{\varepsilon+\hbar\omega,\ell',m_\ell'} \rangle \right|^2\right\}
\end{align}
with $k_\varepsilon=\sqrt{2m_\text{e}\varepsilon}/\hbar$.

The $g_\text{Drude}$ factor introduces a Drude-like behavior in the low-frequency part of the dynamic conductivity and spectral opacity. 
Figure~\ref{fig_free_free} shows the effect of the correcting function $g_\text{Drude}$ in the case of silicon at 5 eV temperature and 2.36 g.cm$^{-3}$ matter density. This correction has a strong impact below the plasma frequency.

At low frequencies, the complex refraction index may also have a significant imaginary part. The assumption $n_\omega^\text{ref}=1$, often used in the dielectric regime ($\omega >> \omega_\text{P}$), is no longer valid, and a more realistic estimate is required. In~\cite{Perrot96}, a simple estimate obtained from the Drude formula is used. However, since $\text{Im}(\chi_\omega)$ is known, one can also use the Kramers-Kronig relations to obtain $\text{Re}(\chi_\omega)$ (see, for instance, \cite{LandauStatisticalPhysics}, \textsection 123):
\begin{align}
\text{Re}(\chi_\omega)
=
\frac{1}{\pi}\mathcal{PP}\int_{-\infty}^{+\infty}
d\omega'\left\{
\frac{\text{Im}(\chi_{\omega'})}
{\omega'-\omega}
\right\}
\label{eq_Kramer_Kronig}
\end{align}
Then, one uses Equations~\eqref{eq_real_nref} and \eqref{eq_k_abs} to obtain the absorption coefficient.

\begin{figure}[t]
\begin{center}
\includegraphics[width=8cm] {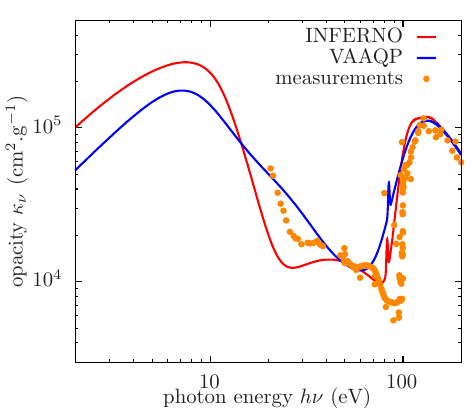} 
\end{center}
\caption{Opacity of cold silicon at solid density (2.36 g.cm$^{-3}$) in the visible-to-XUV range, typical of the L and M edges. Results from INFERNO and VAAQP were obtained at a temperature of 2.5 eV, using the correction $g_\text{Drude}$ of Equation~\eqref{eq_g_Drude} and the refraction index obtained from the Kramers-Krönig Equation~\eqref{eq_Kramer_Kronig}. Comparison with measurements of the opacity of cold silicon from \cite{Henkereport}.  
\label{fig_si_cold_opacity}}
\end{figure}

Figure~\ref{fig_si_cold_opacity} displays the results of the present approach~\cite{Piron18}, using the heuristic coefficient $g_\text{Drude}$ and the refraction index obtained from the Kramers-Kronig relation, for the case of silicon at solid density and 2.5 eV temperature, using the VAAQP and INFERNO models. Rather good agreement is found with measurements performed on cold solid silicon~\cite{Henkereport}. In fact, with cold solid silicon being a metal, it is not so surprising that plasma models can give a reasonable description of it.

The method of~\cite{Perrot96,Johnson06,Kuchiev08,Johnson09} enables a smooth transition to the static collective behavior of the plasma, accounting for electron-ion collisions, but does not account for the dynamic screening by electrons, which notably yields the collective plasma oscillations. This is the subject of next section.

\section{Self-Consistent Linear Response \label{sec_sc_linear_response}}
Making the independent-particle approximation of Equation~\eqref{eq_indep_part_approx}, one disregards the excitation of collective electron modes. In any approach based on an effective system of independent particles, the collective behavior is accounted for through the self-consistency of the effective potential. When using Equation~\eqref{eq_indep_part_approx}, the static electron density and atomic potential are calculated self-consistently, but the frequency-dependent perturbed density and potential are not.

A relevant approach to account for the collective behavior in the response of the ion is the self-consistent linear-response theory. This was the subject of C. Caizergues' Ph.D. thesis. 

Using the formalism of TD-DFT~\cite{Stott80b,Zangwill80,Runge84,Dhara87,Gosh88} to calculate the linear response leads to a self-consistent scheme for the frequency-dependent density perturbation and induced potential:
\begin{align}
&\delta n_\omega(\vec{r})
=
\int d^3r'\left\{\mathcal{D}_\omega^\text{R}(\vec{r},\vec{r}')
\delta v_{\text{pert},\omega}(\vec{r}')\right\}
=
\int d^3r'\left\{\mathcal{D}_{\omega}^{\text{R,indep}}(\vec{r},\vec{r}')
\left(\delta v_{\text{ext},\omega}(\vec{r}')+\delta v_{\text{ind},\omega}(\vec{r}')\right)\right\}
\label{eq_LR_TDDFT1}
\\
&\delta v_{\text{ind},\omega}(\vec{r})
=\ee^2
\int d^3r'\left\{
\frac{\delta n_\omega(\vec{r}')}{|\vec{r}-\vec{r}'|}\right\}
+\left.\frac{\partial \mu_\text{xc}(n)}{\partial n}\right|_{n(\vec{r})} \delta n_\omega(\vec{r})
\label{eq_LR_TDDFT2}
\end{align}
Here, $\delta v_{\text{ind},\omega}$ corresponds to the potential induced by the perturbation of the density $\delta n_\omega$.
In the above equations, we limit ourselves to the adiabatic local density approximation of the exchange-correlation term.

\subsection{Self-Consistent Linear Response of the Thomas-Fermi Ion at Finite Temperature}
As a first step we can consider the self-consistent linear response of the semiclassical version of the VAAQP model, which is equivalent to the usual TF model (see \cite{PironPhD}). Just as the TF model can be seen as an hydrostatic model of a charged ideal gas, the dynamic behavior of the system can be addressed through hydrodynamic equations. The hydrodynamics of a charged ideal gas of electron is usually referred to as Bloch's hydrodynamics \cite{Bloch33}. 

The study of the self-consistent linear response of the TF model was the first part of C. Caizergues' Ph.D. work. In this effort, we have benefited from the pioneering paper \cite{Ball73} and also from the work performed by K. Ishikawa during his Ph.D. thesis \cite{IshikawaPhD,Ishikawa98,Ishikawa98b}. The system of interest in the latter study was the TF model of an impurity in a jellium, but the numerical methods for solving the equations of Bloch-hydrodynamics can be adapted to the VAAQP model (see \cite{CaizerguesPhD,Caizergues14}).

As regards the formalism, the study of the semiclassical version of the model \cite{Caizergues14} allowed us to derive the following relation in the framework of Bloch's hydrodynamics:
\begin{align}
\int d^3r\left\{z\delta n_\omega(\vec{r})\right\}
=\frac{1}{m_\text{e}\omega^2\left(1-\frac{\omega_\text{P}^2}{\omega^2}\right)}
\left(
-\int  d^3r \left\{
\delta n_\omega(\vec{r})
\frac{\partial v_\text{trial}(\vec{r})}{\partial z}
\right\}
+\int  d^3r \left\{
\delta v_{\text{ind},\omega}(\vec{r})
\frac{\partial q(\vec{r})}{\partial z}
\right\}
\right)
\label{eq_sum_ehrenfest}
\end{align}
where $z$ denotes the projection of $\vec{r}$ along the $z$-axis.
This relation was derived previously in the quantum framework in \cite{Blenski06} and plays the same role as the switching between the length and acceleration form of the dipole matrix element, in the context of the self-consistent linear response. 

\begin{figure}[t]
\begin{center}
\includegraphics[width=8cm] {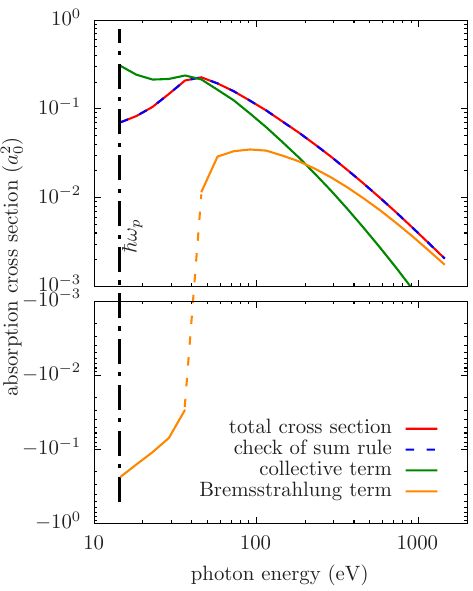} 
\end{center}
\caption{Photoabsorption cross-section of Aluminum at temperature $10$ eV and matter density $2.7$ g.cm$^{-3}$, according to the self-consistent linear response in the Thomas-Fermi approximation. The photoabsorption was calculated only above the plasma frequency $\omega_\text{P}$.
\label{fig_TF_LR_Al}}
\end{figure}

The factor in front of the right-hand side stems from the contribution of the homogeneous medium and results in a singularity at the plasma frequency $\omega_\text{P}$, related to $n_\text{e}$. This is due to the accounting for the induced field in the homogeneous plasma response, which yields the collective resonant plasma oscillation. The present model disregards collisions with other ions, which should lead to a saturation of this resonant behavior. 

If one considers the $\omega >> \omega_\text{P}$ limit, then the first term on the right-hand side of Equation~\eqref{eq_sum_ehrenfest} may be recovered from Equation~\eqref{eq_accel_form}. For that reason, we call this contribution ``Bremsstrahlung'' term.  The second term is purely due to the accounting for the induced potential, that is, for self-consistency in the dynamic behavior of the displaced-electron density. Let us call it ``collective'' term. Using Equation~\eqref{eq_sum_ehrenfest} allows one to make a distinction among those two contributions, and evaluate the role of collective phenomena in the photoabsorption.

Figure~\ref{fig_TF_LR_Al} shows the photoabsorption cross-section of Aluminum at temperature $10$ eV and solid density, as calculated from the TF self-consistent linear response. The respective contributions of the Bremsstrahlung and collective terms are also shown. As can be seen on this figure, the collective response may be non-negligible even far from the plasma frequency, and can even dominate close to it. Close to the plasma frequency, we also obtain a negative Bremsstrahlung contribution to the absorption, which may be viewed as induced Bremsstrahlung emission.

The TF model is a relevant theoretical test-bed for the VAAQP model. However, this model leads to unphysical results for the radiative properties, because it lacks the quantum shell structure of the atom, and exhibit an unphysical behavior at high frequencies~\cite{Ishikawa98b}. It is therefore of limited practical use. Whether the above conclusions about the importance of the collective term can be extended to the quantum VAAQP model is an open question.

\subsection{Self-Consistent Linear Response of the VAAQP Atom}

In the quantum framework, the self-consistent linear-response approach should enable one to account for the channel mixing between bound-bound, bound-continuum and continuum-continuum channels, in addition to describe the coupling with collective excitation modes. Let us mention that, although the formalism is different, as well as the approximation framework, such channel-mixing effects are of same nature than those addressed in quantum defect theory \cite{Seaton66,Seaton83}.

In the pioneering studies of \cite{Zangwill80,Stott80b,Zangwill84}, Equations~\eqref{eq_LR_TDDFT1} and \eqref{eq_LR_TDDFT2} are solved to obtain the photoabsorption cross-section in cases that do not involve continuum-continuum channels. In particular, \cite{Zangwill80} regards the photoabsorption cross-section of neutral rare gases, and it is showed that channel mixing between bound-bound and bound-continuum contributions have a significant impact near the photo-ionization edge. 

For a plasma, the contribution of continuum-continuum transitions causes difficulties since they involve non-localized wave functions. The study of the self-consistent linear response of the quantum VAAQP model, accounting for continuum electrons, and using the methods described in \cite{Zangwill80,Zangwill84,MahanSubbaswamy} was the second part of C. Caizergues' Ph.D. work. 

Recently, another study regarding the  application of the self-consistent linear response to plasmas was performed \cite{Gill21}, but disregarding the continuum-continuum channels.

\begin{figure}[t]
\begin{center}
\includegraphics[width=8cm] {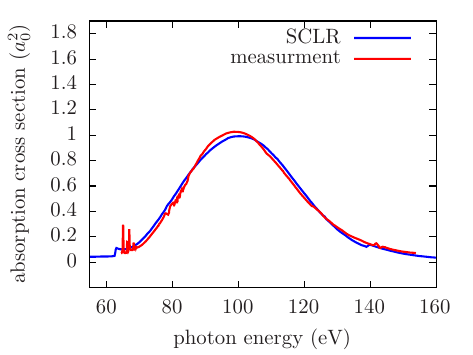} 
\end{center}
\caption{Photoabsorption cross-section of neutral Xe in the XUV region typical of the N-edge. Comparison between the self-consistent linear response of the DFT atom, using Gunnarsson-Lundqvist exchange-correlation term (same model as in~\cite{Zangwill80}) and measurements from~\cite{Haensel69}.
\label{fig_LR_xe}}
\end{figure}

As a validation step, we attempted to recover the results of \cite{Zangwill80} on neutral atoms of rare gases. Figure~\ref{fig_LR_xe} displays the result of a self-consistent dynamic linear response calculation using the same model as in~\cite{Zangwill80}, on a case of application considered in this paper. The results are in close agreement both with those of~\cite{Zangwill80} and with the measurement of~\cite{Haensel69} on liquid xenon and exhibit the significant impact of channel mixing (see \cite{Caizergues16} for additional examples).

In the case of the VAAQP ion, the cluster expansion of Equation~\eqref{eq_cluster_chi}, leads to the subtraction of the non-integrable contribution of the homogeneous medium Equation~\eqref{eq_cluster_Deltachi1_D}. This enables the accounting for continuum-continuum channels in the first-order susceptibility. From the corresponding self-consistent linear-response formalism, Equation~\eqref{eq_sum_ehrenfest} was derived in~\cite{Blenski06}.

On the quantum version of the VAAQP model, although progress was achieved in the understanding of the problem, the application of the self-consistent linear-response approach still leads to inconclusive results~\cite{Caizergues16}. In particular, the direct check of Equation~\eqref{eq_sum_ehrenfest} from an implementation of the self-consistent linear response, using the methods described in~\cite{Zangwill80,MahanSubbaswamy} failed. The self-consistent linear response in the context of dense plasmas, accounting for the continuum, remains an open problem, at least from an implementation standpoint. 

The work performed during C. Caizergues Ph.D. thesis nevertheless allowed us to test various numerical methods and formalisms for the quantum model. It also confirmed the relevance of the sum rule of Equation~\eqref{eq_sum_ehrenfest}, by checking it in the TF case. This ultimately leads us to suspect the boundary conditions used in the solution of the quantum problem.

\section{Research Prospects}
Heuristic approaches to collective effects are known~\cite{Perrot96,Johnson06,Kuchiev08,Johnson09}. However, all of them are in fact very similar, and a better-founded approach would be of great interest. First, these approaches only provides a smooth bridging with the Ziman direct-current conductivity which is not necessarily relevant for the whole range of low frequencies. Moreover, the electron density that is used as a free electron density in this context corresponds at best to the asymptotic electron density, in the VAAQP model, or to some arbitrary definition in other models. In either case, weakly bound electrons are disregarded, although they may participate significantly to the conduction.

During the Ph.D. thesis of Clément Caizergues progress was achieved on the linear response of the TF model, which corresponds to the semiclassical version of the VAAQP model. The quantum self-consistent linear-response was successfully applied as regards bound electrons and their related contribution to radiative properties, recovering the results of \cite{Zangwill80,Zangwill84,Doolen87}. 

However, despite a significant theoretical and numerical effort~\cite{Caizergues16}, the study of the self-consistent linear response of the quantum version of VAAQP was inconclusive. The work performed during this Ph.D. thesis seems to indicate an issue with the boundary conditions. Further work on this subject will have to give a fresh look to this particular problem.

%-------------------------
\chapter{Dense Plasmas as Ion Fluids\label{ch_ion_correlations}}
%-------------------------
%Cette étude s'inscrit dans le contexte de nos recherches de modèles variationnels d'atome dans un plasma. Après avoir franchi une étape avec le modèle VAAQP et ses applications [1-11], nous cherchons à construire un modèle variationnel qui inclurait une description des corrélations ion-ion allant au delà de l'hypothèse de la cavité. Un tel modèle permettrait, tout en restant de le cadre d'un modèle d'atome, une description unifiée de la thermodynamique du plasma. On éviterait alors le recours à un traitement heuristique et séparé de la contribution ionique aux grandeurs thermodynamiques. D'autres groupes de recherche travaillent sur des sujets proches [12-14], mais ce problème reste actuellement une question ouverte. Dans cette recherche, nous avons besoin de formulations variationnelles pour les modèles de fluides classiques, à une ou plusieurs composantes, avec des potentiels d'interaction arbitraires. Nous entendons ici par formulation variationnelle, une dérivation du modèle procédant par minimisation vis-à-vis de la (ou des) fonction de corrélation à 2-corps, en partant d'une fonctionnelle d'énergie libre du système.

\section{Difficulty of Dense-Plasma Atomic Modeling beyond the Cavity Hypothesis}
All  models of pressure-ionized plasma that are described in Section~\ref{sec_pressure_ionized_plasmas} focus on the description of the electronic structure of a particular ion, assuming that the surrounding ions will either restrict its spatial extension to the WS cell (Rozsnyai's model, INFERNO model) or will interact with it as a spread-out uniform medium, not entering the WS sphere (AJCI, VAAQP). In these models, a key function is the inhomogeneous density $n(r)$ of the electron cloud associated with the ion. Of course such spatial inhomogeneity of the electron density around a nucleus is in principle referring to correlations among the respective positions of nuclei and electrons.

For a $M$-component classical fluid of particles interacting through pair potentials, obtaining the $M(M+1)/2$ pair distribution functions $g_{\alpha,\gamma}(r)$, where $\alpha$, $\gamma$ label species, gives access to all statistical averages of observables (see, for instance,~\cite{LandauStatisticalPhysics}). Each pair distribution function $g_{\alpha,\gamma}(r)$ expresses the probability of finding a particle of species $\gamma$ at a distance $r$ of a particle of species $\alpha$. Percus' picture (or Percus' trick) is a reasoning that relates the pair distribution functions to densities of fictitious systems with a particular particle placed at the origin~\cite{Percus64}.  
\begin{align}
n_\alpha g_{\alpha,\gamma}(r)=n_\alpha^\text{inhom.}\{\underline{v}_{\alpha'}=\underline{u}_{\alpha',\gamma},r\}\label{eq_Percus}
\end{align}
where $n_\alpha^\text{inhom.}\{\{\underline{v}_{\alpha'}\},r\}$ is the density of $\alpha$-particles in a $M$-component classical fluid rendered inhomogeneous by the set of external potentials $v_{\alpha'}$ acting on each species $\alpha'$, respectively.
Using such equivalence, one can use approaches to 1-particle densities of inhomogeneous systems to address pair distribution functions of homogeneous systems.

However, to correctly describe the electronic structure of an ion, one has to resort to quantum mechanics. Quantum behavior is at the origin of the stability of matter with long-range attractive potentials \cite{Dyson67,Lenard68,Lebowitz69,Lieb72,Lieb76}, and notably results in the shell structure. It turns out that Percus' relation \eqref{eq_Percus} is not valid in the quantum-mechanical context. This may be seen as a consequence of the impossibility of separating the kinetic part of the partition function from its configurational (or interaction) part. For this reason, a practical approach to the correlation functions in quantum mechanics remains a long-standing, open problem (see, for instance,~\cite{Percus05}).

In order to circumvent this problem, an idea is to keep the framework of classical statistical mechanics and include approximate accounting for the quantum behavior in the interaction potentials~\cite{Lado67}, or resorting to an effective temperature~\cite{Dharma-wardana00,Perrot00b}. Approaches of this kind eventually led to the classical-mapping approach to quantum systems~\cite{Dufty13,Dutta13}. However, such methods may only give an approximate accounting for the quantum behavior of the system.

Another kind of approach is to generalize Percus' trick (see, for instance,~\cite{Chihara78}). Among the goals of such an effort is the quantum generalization of the hyper-netted chain model, which may be derived in the classical framework using the Percus picture and the classical DFT. As was pointed out by Chihara in~\cite{Chihara91}, if one assumes that nuclei behave as classical particles, it is even possible to partially apply the Percus trick. This requires one to supplement the model with assumptions about certain correlation functions. This approach led to the ``quantum hyper-netted chain'' (QHNC) model. 

%However, the DFT-based reasoning for deriving the HNC equations resorts to a functional Taylor expansion around a reference homogeneous medium at a given density. Even if this derivation leads to the HNC equations, it does not give access to the HNC free energy of the homogeneous system (see, for instance,~\cite{DeWitt88} and~\cite{Piron19a} appendix). This is an issue for the modeling of the thermodynamic equilibrium of a plasma, which includes the determination of the free-electron density. For that reason, the QHNC approach was first used in the modeling of electrons in metals~\cite{Anta00}, considering given free-electron densities. The QHNC was also applied to the modeling of dense plasmas~\cite{Starrett12,Starrett13,Starrett14,Chihara16}, in order to account for the ion-ion correlations. However, in this case it requires an additional assumption to set the density of free electrons.

In order to better point out the limitations of deriving the HNC integral equations through classical DFT, let us recall briefly the main steps. Nonlinear DH and DH models can also be derived similarly (see e.g., \cite{Piron19a} appendix), with the same limitations. For the sake of simplicity, we limit ourselves to a 1-component system.

In the finite-temperature classical DFT, one writes the grand-potential $\Omega$ for an inhomogeneous system of particles in an external potential $v_\text{ext}(\vec{r})$, viewed as a unique functional of the particle density $n(\vec{r})$:
\begin{align}
\Omega\left\{\underline{n},\underline{v}_\text{ext};\mu,T\right\}=&
{F}_\text{id}\left\{\underline{n};T\right\}
+{F}_\text{ex}\left\{\underline{n};T\right\}
+\int d^3r\left\{n(\vec{r})\left(v_\text{ext}(\vec{r})-\mu\right)\right\}
\label{eq_app_omega}
\end{align}
where ${F}_\text{id}$ is the ideal-gas part of the intrinsic free energy, ${F}_\text{ex}$ is the excess free energy, and $\mu$ is the chemical potential. 
The ideal-gas part of the intrinsic free-energy is written as:
\begin{align}
{F}_\text{id}\left\{\underline{n};T\right\}
=\frac{1}{\beta}\int  d^3r\left\{n(\vec{r})\left(\ln\left(n(\vec{r})\Lambda^3\right)-1\right)\right\}
\end{align}
$\Lambda=h/\sqrt{2\pi mk_BT}$ being the classical thermal length.
The excess free energy of the inhomogeneous system is approached using a second-order expansion around a \emph{homogeneous reference system}, of density $n_0$, with $n(\vec{r})$ tending to $n_0$ far from the origin.
\begin{align}
{F}_\text{ex}\left\{\underline{n};T\right\}
=&{F}_\text{ex}\left\{n_0;T\right\}
+\int d^3r\left\{(n\left(\vec{r})-n_0\right)
\left.\frac{\delta {F}_\text{ex}}{\delta n(\vec{r})}\right|_{n_0}\right\}
\nonumber\\
&+\frac{1}{2}\int d^3rd^3r'\left\{(n\left(\vec{r})-n_0\right)(n\left(\vec{r}')-n_0\right)
\left.\frac{\delta^2 {F}_\text{ex}}{\delta n(\vec{r})\delta n(\vec{r}')}\right|_{n_0}\right\}
\label{eq_2nd_order_Fex}
\end{align}
For the total intrinsic free energy ${F}$, we have the relation:
\begin{align}
\frac{\delta{F}}{\delta n(\vec{r})}
=
\frac{\delta}{\delta n(\vec{r})}
\left({F}_\text{id}+{F}_\text{ex}\right)
=\mu-v\left\{n(\vec{r}');\vec{r}\right\}
\end{align}
where $v\left\{n(\vec{r}');\vec{r}\right\}$ is the external potential such that $n(\vec{r}')$ is the equilibrium density (trial potential).
In the case of the homogeneous system of density $n_0$, we then have:
\begin{align}
\left.\frac{\delta {F}_\text{ex}}{\delta n(\vec{r})}\right|_{n_0}
=\mu_0-\frac{1}{\beta}\ln\left(n_0\Lambda^3\right)
\end{align}
$\mu_0$ being the chemical potential of the homogeneous system.

We now come to the hypothesis of the HNC model. Let us define the second-order direct correlation function of the homogeneous system: $c(|\vec{r}-\vec{r}'|)$ as:
\begin{align}
c(|\vec{r}-\vec{r}'|)
=-\left.\frac{\delta^2 \beta {F}_\text{ex}}{\delta n(\vec{r})\delta n(\vec{r}')}\right|_{n_0}
\end{align}
In the exact many-body problem, this function fulfills the Ornstein-Zernike (OZ) relation:
\begin{align}
h(r)=c(r)+n_0\int d^3r'\left\{h(r')c(|\vec{r}-\vec{r}'|)\right\}
\label{eq_ornstein_zernike}
\end{align}
which relates it to the exact correlation function $h(r)\equiv g(r)-1$ (see, for instance \cite{Percus64,HansenMcDonald}). In the HNC model, we consider this relation as a definition for $c(r)$, holding with the approximate $h(r)$.

Minimizing $\Omega$ with respect to $n(\vec{r})$, for a given external potential, we get:
\begin{align}
\ln\left(\frac{n(\vec{r})}{n_0}\right)=&\beta\left(\mu-\mu_0-v_\text{ext}(\vec{r})\right)
+n_0\int d^3r'\left\{\left(\frac{n(\vec{r}')}{n_0}-1\right)c(|\vec{r}-\vec{r}'|)\right\}
\end{align}
One immediately sees that, if $n(\vec{r})$ tends to $n_0$ far from the origin, we have $\mu=\mu_0$.

Finally, we use the Percus trick of Equation~\eqref{eq_Percus}, that is: we consider the density $n(\vec{r})$ around a particle of the homogeneous fluid fixed at the origin. This density is directly related to the pair distribution function of the homogeneous fluid:
\begin{align}
v_\text{ext}(\vec{r})\equiv u(r)\ ;\ \frac{n(\vec{r})}{n_0}=g(r)\ ;\ n_0\equiv n_\text{i}
\label{eq_app_percus_trick}
\end{align}
where $u(r)$ is the pair interaction potential. We denote by $n_\text{i}$ the particle density of the homogeneous fluid, and get:
\begin{align}
\ln\left(g(r)\right)&=-\beta u(r)
+n_\text{i}\int d^3r'\left\{h(r')c(|\vec{r}-\vec{r}'|)\right\}\\
&=-\beta u(r)+h(r)-c(r)\label{eq_hnc_closure}
\end{align}
where the second line is obtained using the OZ relation, and corresponds to the HNC closure relation. The HNC model just corresponds to Equations~\eqref{eq_hnc_closure} and \eqref{eq_ornstein_zernike}. Such a derivation is, for instance, presented in \cite{HansenMcDonald}.

Let us now go back to the expression of the grand-potential of Equation~\eqref{eq_app_omega}, using the approximation of Equation~\eqref{eq_2nd_order_Fex}. This expression depends on the excess free energy of the homogeneous reference system ${F}_\text{ex}\left\{n_\text{i};T\right\}$, which is not known. In itself, the DFT approach to the Percus picture give access neither to the grand-potential of a homogeneous system, nor to its free energy. One may only evaluate the increment of the grand potential, which corresponds to the excess chemical potential (see, for instance, \cite{HansenMcDonald}). For that reason, Chihara's methods was applied to the problem of electrons in metals, with given free-electron density, notably in \cite{Anta00}. Applications to the modeling of a plasma resort to a supplementary condition, in order to set the free-electron density \cite{Starrett14,Chihara16}.

In principle, in a model of plasma in thermodynamic equilibrium, the ionization state of the plasma should stem from its equilibrium state. The thermodynamic equilibrium state of a system is best expressed by minimizing its free energy. This is the reason why the trail we followed to include ion-ion correlations in an atomic structure model was not that of Chihara. We rather tried building an approximate free energy of a homogeneous plasma and then minimize it. This required in particular to have generalized free-energy functionals for models of classical fluids with arbitrary interaction potentials.

As regards the electronic structure part of the model, the VAAQP model was seemingly a good starting point, because it starts from an approximate free-energy functional and also enables the accounting for the ion surroundings in the whole space rather than within an ion cell.

%-------------------------
\section{Generalized Free Energy of Classical Fluids\label{sec_classical_fluids}}
%-------------------------
Relating the free energy ${F}_\text{ex}\left\{n_\text{i};T\right\}$ to the pair distribution function of the homogeneous medium may be done through the Debye-Kirkwood charging relation.
\begin{align}
&\dot{F}_\text{ex,eq}(n_\text{i},T)=\frac{n_\text{i}}{2}\int_0^1 d\lambda \int d^3r
\left\{g_\text{eq}^{\lambda}(r)u(r)\right\}\hspace{1cm}\text{(for short-ranged $u(r)$)}\\
&\dot{F}_\text{ex,eq}(n_\text{i},T)=\frac{n_\text{i}}{2}\int_0^1 d\lambda \int d^3r
\left\{h_\text{eq}^{\lambda}(r)u(r)\right\}\hspace{1cm}\text{(divergence-free, for long-ranged $u(r)$)}
\label{eq_charging}
\end{align}
where $\dot{F}_\text{ex,eq}$ is the excess free energy per particle. In the case of a long-range potential, the free energy per particle has a logarithmic divergence and we rather use the divergence-free (or renormalized) free-energy. $\lambda$ is a charging parameter, which multiplies the interaction potential $u(r)$. $g_\text{eq}^{\lambda}(r)=h_\text{eq}^{\lambda}(r)+1$ is the equilibrium pair distribution function of the system with interaction potential $\lambda u(r)$.
 
This relation rigorously relates the exact excess free energy to the exact pair distribution function. It expresses a gradual switching of \emph{all interactions} in the system. It is not to be confused with the method of gradually switching the interaction potentials associated to a \emph{single particle}. The latter method in fact leads to the derivation of previous section and give access to the excess chemical potential.

Using the charging relation \eqref{eq_charging}, one can build an approximate free energy, starting from the integral equation of an approximate model of fluid. Other routes to the free energy exist. Integration over temperature, starting from the internal energy is an alternative way (see, for instance, \cite{LandauStatisticalPhysics}, \textsection 78). Using the chemical potential and the virial pressure is also possible (see, for instance, \cite{Iyetomi86}, Eq.~28). The consistency among the obtained thermodynamic quantities may be checked \textit{a posteriori}, as well as the fulfilment of virial theorem, by differentiation.

For the sake of our application, the free energy should ideally take as input an arbitrary interaction potential. In a plasma model, the effective interaction potential among ions would in principle stem from the equilibrium condition. Moreover, in such a functional, the pair distribution function, which describes the structure of the ion fluid, should play the role of an internal degree of freedom. 
In brief, for a given model of classical fluid, our interest is in a functional that, for a given arbitrary interaction potential, is minimal and equal to the corresponding free energy when the model equations are fulfilled.

A generalized free-energy functional of this kind was proposed by Morita and Hiroike in 1960 for the HNC model \cite{MoritaHiroike60}. An alternative derivation was later proposed by Lado \cite{Lado73} and subsequently extended to the case of two-component fluids \cite{Lado73b,Enciso87}. The expression they obtained for the generalized, divergence-free excess-free-energy functional in the 1-component case is:
\begin{align}
\dot{F}_\text{ex}^\text{HNC}(\underline{h},\underline{u};n_\text{i},T)
=&\frac{n_\text{i}}{2\beta}\int d^3r \left\{h(r)\beta u(r)+(h(r)+1)\ln\left(h(r)+1\right)
-h(r)-\frac{h(r)^2}{2} \right\}
\nonumber\\
&+\frac{1}{2\beta n_\text{i}}\int \frac{d^3k}{(2\pi)^3}
\left\{ n_\text{i} h_{k}-\ln(1+n_\text{i} h_{k}) \right\}
\label{eq_HNC_functional}
\end{align}

Surprisingly, it seems that, prior to our work on the subject, no free-energy functional of this kind had been proposed for the DH model with arbitrary interaction potential. The integral equation of this model is:
\begin{align}
h(r)=-\beta u(r)-n_\text{i}\int d^3r'\left\{h(r')\beta u(|\vec{r}-\vec{r}'|)\right\}
\label{eq_DH_integral_equation}
\end{align}
which may be seen as the OZ Equation~\eqref{eq_ornstein_zernike} with the DH closure: $c(r)=-\beta u(r)$.

The DH model is only relevant to the limit of weak coupling, and does not account for the correlations beyond the mean-field approximation. It nevertheless remains of permanent theoretical interest in plasma physics, for it gives insight into the screening phenomenon and the decay of correlation functions. It brings useful qualitative information both for systems with repulsive interactions only, and for systems including attractive long-range potentials, since it circumvent the classical Coulomb catastrophe. For these reasons we had a strong interest in a generalized free-energy functional for the DH model.

Moreover, in the study of \cite{MoritaHiroike60,Lado73}, the focus was put on the derivation of the equilibrium free energy. The possibility of extending \textit{a posteriori} the expression in order to obtain a generalized functional of $g(r)$ or $h(r)$ was mentioned, but the result was not derived in the form of a generating functional for the model equations.

In \cite{Piron16}, we proposed a first expression for a generalized free-energy functional of the DH model. In this derivation, we constrained the functional to be a generating functional of the DH equation from the beginning. Extension to two-component systems followed in \cite{Blenski17}, but the calculations were too tedious to address general multi-component systems. The expression we obtained, in the 1-component case, for the generalized, divergence-free, excess-free-energy functional is:
\begin{align}
&\dot{F}^{\text{DH}}_\text{ex}\left\{\underline{h},\underline{u};n_\text{i},T\right\}
=
\frac{1}{\beta}
\int \frac{d^3k}{(2\pi)^3}\left\{
\left(1+\frac{1}{n_\text{i}\beta u_k}\right)
\left(1-\frac{\ln\left(1+n_\text{i} \beta u_k\right)}{n_\text{i}\beta u_k}\right)
h_k\left(\frac{h_k}{2}+\beta u_k+\frac{n_\text{i}\beta}{2}h_k u_k\right)
\right\}
\label{eq_DH_functional_2016}
\end{align}
We also showed that in the DH case, as in the case of HNC, starting from the charging relation allows the obtained free-energy to fulfill the virial theorem.

In \cite{Piron19a}, we proposed a simpler derivation of a generalized free-energy functional for the DH model. The latter derivation was closer to that of Lado \cite{Lado73}. We also showed the link with the method proposed by Olivares and McQuarrie \cite{Olivares76} to build generating functionals of integral equations. This allowed us to easily extend the derivation to general multi-component systems. In the 1-component case, the expression we obtained is:
\begin{align}
\dot{F}_\text{ex}^\text{DH}(\underline{h},\underline{u};n_\text{i},T)
=&\frac{n_\text{i}}{2\beta}\int d^3r \left\{h(r)\beta u(r) \right\}
+\frac{1}{2\beta n_\text{i}}\int \frac{d^3k}{(2\pi)^3}
\left\{ n_\text{i} h_{k}-\ln(1+n_\text{i} h_{k}) \right\}
\label{eq_DH_functional_2019}
\end{align}

It can be shown easily that Equations~\eqref{eq_DH_functional_2016} and \eqref{eq_DH_functional_2019} define two distinct functionals of $h(r)$,$u(r)$, which become identical functionals of $u(r)$ when $h(r)$ fulfills the DH integral equation.

These studies on the DH generalized free-energy functionals helped us to point out the non-unique character of generalized free-energy functionals in the classical theory of fluids. We showed in \cite{Piron19a} that the Olivares-McQuarrie formalism \cite{Olivares76} is a well-suited framework to explain this nonuniqueness.

The physical interpretation of generalized free-energy functionals for pair distribution functions that do not fulfill the integral equation of the model is unclear. In statistical physics, the meaning of generalized thermodynamic potentials for out-of-equilibrium probability distributions is found in their time-evolution through Markovian dynamics, and Boltzmann's H theorem. Gaining further insight on the free-energy functionals of fluid models would maybe require to have dynamic versions of the considered fluid models.

In practice, the generalized free-energy functionals of the HNC and DH models offer variational formulations of the corresponding fluid integral equations, together with a expression of the equilibrium free-energy for any interaction potential.

Finally, in addition to the specific context of our search for a variational atomic model of plasma including the ion-fluid structure, such work may have other applications. Some theoretical approaches to plasma modeling are based on the calculation of corrections, using the DH model as a starting point \cite{Abe59,DeWitt65}. Some models having direct practical applications are also based on the DH approach \cite{KidderDeWitt61,Vieillefosse81}. Moreover, in the physics of colloïds, the DH model also remains of practical interest, since improved versions of this model are used \cite{Nordholm84,Penfold90}.

\section{VAMPIRES Model\label{sec_VAMPIRES}}
A way towards an improved modeling of pressure-ionized plasma is to couple an atomic model of plasma to a classical model of fluid through its interaction potentials, without making a point-like-ion hypothesis to split the problem. 

In such a model, one should account for the impact of the ion-fluid structure on the electronic structure of ions but also for the effect of the electronic structure on the  interaction potentials in the ion fluid. The interaction potentials are then to be determined self-consistently with the electronic structure. Ideally, they should be seen as thermodynamic averages, obtained from the minimization of the total free energy. 

A preliminary work on a model accounting for both the bound electrons of an ion and the ion fluid structure was described in~\cite{Piron19b}. In this model, continuum electrons are excluded from the ion electronic structure, as in an isolated-ion model, and constitute a species of a classical fluid. This classical fluid may be treated either through the Debye-Hückel model (thus avoiding the Coulomb collapse) or by neglecting the polarization of continuum electrons, as in an OCP. Applicability of this model has obvious limitations, due to its crude treatment of continuum electrons. However, this model formally introduces the screening of the effective potential in the electronic structure. In the DH case, when bound electrons are localized in a small region compared to the Debye length, this model yields the point-like DH correction of Equation~\eqref{eq_DH_AA_correction}.

The variational atomic model of plasma with ion radial correlations and electronic structure (VAMPIRES)~\cite{BlenskiPiron23} is both an atom-in-jellium model of the ion electronic structure and a statistical model of ion fluid. In this model, the continuum electrons are treated quantum-mechanically, as a part of the electronic structure that is partially shared among ions. This model stems from the minimization of an approximate free energy, and we showed that it fulfills the virial theorem of Equation~\eqref{eq_virial_theorem}. 

Let us consider the free energy of $N_\text{i}$ nuclei of atomic number $Z$ and $N_\text{i} Z$ electrons, in a large volume $V$ and at a fixed temperature $T$. The nuclei are approximated by indistinguishable classical particles, which allows us to write (see~\cite{Kirkwood33, Zwanzig57}, and~\cite{BlenskiPiron23} for the present generalized form):
\begin{align}
&F_\text{eq}(N_\text{i},V,T)
\nonumber\\
&=\underset{\underline{w}}{\text{Min}}\,\iint_V \frac{d^3R_1...d^3P_{N_\text{i}}}{N_\text{i} !h^{3N_\text{i}}}\left\{
w(\vec{R}_1...\vec{P}_{N_\text{i}})
\left({\sum_{j=1}^{N_\text{i}} \frac{\vec{P}_j^2}{2m_\text{i}}}
+F_\text{eq}^\text{e}(\vec{R}_1...\vec{R}_{N_\text{i}};N_\text{i},V,T)
+{\frac{1}{\beta}\ln\left(w(\vec{R}_1...\vec{P}_{N_\text{i}})\right)}
\right)
\right\}
\nonumber\\
&\text{\hspace{1cm}s. t. } \iint_V \frac{d^3R_1...d^3P_{N_\text{i}}}{N_\text{i}!h^{3N_\text{i}}}\left\{w(\vec{R}_1...\vec{P}_{N_\text{i}})\right\}=1
\label{eq_VAMPIRES_general_minimiz}
\end{align}
where $w(\vec{R}_1...\vec{P}_{N_\text{i}})$ denotes the probability distribution of the nuclei classical many-body states $(\vec{R}_1...\vec{P}_{N_\text{i}})$, and where $F_\text{eq}^\text{e}$ is the equilibrium free energy of a system of electron with a fixed configuration $(\vec{R}_1...\vec{R}_{N_\text{i}})$ of the nuclei (Hamiltonian $H_\text{static}$ of Equation~\eqref{eq_H_static}), plus the nucleus-nucleus interaction energy. The constraint in Equation~\eqref{eq_VAMPIRES_general_minimiz} simply enforces the correct normalization of the probability.

Electrons are modeled quantum-mechanically, using a finite-temperature DFT formalism~\cite{Hohenberg64,KohnSham65a,Mermin65}. That is, we obtain $F_\text{eq}^\text{e}$ from the following minimization:
\begin{align}
F_\text{eq}^\text{e}(\vec{R}_1...\vec{R}_{N_\text{i}};N_\text{i},V,T)=&\underset{\underline{n}}{\text{Min}}\,\left[{F^0\left\{\underline{n};V,T\right\}}
+W_\text{direct}\left\{\underline{n};\vec{R}_1...\vec{R}_{N_\text{i}};N_\text{i}\right\}
+{F^\text{xc}\left\{\underline{n};V,T\right\}}\right]
\nonumber\\
&~\text{ s. t. } \int_V d^3r\left\{n(\vec{r})\right\}=ZN_\text{i}
\label{eq_VAMPIRES_DFT_minimiz}
\end{align}
$n(\vec{r})$ is the electron density; $F^0$ denotes the kinetic-entropic contribution to the free energy of a non-interacting electrons gas of density $n(\vec{r})$; and $W_\text{direct}$ denotes the total direct-interaction energy, which includes the nucleus-nucleus contribution. $F^\text{xc}$ is the contribution of exchange and correlations. The constraint corresponds to the neutrality condition of the nuclei-electron system.

Like in the VAAQP model, we assume that the equilibrium electron density $n(\vec{R}_1 ... \vec{R}_{N_\text{i}};\vec{r})$ for a system of $N_\text{i}$ nuclei is correctly described using a first-order cluster expansion (see Equation~\eqref{eq_cluster_density}). This leads us to the following Ansatz for the electron density:
\begin{align}
n(\vec{R}_1...\vec{R}_{N_\text{i}};\vec{r})
\approx n_\text{e}+\sum_{j=1}^{N_\text{i}} q(|\vec{r}-\vec{R}_j|)
\label{eq_ansatz_VAMPIRES}
\end{align}
The system is seen as a set of ions, that is: nuclei, each with its spherical cloud of displaced electrons, sharing a common uniform background of free electrons. The minimization with respect to the electron density $n(r)$ is thus performed within a particular class of functions and consists of minimization with respect to the two parameters of the Ansatz, namely, $n_\text{e}$ and the function $q(r)$.

The neutrality condition of Equation~\eqref{eq_VAMPIRES_DFT_minimiz} can be rewritten using Equation~\eqref{eq_ansatz_VAMPIRES}, as:
\begin{align}
\frac{n_\text{e}}{n_\text{i}}+\int_V d^3r\left\{q(\vec{r})\right\}=Z
\label{eq_VAMPIRES_neutrality}
\end{align}
Still using Equation~\eqref{eq_ansatz_VAMPIRES}, $W_\text{direct}$ can be written as:
\begin{align}
W_\text{direct}
={\frac{1}{2}\sum_{i=1}^{N_\text{i}}\sum_{\substack{j=1\\j\neq i}}^{N_\text{i}}v_\text{ii}\left\{\underline{q};|\vec{R}_i-\vec{R}_j|\right\}}
+{N_\text{i}\,W_\text{intra}\left\{\underline{q};V\right\}}
+{N_\text{i}\,W_\text{bg}\left\{\underline{q},n_\text{e};V\right\}}
\end{align}
with the definitions:
\begin{align}
&v_\text{ii}\left\{\underline{q};R,V\right\}
=
\frac{Z^2\ee^2}{R}
-2Z\ee^2\int_V d^3r \left\{ \frac{q(r)}{|\vec{r}-\vec{R}|} \right\}
+\ee^2\int_V d^3r d^3r' \left\{ \frac{q(r)q(r')}{|\vec{r}-\vec{r}'+\vec{R}|} \right\}
\nonumber\\
&W_\text{intra}\left\{\underline{q};V\right\}=
-Z\ee^2\int_V d^3r \left\{ \frac{q(r)}{r} \right\}
+\frac{\ee^2}{2}\int_V d^3r d^3r' \left\{ \frac{q(r)q(r')}{|\vec{r}-\vec{r}'|} \right\}
\nonumber\\
&W_\text{bg}\left\{\underline{q},n_\text{e};V\right\}=
n_\text{e}\ee^2\int_V d^3rd^3r'\left\{ \frac{q(r)}{|\vec{r}-\vec{r}'|}\right\}
+\ee^2\left( \frac{n_\text{e}^2}{2n_\text{i}}-n_\text{e}Z \right)\int_V d^3r \left\{ \frac{1}{r} \right\}
\end{align}
$v_\text{ii}$ plays the role of an ion-ion interaction potential, $W_\text{intra}$ corresponds to an intra-ion interaction energy, and $W_\text{bg}$ gathers all terms related to interactions with the electron homogeneous background. 
In VAAQP, we make a specific hypothesis on the electrostatic interaction term pertaining to an ion, $\Delta F_1^\text{el}$, in order to introduce the cavity (see Equation~\eqref{eq_VAAQP_dF1el}). In the VAMPIRES model, we do not introduce any additional hypothesis in $W_\text{intra}$. The interaction terms just follow from the cluster expansion of the electron density and the statistical treatment of the ion fluid.

Electron terms $F^0$ and $F^\text{xc}$ are approximated using a first-order cluster expansion, as in VAAQP: 
\begin{align}
F^{\bullet}&\left\{\underline{n}(\vec{r})=n_\text{e}+\sum_{i=1}^{N_\text{i}}q(|\vec{r}-\vec{R}_i|);V,T\right\}
= F^{\bullet}\left\{\underline{n}(\vec{r})=n_\text{e};V,T\right\}
+\sum_{i=1}^{N_\text{i}} \Delta F_1^{\bullet}\left\{\underline{q},n_\text{e},\vec{R}_i;V,T\right\}
\end{align}
\begin{align}
\Delta F_1^{\bullet}\left\{\underline{q},n_\text{e},\vec{R};V,T\right\}
=&F^{\bullet}\left\{\underline{n}(\vec{r})=n_\text{e}+q(|\vec{r}-\vec{R}|);V,T\right\}-F^{\bullet}\left\{\underline{n}(\vec{r})=n_\text{e};V,T\right\}
=\Delta F_1^{\bullet}\left\{\underline{q},n_\text{e};V,T\right\}
\end{align}
where the $^{\bullet}$ symbol is to be replaced by either the $^0$ or the $^\text{xc}$ label.
$\Delta F_1^0\left\{\underline{q},n_\text{e};V,T\right\}$ is the kinetic and entropic contribution to the free energy of non-interacting electrons in a trial potential $v_\text{trial}\left\{\underline{q},n_\text{e};r;T\right\}$, which yields the electron density $n(r)=n_\text{e}+q(r)$, minus the contribution of the homogeneous background (see Equation~\eqref{eq_VAAQP_dF10}).
$\Delta F_1^\text{xc}\left\{\underline{q},n_\text{e};V,T\right\}$ is the exchange-correlation contribution to the free energy of a system of electrons having density $n(r)=n_\text{e}+q(r)$, minus the contribution of the homogeneous background (see Equation~\eqref{eq_VAAQP_dF1xc}).

At this point, the minimization of Equation~\eqref{eq_VAMPIRES_general_minimiz} becomes

\begin{align}
F_\text{eq}(N_\text{i},V,T)=&
\underset{\underline{q},n_\text{e}}{\text{Min}}
\left[F^0\left\{n_\text{e};V,T\right\}
+F^\text{xc}\left\{n_\text{e};V,T\right\}
+F_\text{eq}^\text{i}
\left\{
\underline{v}(R)=v_\text{ii}\left\{\underline{q};R,V\right\}
;N_\text{i},V,T
\right\}
\right.\nonumber\\&\left.
+N_\text{i}\left( 
\Delta F_1^0\left\{\underline{q},n_\text{e};V,T\right\}
+\Delta F_1^\text{xc}\left\{\underline{q},n_\text{e};V,T\right\}
+W_\text{intra}\left\{\underline{q},V\right\}
+W_\text{bg}(\underline{q},n_\text{e};V)
\right)\right]
\nonumber\\
&\text{s.\,t. }\frac{n_\text{e}}{n_\text{i}}+\int_V d^3r\left\{q(r)\right\}=Z
\label{eq_free_energy_one-ion}
\end{align}
where $F_\text{eq}^\text{i}\left\{\underline{v};N_\text{i},V,T\right\}$ gathers the nuclei kinetic energy and entropy, as well as ion-ion interaction terms. This corresponds to the free energy of a one-component classical fluid of ions, interacting through the potential $v_\text{ii}$:

\begin{align}
F_\text{eq}^\text{i}\left\{\underline{v};N_\text{i},V,T\right\}
=&\underset{\underline{w}}{\text{Min}}
\int_V \frac{d^3R_1...d^3P_{N_\text{i}}}{N_\text{i}!h^{3N_\text{i}}}\left\{
w(\vec{R}_1...\vec{P}_{N_\text{i}})
\left(
\sum_{j=1}^{N_\text{i}} \frac{P_j^2}{2m_\text{i}}
+\frac{1}{2}\sum_{i=1}^{N_\text{i}}\sum_{\substack{j=1\\j\neq i}}^{N_\text{i}}v(|\vec{R}_i-\vec{R}_j|)
\right.\right.\nonumber\\
&\left.\left.\vphantom{\sum_{\substack{j=1\\j\neq i}}^{N_\text{i}}}\hspace{8cm}
+\frac{1}{\beta}\ln\left(w(\vec{R}_1...\vec{P}_{N_\text{i}})\right)
\right)
\right\}
\nonumber\\
&\text{s.\,t. }\int_V \frac{d^3R_1...d^3P_{N_\text{i}}}{N_\text{i}!h^{3N_\text{i}}}\left\{w(\vec{R}_1...\vec{P}_{N_\text{i}})\right\}=1\\
\equiv&\underset{\underline{w}}{\text{Min}}\,
F^\text{i}\left\{\underline{w},\underline{v};N_\text{i},V,T\right\}
\text{ s.\,t. }\int_V \frac{d^3R_1...d^3P_{N_\text{i}}}{N_\text{i}!h^{3N_\text{i}}}\left\{w(\vec{R}_1...\vec{P}_{N_\text{i}})\right\}=1
\label{eq_general_classical_fluid}
\end{align}

In the thermodynamic limit, the free energy per ion $\dot{F}_\text{i}={F}_\text{i}/N_\text{i}$ of such a system has a logarithmic divergence because $v_\text{ii}$ has a Coulomb tail. However, as in a usual OCP model, this divergence is cancelled by an opposite-sign divergence in $W_\text{bg}$. We therefore group these terms together, which renormalizes the free energy. We use either the HNC or the DH model to approximate the resulting divergence-free ion-fluid free energy per ion, as a functional of the interaction potential $v(r)$.
\begin{align}
\dot{F}^\text{i}_\text{eq}+{W_\text{bg}}
\approx \dot{F}_\text{id, i}(n_\text{i},T)
+\dot{F}_\text{ex, eq}^\text{approx}\left\{\underline{v};n_\text{i},T\right\}
\end{align}
$\dot{F}_\text{ex, eq}^\text{approx}$ is either the HNC or the DH divergence-free excess free energy per ion.
Such approximate equilibrium free energy may be written as the minimum of a generalized free-energy functional of the radial correlation function $h(r)=g(r)-1$ (see previous section):
\begin{align}
\dot{F}_\text{ex, eq}^\text{approx}\left\{\underline{v};n_\text{i},T\right\}
=\underset{\underline{h}}{\text{Min}}\,\dot{F}_\text{ex}^\text{approx}\left\{\underline{v},\underline{h};n_\text{i},T\right\}
\end{align}
with the minimum occurring for the $h(r)$ fulfilling the equations of the approximate model, either HNC or DH.
In the DH case, $\dot{F}_\text{ex}^\text{approx}$ is given by Equation~\eqref{eq_DH_functional_2019}, whereas it is given by Equation~\eqref{eq_HNC_functional} in the case of HNC.

Finally, the VAMPIRES model is based on the minimization of the following approximate free energy per ion $\dot{F}\{\underline{h},\underline{q},n_\text{e}\}$, with the neutrality constraint:
\begin{align}
\dot{F}\{\underline{h},\underline{q},n_\text{e};n_\text{i},T\}
=&
\dot{F}_\text{id,i}(n_\text{i},T)
+\dot{F}_\text{ex}^\text{i\,approx}\left\{\underline{h},\underline{v}(R)
=v_\text{ii}\left\{\underline{q},R\right\};n_\text{i},T\right\}
\nonumber\\&
+\frac{f^\text{F}_\text{e}(n_\text{e};T)}{n_\text{i}}
+\frac{f_\text{xc}(n_\text{e};T)}{n_\text{i}}
+\Delta F_1^0\left\{\underline{q},n_\text{e};T\right\}+\Delta F_1^\text{xc}\left\{\underline{q},n_\text{e};T\right\}+W_\text{intra}\left\{\underline{q}\right\}
\\
\dot{F}_\text{eq}(n_\text{i},T)
=&\underset{\underline{h},\underline{q},n_\text{e}}{\text{Min}}\,\dot{F}\{\underline{h},\underline{q},n_\text{e};n_\text{i},T\}
~\text{ s. t. }\frac{n_\text{e}}{n_\text{i}}+\int d^3r\left\{q(r)\right\}=Z
\end{align}

The minimization with respect to $h(r)$ leads to the fluid integral equations, that is, the Ornstein-Zernike relation with the closure relation corresponding to the chosen approximate model:
\begin{align}
&h(r)=c(r)
+n_\text{i}\int d^3r'\left\{
c(|\vec{r}'-\vec{r}|)h(r') 
\right\}
\\
&c(r)=
\begin{cases}
-\beta v_\text{ii}(r)-\ln(h(r)+1)+h(r)&\text{(HNC)}
\\
-\beta v_\text{ii}(r)&\text{(DH)}
\end{cases}
\end{align}

The minimization with respect to $q(r)$ includes that on $n_\text{e}$, which is expressed as a functional $n_\text{e}\{\underline{q};n_\text{i}\}$ using the neutrality constraint. It yields:
\begin{align}
0=&-v_\text{trial}\left\{\underline{q},n_\text{e};r;T\right\}
+\mu_\text{xc}\left(n_\text{e}+q(r),T\right)-\mu_\text{xc}\left(n_\text{e},T\right)+v_\text{el}\left\{\underline{h},\underline{q};r\right\}
\nonumber\\
&-n_\text{i}\int d^3r' \left\{
-v_\text{trial}\left\{\underline{q},n_\text{e};r';T\right\}
\right.\left.
+\mu_\text{xc}\left(n_\text{e}+q(r'),T\right)-\mu_\text{xc}\left(n_\text{e},T\right) \right\}\label{eq_VAMPIRES_SCF_intermediary}
\end{align}
where we have defined:
\begin{align}
&v_\text{el}\left\{\underline{h},\underline{q};r\right\}
\equiv v_\text{intra}\left\{\underline{q};r\right\}
+n_\text{i}\int d^3r' \left\{h(r')v_\text{intra}\left\{\underline{q};|\vec{r}'-\vec{r}|\right\}\right\}
\label{eq_VAMPIRES_vel}
\\
&v_\text{intra}\left\{\underline{q};r\right\}
=\frac{-Z\ee^2}{r}+\ee^2\int d^3r'\left\{\frac{q(r')}{|\vec{r}-\vec{r}'|}\right\}
\end{align}
In order to solve Equation~\eqref{eq_VAMPIRES_SCF_intermediary}, we define the distribution $\tilde{v}_\text{el}$ such that: 
\begin{align}
v_\text{el}(r) = \tilde{v}_\text{el}\left\{\underline{v}_\text{el},r\right\}	- n_\text{i}\int d^3r'\left\{\tilde{v}_\text{el}\left\{\underline{v}_\text{el},r'\right\}\right\}
\label{eq_VAMPIRES_eq_vel_veltilde}
\end{align}
We thus obtain from Equation~\eqref{eq_VAMPIRES_SCF_intermediary} the following electron self-consistent equation:
\begin{align}
v_\text{trial}\left\{\underline{q},n_\text{e};r;T\right\}=\tilde{v}_{\text{el}}(r)
+\mu_\text{xc}\left(n_\text{e}+q(r),T\right)-\mu_\text{xc}\left(n_\text{e},T\right)
\end{align}
where $\tilde{v}_{\text{el}}(r)$ is a shorthand notation for $\tilde{v}_\text{el}\left\{\underline{v}_\text{el}\left\{\underline{h},\underline{q}\right\},r\right\}$. 
From Equation~\eqref{eq_VAMPIRES_eq_vel_veltilde}, $\tilde{v}_\text{el}$ may be expressed in the Fourier space as:
\begin{align}
&\tilde{v}_{\text{el},k} = 
\begin{cases}			
v_{\text{el},k}=- \frac{4\pi e^{2}}{k^{2}}(Z-q_k)(1+n_\text{i}h_k) & \text{if } 	\vec{k}\neq 0\\
0 					  &	\text{if }  \vec{k}= 0
\end{cases}	
\end{align}
The difference between $\tilde{v}_\text{el}$ and ${v}_\text{el}$ only impacts on integrals of product of $\tilde{v}_\text{el}$ with a function that is not regular at $\vec{k}=0$ in the Fourier space. For instance, we have:
\begin{align}
\int d^3r\left\{\tilde{v}_\text{el}(r)\right\}=0
\neq
\int d^3r\left\{{v}_\text{el}(r)\right\}=-\frac{1}{\beta n_\text{e}}
\end{align}
where the last equality may be shown from the equations of the model.

Thermodynamic quantities are rigorously derived from the equilibrium free energy $\dot{F}_\text{eq}$. Especially, the pressure is given by the following expression:
\begin{align}
P_\text{thermo}(n_\text{i},T)=&n_\text{i} k_\text{B} T
+\left.n_\text{i}^2
\frac{\partial {A}^\text{i\,approx}}{\partial n_\text{i}}
\right|_\text{eq}
+
n_\text{e} \big(\mu(n_\text{e},T)+\mu_\text{xc}(n_\text{e},T)\big) - f_0(n_\text{e},T) - f_\text{xc}(n_\text{e},T)
\end{align}
In this formula, the first two terms correspond to the pressure of the ion fluid (ideal-gas and excess contributions), while the next four terms correspond to the pressure of the uniform electron gas, as in the VAAQP model. This means that displaced electrons only contribute to the pressure through the ion-fluid excess term. From the expression of the virial pressure, it can also be shown that the virial theorem is fulfilled in the VAMPIRES model.

\begin{figure}[h]
\includegraphics[width=8cm] {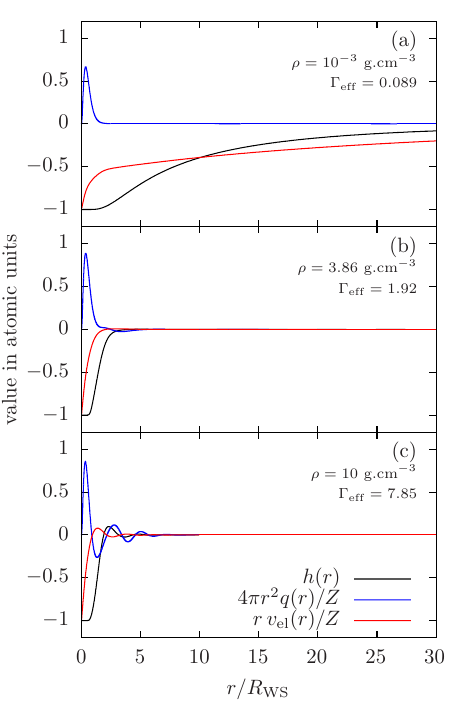} 
\hspace{1cm}
\includegraphics[width=8cm] {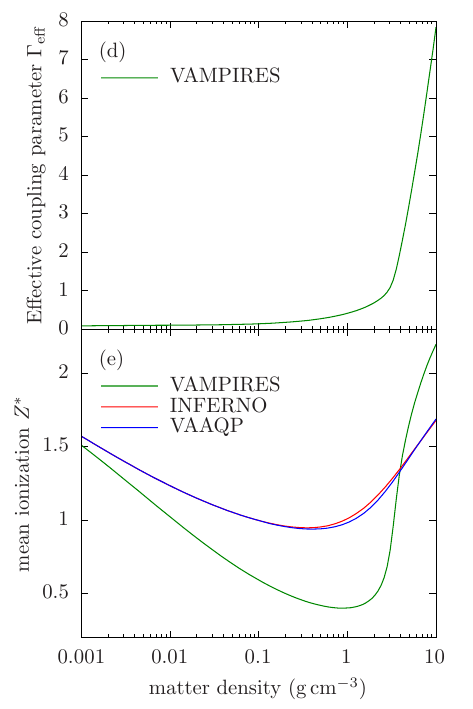} 
\caption{Results from the VAMPIRES model for lithium at 10 eV temperature. Pair correlation function $h(r)$, electron density $4\pi q(r)/Z$, and electrostatic potential $r\,v_\text{el}(r)/Z$, for various matter densities ({a}-{c}). Mean ionization $Z^*$ ({e}) and effective coupling parameter $\Gamma_\text{eff}$ ({d}) as functions of the matter density.
\label{fig_Li_VAMPIRES}}
\end{figure}

Figure~\ref{fig_Li_VAMPIRES} presents results from the VAMPIRES model for lithium at 10 eV temperature. First, one sees from Figure~\ref{fig_Li_VAMPIRES}e, which displays the mean ionization as a function of density, that the accounting for ion-ion correlations in the model yields the qualitative behavior of pressure ionization.

In order to quantify ion-ion coupling in this model, the usual coupling parameter $\Gamma=\beta Z^{*\,2}\ee^2/R_\text{WS}$ is not relevant. The ion-ion potential is not purely Coulombic, and the ion charge $Z^*$ corresponds to an asymptotic limit, which, in general, is not relevant to the WS radius. Consequently, we use an effective coupling parameter $\Gamma_\text{eff}=-\beta \dot{U}_\text{ex}^\text{approx}$, which really corresponds to the ratio of the ion-fluid interaction energy to the thermal energy. 

Figure~\ref{fig_Li_VAMPIRES}\,a--c present the ion-ion pair correlation function $h(r)$, electron-cloud density $q(r)$, and ion effective electrostatic potential $v_\text{el}(r)$, for three values of matter densities corresponding to weak, moderate, and strong coupling, respectively. In each case, close to the central nucleus, one sees a sharp peak in the electron linear density, which corresponds to the bound electronic structure of the ion. 

For a weakly coupled plasma (case of Figure~\ref{fig_Li_VAMPIRES}\,a), the range of the potential $v_\text{el}$ seen by the electrons extends far beyond the WS sphere. $v_\text{el}$ variations may be decomposed into two regions. Close to the nucleus, the steep variation is related to the ``internal'' screening by the bound electronic structure. Far from the nucleus, the longer-range decay is related to both a tail of weakly displaced electrons and the DH-like decay of the ion-ion correlation function.

Moreover, one can see in Figure~\ref{fig_Li_VAMPIRES}\,e that the mean ionization in these cases is lower than in INFERNO or VAAQP. However, in VAMPIRES, some of the electrons that do not participate in the background density $n_\text{e}$, which defines $Z^*$, may in fact be weakly displaced and play a role similar to the background electrons in an observable quantity. Especially, these electrons may interact significantly with the surrounding ions. 

For a moderately coupled plasma (in the case of Figure~\ref{fig_Li_VAMPIRES}\,b), the ion-ion pair correlation function has the shape of a cavity, resembling the WS cavity assumed in VAAQP. In such situations, results from the VAMPIRES model are indeed close to those of VAAQP and of INFERNO. One may check in the figure that the range of $v_\text{el}$ is close to $R_\text{WS}$.

For a strongly coupled plasma (case of Figure~\ref{fig_Li_VAMPIRES}\,c), the ion-ion pair correlation function exhibits oscillations beyond the WS radius, which is typical of liquid-like behavior. It is easily understood from Equation~\eqref{eq_VAMPIRES_vel} that the correlation peaks of $h(r)$ draw some electrons. They are also ``dressed'' with the ion electron cloud density. This generates repulsive features between the central nucleus and the first correlation peak and between the successive correlations peaks because of the potential overlap of electron clouds. Consistently, electrons are displaced away from these regions of potential overlap. As a consequence, $v_\text{el}$ has a zero inside the WS sphere, and its effective range is thus shorter than $R_\text{WS}$.

In this model, it seems that the pressure ionization phenomenon goes along with the switching to the liquid-like regime. This is illustrated in Figure~\ref{fig_Li_VAMPIRES}d,e. The increase in the mean ionization is connected to a sharp increase in the coupling parameter. Across the pressure ionization edge, the plasma switches from a moderate-coupling to a strong-coupling regime, with the related feedback on the range of $v_\text{el}$. In addition to decreasing the value of $R_\text{WS}$, the range of $v_\text{el}$ switches from longer than $R_\text{WS}$ to shorter than $R_\text{WS}$. This explains why pressure ionization leads to a steeper increase in the mean ionization in this model than in VAAQP or INFERNO.

In the VAMPIRES model, the pressure-ionization phenomenon, as well as the switching from the Debye-Hückel-scale to the WS-scale decay, stems from a first-principle accounting for the structure of the ion fluid. At the same time, at each thermodynamic condition, the ionization state of the plasma is obtained from the condition of thermodynamic equilibrium.

One of the known modeling issues for such a model is that the electrons of an ion feel the surrounding ions through their average distribution, given by the pair correlation function. Among the possible outcomes of such a modeling effort could be that this static picture breaks down at some thermodynamic conditions.

%Work on the VAMPIRES model is still in progress, and application studies are still to be performed. Therefore, the following sections will only focus on applications of cavity-based models, which may, however, give relevant information.

\section{Research Prospects}

Accounting for the ion structure of the fluid in the modeling of the atomic structure is the only way to better address pressure ionized plasma, for it is the interactions between the electron clouds of the various ions that yield the pressure ionization phenomenon. Whether a statistical description based on the pair distribution function (so in a sense, on the average environment of an ion) is sufficient to  correctly model the pressure ionization is unknown. An alternative description could be that of ion-ion collisions, which is adopted for instance in line shape calculations. However, such representation of ion-ion interactions resorts to some assumed definition of the ion. 

The VAMPIRES model is in my opinion a notable step towards the accounting of the ion-fluid structure in atomic structure calculations. Extensions of this model to the cases of mixture and detailed modeling are already in progress.

In the limit of weak coupling, a thorough study of the transition of the VAMPIRES model to the ideal-gas limit may shed some new light on the long-standing controversy related to continuum-lowering models.

However, some aspect of this model also remain unsatisfactory. For instance, results stemming from this model for strong-coupling situations are puzzling, and we still need to gain insight on this model and assess its weaknesses critically.

Of course, it is to be expected that such research effort results in the conclusion that atomic modeling is not relevant at some conditions. However, atomic description of the matter is such a powerful theoretical tool that it deserves to be extended as far as possible. In this topic, molecular-dynamics simulations may constitute a complementary approach that could help to address some issues.

{ 
\newpage
\thispagestyle{empty}~
}

%-------------------------
\chapter{Atomic Physics of Non-LTE Plasmas\label{ch_NLTE_plasmas}}
%-------------------------

%--------------------------------------------------
\section{Collisional-Radiative Modeling of Plasmas\label{sec_CR_modeling}}
%--------------------------------------------------

\subsection{Ideal Plasma as a Premise of Collisional-Radiative Modeling}
For systems out of equilibrium, the fundamental postulate of statistical physics is the evolution according to a stationary Markovian dynamics. This probabilistic postulate relates to entropy production in the system (i.e.\ second principle), just as the postulate of equiprobability of microstates relates to entropy maximization at equilibrium. As a consequence of the stationary Markovian dynamics of the system, evolution of the probability distribution function for a statistical ensemble is solution of a master equation.

\begin{align}
&\frac{dP_\Xi(t)}{dt}=\sum_{\Xi'}P_{\Xi'}(t)\tau_{\Xi'\rightarrow \Xi}-P_\Xi(t) \sum_{\Xi'}\tau_{\Xi\rightarrow \Xi'}\\
&\frac{d}{dt}\bar{P}(t) = \bar{\bar{\mathcal{T}}} \bar{P}(t)\text{ under matrix form}
\end{align}
Here $\Xi$ denotes a microstate of the system as a whole. 
By construction, the rate matrix $\bar{\bar{\mathcal{T}}}$ preserves the normalization of the probability distribution $P_\Xi$ associated to the ensemble. 

When choosing a particular set of external thermodynamic variables imposed to the system, one requires the rate matrix to have the relevant equilibrium statistical distribution as a stationary solution. It is the microcanonical distribution for an isolated system, the canonical distribution if the system has a thermostat, etc. Usually, this is done using the stronger condition of detailed balance of probability fluxes \cite{vanKampenbook}.
\begin{align}
&P_{\Xi'}^\text{eq}\tau_{\Xi'\rightarrow \Xi}=P_\Xi^\text{eq}\tau_{\Xi\rightarrow \Xi'}
\end{align}
where $P_\Xi^\text{eq}$ denotes the equilibrium distribution related to the chosen ensemble.

In the context of NLTE plasmas, one often postulate as a premise that the plasma is modeled as a collection of \emph{independent} ions and continuum electrons, seen as free electrons. This corresponds to the picture of an ideal plasma described in Section~\ref{sec_ideal_plasmas}. It usually goes along with the isolated-ion model of the atomic structure, which disregards the continuum electrons. 

Using such a model, we can switch from a description of the system in terms of general many-ion-many-electron-many-photon microstates $\Xi$ to a tremendously simpler description in terms of independent atomic states $\Psi$, free-electron states $\vec{k},s$, and photon states $\vec{k},s$. These various particles are then described using separate probability distributions. 

The notion of probabilistic atomic processes that may occur in the plasma also stems from this modeling framework. One considers that interactions of ions with photons are related to radiative atomic processes, and interactions of ions with free electrons are related to collisional atomic processes. This leads to the collisional-radiative (CR) modeling of the probability distribution $P_\Psi$ of the ion atomic states $\Psi$. In the following, $\Psi$ will indeed denote an energy level rather than a single atomic state. Such an energy level may be a fine-structure level or the mean energy of a statistical object gathering several fine-structure levels, as for instance a configuration or a superconfiguration.

Radiative processes are characterized by the absorption or the emission of radiation by an ion, changing its state $\Psi$. The strong distinction between bound and continuum (free) electrons implies that bound-bound, bound-free and free-free channels are treated separately. We thus distinguish among photo-excitation (PE), line emission (LE, spontaneous and stimulated), photo-ionization (PI), and radiative recombination (RR, spontaneous and stimulated).
\begin{align}
&^Z\text{X}^{Z^*+}_\Psi(E_\Psi)+\gamma(h\nu) \xrightleftharpoons[\text{LE}]{\text{PE}} {}^Z\text{X}^{Z^*+}_{\Psi'}(E_{\Psi'}=E_\Psi+h\nu)
\label{eq_LEPE}\\
&^Z\text{X}^{Z^*+}_\Psi(E_\Psi)+\gamma(h\nu) \xrightleftharpoons[\text{RR}]{\text{PI}} {}^Z\text{X}^{(Z^*+1)+}_{\Psi'}(E_{\Psi'})+\text{e}^-(\varepsilon=h\nu-E_{\Psi'}+E_{\Psi})
\label{eq_RRPI}
\end{align}
Free electrons being considered as a separate species, and not as a part of the ion atomic structure, free-free processes do not play a direct role in the CR atomic modeling.
 
Processes involving the interaction between a free electron and an ion, without the emission of a real photon, are called collisional processes. The interaction between a free electron and an ion is modeled as a Coulomb interaction. Coulomb interaction being a two-body observable, the Slater selection rules result in the possibility for both one- and two-electron jumps between the initial and final atomic states. The one-electron jumps correspond to collisional excitation (CE), collisional de-excitation (CD), collisional ionization (CI), and 3-body recombination (3R).
\begin{align}
&^Z\text{X}^{Z^*+}_\Psi(E_\Psi)+\text{e}^-(\varepsilon) \xrightleftharpoons[\text{CD}]{\text{CE}} {}^Z\text{X}^{Z^*+}_{\Psi'}(E_{\Psi'})+\text{e}^-(\varepsilon'=\varepsilon-E_{\Psi'}+E_\Psi)
\label{eq_CDCE}\\
&^Z\text{X}^{Z^*+}_\Psi(E_\Psi)+\text{e}^-(\varepsilon) \xrightleftharpoons[\text{3R}]{\text{CI}} {}^Z\text{X}^{(Z^*+1)+}_{\Psi'}(E_{\Psi'})+\text{e}^-(\varepsilon')+\text{e}^-(\varepsilon''=\varepsilon+E_\Psi-E_{\Psi'}-\varepsilon')
\label{eq_3RCI}
\end{align}
Accounting for the energy conservation, the only two-electron processes that possibly occur are the auto-ionization (AI, called Auger effect in some context) and dielectronic capture (DC).
\begin{align}
&^Z\text{X}_{\Psi}^{Z^*+}(E_{\Psi}) \xrightleftharpoons[\text{DC}]{\text{AI}} {}^Z\text{X}^{(Z^*+1)+}_{\Psi'}(E_{\Psi'}=E_\Psi-\varepsilon)+\text{e}^-(\varepsilon) 
\label{eq_DCAI}
\end{align}

At this level of approximation, the CR model is connected to the modeling of the whole macroscopic plasma through the energy distributions of the photons and free electrons, which may be subject to non-local transport phenomena. In some situation, one may also consider ion-ion collisional processes (for instance, ionization by ion-ion collision or charge-exchange processes). In the latter case, the velocity distribution of ions also enters into play.

For many applications we are dealing with, the mean free path of ions and free-electrons is much shorter than that of photons. We then assume that free-electrons are locally in equilibrium at a temperature $T_\text{e}$ and that the velocity distribution of ions is also Maxwellian, with a temperature $T_\text{i}$. One thus focuses on the description of the photon energy distribution (local radiation field) and of the probabilities of the various atomic states of ions. 

This leaves us with a master equation for the probabilities of the ion atomic states $P_\Psi$, where the transition rates between atomic states $\tau_{\Psi\rightarrow \Psi'}$ depend on the local free-electron temperature $T_\text{e}$ and density $n_\text{e}$, and on the local radiative intensity averaged over polarization and direction $I_\omega$ (we assume isotropic distribution of the ions).

The CR master equation for the probabilities of the ion atomic states $\Psi$ then writes:
\begin{align}
&\frac{dP_\Psi(t)}{dt}=\sum_{\Psi'}P_{\Psi'}(t)\tau_{\Psi'\rightarrow \Psi}\left\{I_\omega,n_\text{e},T_\text{e}\right\}-P_\Psi(t) \sum_{\Psi'}\tau_{\Psi\rightarrow \Psi'}\left\{I_\omega,n_\text{e},T_\text{e}\right\}\label{eq_CR_atomic}\\
&\frac{d}{dt}\bar{P}(t) = \bar{\bar{T}} \bar{P}(t)\text{ under matrix form}\label{eq_CR_atomic_matrix}
\end{align}

Let us remark that extensions to improve over the LTE hypothesis for free electrons is an active field of research \cite{Le19}. However, the consistency between the probabilistic postulate of statistical physics and the partially or fully deterministic treatment of the transport equation is a highly nontrivial issue, also related to the treatment of Bremsstrahlung in the classical kinetic theory of plasma (see, for instance \cite{Martinez19}).

When $I_\omega$, $n_\text{e}$ and $T_\text{e}$ do not vary appreciably over the lifetime of the atomic states, it is justified to make a quasi-stationary approximation. One is then interested in the stationary distribution $P_\Psi^\text{CRE}$ of equation \eqref{eq_CR_atomic}, at given $I_\omega,n_\text{e},T_\text{e}$, which is called \emph{collisional radiative equilibrium} (CRE). To find the CRE requires solving a set of linear equations. On the other hand, calculating the evolution of $P_\Psi(t)$ with time requires one to solve the set of coupled ordinary differential equations, which can be stiff due to the range spanned by the lifetimes of the various atomic states.

Normalization of the probability distribution is included in the initial condition for the time-dependent problem. Since the rate matrix preserves the normalization, one of the equation (or matrix row) may be replaced by the normalization condition. When considering the stationary case, this property of the rate matrix results in redundancy of the equation set, since the system has a solution for any possible choice of probability normalization. One of the equation has to be replaced by the chosen normalization condition in order to remove the indeterminacy.

Let us finally remark that, in many contexts, the input to the problem of determining the microscopic state of the plasma is the matter density, or the ion density $n_\text{i}$, rather than the free-electron density $n_\text{e}$. Modeling of the CRE of the plasma then includes determining its mean ionization. One has to solve self-consistently for the stationary state of equation~\eqref{eq_CR_atomic_matrix} together with the neutrality condition:
\begin{align}
\sum_{\Psi} P_\Psi Z^*_\Psi = \frac{n_\text{e}}{n_\text{i}}
\end{align}
This requires in practice to iterate the calculations of the probability distribution $P_\Psi$ and of the free-electron density $n_\text{e}$.

\subsection{Relation to Radiation Transfer and Hydrodynamics}
The radiative  properties (emissivity, opacity), thermodynamic  properties (pressure, internal energy, specific heat) and transport properties (heat conductivity) of the non-LTE plasma depend on the probability distribution of the atomic states $P_\Psi$. These macroscopic properties are inputs for the hydrodynamic equations, that allow to determine $n_\text{i}$, $T_\text{e}$, and for the radiation transfer equation, that allows to determine $I_\omega$. In principle, one has to solve all these equations and the CR equation self-consistently. In practice, an operator splitting is frequently performed and one only cares about achieving some degree of consistency among radiation transfer, heat transfer and CR modeling. This problem is already of tremendous complexity.

When the radiation transfer of interest is not contributing significantly to heat transfer, one may consider $n_\text{e}$, $T_\text{e}$ as fixed functions of space. This is often the case when considering, for spectroscopic purposes, the radiation transfer of a few lines that are escaping from a plasma. These lines may be partially reabsorbed, while not contributing significantly to the heat transfer. To deal with this problem, there are many approaches offering various degrees of approximation (for syntheses, see \cite{Kalkofenbook,HubenyMihalasbook}).
 
%For instance, the escape-factor method renders implicit the dependency of the radiation field by resorting to net rates of emission and absorption of photon at the level of a finite-size cell in the CR problem. This method is rather simple to implement but is mostly limited to the accounting for line reabsorption within a sole homogeneous cell, or a small set of cells. It nevertheless requires to evaluate the line profiles in addition to the rates and resorts to an assumption on the cell shape. This model is often applied with the simplest model for $n_\text{e}$, $T_\text{e}$: an homogeneous cell in a optically thin medium with no incoming radiation field (i.e. $I_\omega=0$).

%A common way of dealing with the coupling between CR and radiation-transfer equations is to iterate between solution of the radiation transfer equation and calculation of the radiative properties using the CR model. This principle is often called Lambda-iteration and several methods aim at improving its efficiency. One may cite the core-saturation method of Ribicki \cite{} or the variant proposed by Peyrusse \cite{PeyrusseTRANSPEC}. These models are 

%When considerinig radiation transfer that drives heat transfer, one also has to obtain 

In general, drastic simplifying assumptions are unavoidable in order to limit either the numerical cost of the CR model, or the number of times it has to be solved. The modeling of a non-LTE plasma flow always relies on some tradeoff between simplifications in the modeling of the radiation transfer and in the CR modeling. 

For instance, the state-of-the-art approaches to CR modeling implemented in numerical codes like ATOMIC \cite{Sampson09,Fontes15} or AVERROES \cite{Peyrusse99,Peyrusse00,Peyrusse01} may serve as references for the CR modeling in itself. However, they are only tractable for given $n_\text{e}$, $T_\text{e}$, $I_\omega$, which amounts to completely disregards the radiation transfer problem. This is a limiting case of the tradeoff but, in some situations, it can be relevant to assume a homogeneous cell at fixed $n_\text{e}$, $T_\text{e}$, optically thin and with a known incoming radiation field $I_\omega$. In such a case, all the effort can be put on the CR model.

The relevant tradeoff largely depend on the considered physical system and, of course, there is no guarantee that there exists a tractable and relevant tradeoff for any system.

\subsection{Detailed Balance in the Collisional-Radiative Framework}

The specialization of the canonical detailed-balance relation to the collisional radiative description is obtained by balancing the probability fluxes between atomic states assuming canonical-equilibrium distributions for the photons (Planckian radiation field) and free-electrons (Maxwellian energy distribution). 

For the bound-bound radiative transitions \eqref{eq_LEPE}, this yields the well-known Einstein relations \cite{Einstein17} among the so-called Einstein coefficients:
\begin{align}
g_\Psi B^\text{PE}_{\Psi\rightarrow\Psi'}&=g_{\Psi'} B^\text{LE}_{\Psi'\rightarrow\Psi}\label{eq_det_bal_BB_RAD1} \\
\frac{A^\text{LE}_{\Psi'\rightarrow\Psi}}{B^\text{LE}_{\Psi'\rightarrow\Psi}}
&=\frac{2\Delta E_{\Psi',\Psi}}{(hc)^2}\label{eq_det_bal_BB_RAD2} 
\end{align}
where $\Delta E_{\Psi',\Psi}=E_{\Psi'}-E_{\Psi}$, $A^\text{LE}_{\Psi'\rightarrow\Psi}$ is the spontaneous line emission rate, $B^\text{PE}_{\Psi\rightarrow\Psi'}$ and $B^\text{LE}_{\Psi'\rightarrow\Psi}$ are the absorption and stimulated line emission rates per unit spectral energy density, respectively. 

The extension of these relations to bound-free processes \eqref{eq_RRPI} are the Einstein-Milne relations \cite{Milne24}:
\begin{align}
\sigma^\text{PI}_{\Psi\rightarrow\Psi'}(\nu)&=\frac{g_{\Psi'}}{g_{\Psi}}\frac{16\pi m_\text{e}}{h^3}\varepsilon G^\text{RR}_{\Psi'\rightarrow\Psi}(\varepsilon) \label{eq_det_bal_BF_RAD1}\\
\frac{F^\text{RR}_{\Psi'\rightarrow\Psi}(\varepsilon)}{G^\text{RR}_{\Psi'\rightarrow\Psi}(\varepsilon)}&=\frac{2(h\nu)^3}{(hc)^2}\label{eq_det_bal_BF_RAD2}
\end{align}
where $\sigma^\text{PI}_{\Psi\rightarrow\Psi'}$ is the photoionization cross-section (cross-section with respect to the radiative spectral flux of photons). $F^\text{RR}_{\Psi\rightarrow\Psi'}$ is the spontaneous radiative-recombination cross-section (cross-section with respect to the free electron spectral flux). $G^\text{RR}_{\Psi\rightarrow\Psi'}$ is the stimulated radiative-recombination cross-section per unit photon spectral flux.

Let us remark that, in this context, the detailed balance relations are obtained assuming Dirac-$\delta$ line profiles. As soon as finite-width line profiles are considered, the detailed balance amounts to fulfilling the Kirchhoff's relation at each frequency.
As such, this topic is an open problem of spectral modeling, related to the description of frequency redistribution \cite{Makhrov90,Zemtsov93}. In practice, one often make an assumption for the line profile of a chosen process and infer the line profiles for the other ones so as to fulfill Kirchhoff's relation (see \cite{Busquet03}, and \cite{Gilleron15} appendix).

For collisional bound-bound processes \eqref{eq_CDCE}, the detailed balance yields the Klein-Rosseland relation \cite{Klein21}:
\begin{align}
\sigma^\text{CE}_{\Psi\rightarrow\Psi'}(\varepsilon)
=\frac{g_{\Psi'}}{g_{\Psi}}\left(1-\frac{\Delta E_{\Psi',\Psi}}{\varepsilon}\right)
\sigma^\text{CD}_{\Psi'\rightarrow\Psi}(\varepsilon')
\label{eq_det_bal_BB_COL}
\end{align}
where $\sigma^\text{CE}$ and $\sigma^\text{CD}$ are the cross-sections for collisional excitation and de-excitation, respectively.

For collisional bound-free processes \eqref{eq_3RCI}, the detailed balance yields the Fowler-Nordheim \cite{Fowler28} relation:
\begin{align}
g_{\Psi}\frac{h^3}{16\pi m_\text{e}}\varepsilon \sigma^\text{CI}_{\Psi\rightarrow\Psi'}(\varepsilon,\varepsilon')
=g_{\Psi'}\varepsilon'\varepsilon''\sigma^\text{3R}_{\Psi'\rightarrow\Psi}(\varepsilon',\varepsilon'')
\label{eq_det_bal_BF_COL} 
\end{align}

Finally, for dielectronic rates, the detailed balance condition gives:
\begin{align}
\sigma^\text{DC}_{\Psi'\rightarrow\Psi}=\frac{g_{\Psi'}}{g_{\Psi}}\frac{h^3}{16\pi m_\text{e}}\frac{A^\text{AI}_{\Psi\rightarrow\Psi'}}{\varepsilon}\delta(\varepsilon-\Delta E_{\Psi,\Psi'})\label{eq_det_bal_DIEL} 
\end{align}
where $A^\text{AI}_{\Psi'\rightarrow\Psi}$ is the autoionization rate and $\sigma^\text{DC}_{\Psi'\rightarrow\Psi}$ is the dielectronic capture (resonant) cross-section.

Obviously, when free electrons are assumed to follow a Maxwellian distribution at temperature $T_\text{e}=1/(k_\text{B}\beta_\text{e})$, then one may directly use the detailed balance of probability fluxes to infer the transition rates:
\begin{align}
g_{\Psi}e^{-\beta_\text{e} E_{\Psi}}\tau_{\Psi\rightarrow\Psi'}^\text{CE}
&=g_{\Psi'}e^{-\beta_\text{e} E_{\Psi'}}\tau^\text{CD}_{\Psi'\rightarrow\Psi}\\
g_{\Psi}e^{-\beta_\text{e} E_{\Psi}}\frac{n_\text{e}\Lambda_\text{e}^3}{2}\tau^\text{CI}_{\Psi\rightarrow\Psi'}
&=g_{\Psi'}e^{-\beta_\text{e} E_{\Psi'}}\tau^\text{3R}_{\Psi'\rightarrow\Psi}\\
g_{\Psi}e^{-\beta_\text{e} E_{\Psi}}\frac{n_\text{e}\Lambda_\text{e}^3}{2}\tau^\text{AI}_{\Psi\rightarrow\Psi'}
&=g_{\Psi'}e^{-\beta_\text{e} E_{\Psi'}}\tau^\text{DC}_{\Psi'\rightarrow\Psi}
\end{align}

Due to the canonical detailed-balance imposed on collisional and dielectronic cross-sections, and to the LTE assumption for free electrons, whenever radiative rates are unimportant, the atomic levels are in LTE at temperature $T_\text{e}$. This corresponds to the case of dense, cold matter. Moreover, due to the Einstein-Milne relations for radiative processes, whenever radiative intensity is a Planckian having temperature $T_\text{R}$, and strongly dominates over collisional processes, the atomic levels are in LTE at temperature $T_\text{R}$. This is the case, for instance, of a thin slab heated by an intense black-body radiation. Finally, whenever the radiative intensity is at equilibrium with the free electrons $T_\text{R}=T_\text{e}$, the atomic levels are in equilibrium at the same temperature. 

In practice, the detailed balance relations reduce the number of rates and cross-sections to calculate in order to build the CR matrix $\bar{\bar{T}}$. Many of the rates and cross-sections can be deduced from the rates and cross-sections of their inverse processes.

\subsection{Complexity and Tradeoff between Precision and Completeness\label{sec_completeness_precision_tradeoff}}

Because the atomic processes (\ref{eq_LEPE}-\ref{eq_DCAI}) cannot connect atomic states differing by more than one charge state, the CR matrix $\bar{\bar{T}}$ has a block-tridiagonal structure, except for the line corresponding to the probability normalization condition. More than the total rank of the matrix, it is the rank of the largest block submatrix than determines the complexity of the matrix-inversion problem.

This rank depends both on the completeness of the chosen set of levels and on its level of detail, when a statistical grouping of levels is performed (grouping by configuration, super-configuration...). For that reason the choice of a relevant tradeoff between level of detail and completeness is crucial in CR modeling (see, for illustration, \cite{Ralchenkobook}, Chapter~1 by S. Hansen). This choice depends on the plasma conditions, but often also on the application considered.

With my colleague Franck Gilleron, we started working on non-LTE plasma modeling with a motivation that was twofold. The first motivation was the development of a fast atomic-physics package to be included in radiation-hydrodynamics codes. The second motivation was the analysis of emission-spectroscopy measurements, which are frequently used as a diagnostics to infer plasma conditions in experiments. Over the past years, we developed two numerical tools, implementing different approaches, each suited to a particular category of application.

In this task, we have benefited much from the access to the AVERROES numerical code, developed by Olivier Peyrusse \cite{Peyrusse99,Peyrusse00,Peyrusse01}. This code implements a super-configuration-based CR model, with the possibility of performing a detailed configuration accounting for some chosen super-configurations. It implements a full quantum calculation of the rates and cross-sections, in particular based on the distorted-wave approach for collisional cross-sections.
 
We also benefited much from the NLTE code-comparison workshops, in which I started to participate in 2011. This workshop series was initiated by R. W. Lee in 1996 and continued since \cite{Lee97,Bowen03,Bowen06,Rubiano07,Fontes09,Chung13,Hansen13,Piron17,Hansen20}.
It is a unique opportunity to make thorough comparisons with many numerical tools, implementing various CR models such as ATOMIC \cite{Sampson09,Fontes15}, SCRAM \cite{Hansen07,Hansen11,Hansen11b}, CRETIN \cite{Scott10}, FLYCHK \cite{Chung05}, THERMOS \cite{Nikiforov}, NOMAD \cite{Ralchenko01}, DLAYZ \cite{Gao13} etc. It is also an occasion for people involved in non-LTE plasma modeling to exchange ideas and report openly about their theoretical and numerical issues. For experimentalists or applied physicists, it is also an opportunity to suggest relevant cases of study in order to stimulate a modeling effort and get some feedback. For instance, such an effort was sustained over several workshops in order to address Tungsten for tokamak applications, or Gold for ICF applications. Finally, interpretation of experiments are a recurring source of case studies (see, for instance \cite{Hansen13}) and feedback from the NLTE workshops may also stimulate experimental efforts (see, for instance \cite{Bastiani15,Marley18,Bishel23,MarleyAPIP23}).

\section{Applications to X-ray Emission Spectroscopy\label{sec_NLTE_Xray_spectro}}
Emission spectroscopy of plasmas is of interest in astrophysics, because its main observable  is the radiation emerging from astrophysical objects, in particular stellar atmospheres. It is also of interest as a diagnostics in experiments involving hot plasmas, performed on high-power lasers, pulsed-power devices and tokamaks. In these contexts, interpretation of emission spectra is often used to infer plasma conditions. 

When collisions do not dominate overwhelmingly over radiative processes, the sole fact that significant radiation is emerging, and thus escaping, from the plasma, drives it out of equilibrium.  For that reason, emission-spectroscopy analysis very often requires non-LTE plasma modeling.

In practice, observed radiation typically emerges from regions of the plasma of low to moderate densities, for which collisional processes do not dominate sufficiently to force LTE, and for which physical broadening of lines is weak enough to allow part of the line structure to be observed, especially lines from closed-shell configurations. Consequently, the modeling of emission spectra frequently requires mixing various levels of detail, in order to enable a relevant tradeoff between precision and completeness of the model. This is among the purposes of our standalone numerical code called DEDALE \cite{Gilleron15}.

In order to identify possible candidates for a fast atomic-physics package to be included in a radiation-hydrodynamics code, we needed our own numerical implementations, allowing us to easily test various simplified models and numerical methods. However, comparisons with more sophisticated tools like AVERROES \cite{Peyrusse99,Peyrusse00,Peyrusse01}, ATOMIC \cite{Sampson09,Fontes15} or SCRAM \cite{Hansen07,Hansen11,Hansen11b}, also required us to calculate some quantities, like spectral radiative properties, using more sophisticated approaches.

Besides, the interpretation of emission-spectroscopy experiments in which are directly involved \cite{MoranaPhD} or which are considered in the NLTE workshops, require flexibility in the modeling of the atomic structure. In particular, we needed a possibility to include tabulated data for fine-structure levels, taken from various sources, such as detailed atomic structure code (e.g., Cowan's code \cite{Cowan}, FAC \cite{Gu08} or CFAC \cite{StambulchikCFAC}), or the NIST atomic structure database \cite{NISTASD,Ralchenko20}.

Our DEDALE code is an implementation of a hybrid CR model mixing super-configurations and more detailed levels taken from tabulated data. According to the specific case of application, we can apply various modeling options.

In its most basic version, the model implemented in DEDALE resort to a statistical grouping of levels based on Layzer complexes. Layzer complexes are super-configurations for which supershells are the shells defined by the principal quantum numbers:
\begin{align}
(1)^{Q_1}\,(2)^{Q_2}\,(3)^{Q_3}...(n_\text{max})^{Q_{n_\text{max}}}
\text{ where } &(1)\equiv (1s)\\
&(2)\equiv (2s\,2p)\nonumber\\
&(3)\equiv (3s\,3p\,3d)\nonumber
\end{align}
A semi-relativistic extension of More's screened-hydrogenic model \cite{More82} is used to approach the electronic structure and estimate the mean energies of the Layzer complexes. The rates and cross-sections of radiative and collisional processes, including the dielectronic ones, are obtained from semi-empirical expressions resorting to the screened charges. Radiative rates are obtained from Kramers model \cite{Kramers23,Zeldovich,MoreUCRL91}.
Collisional cross-sections for excitation and ionization are estimated using the approaches of Van Regemorter and Mewe \cite{VanRegemorter62,Mewe72}, and Lotz \cite{Lotz67}, respectively. Dielectronic capture is estimated using the approach of Burgess \cite{Burgess65}, with a correction factor chosen in order to recover values close to the distorted-wave results of AVERROES.
The rates of non-resonant processes are calculated by integration, assuming Maxwellian distribution for the free electrons and accounting for the given radiative intensity for radiative processes.

\begin{figure}[t]
\begin{center}
\includegraphics[width=8cm] {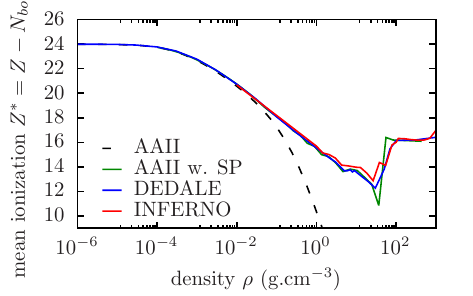} 
\end{center}
\caption{Mean ionization of Iron in thermodynamic equilibrium, defined as $Z$ minus the number of bound electrons, calculated by the average-atom model of isolated ion (AAII), the AAII model with Stewart-Pyatt correction and orbital suppression (AAII w. SP), DEDALE, and INFERNO, for Iron at temperature of 200 eV.
\label{fig_DEDALE_comparison_INFERNO}}
\end{figure}

This basic model is close to those implemented in CRETIN \cite{Scott10} or FLYCHK \cite{Chung05}. It enables fast calculations with a high level of statistical completeness, due to the large super-configurations used. In a quantum calculation of the rates, among the most time-consuming part is the computation of collisional ionization rates, which require double summations. Using semi-empirical formulas allows us to build most of the CR matrix in a relatively short amount of time. 

In order to refine over this approach, we can resort to tabulated data. Tabulated data may be used either to replace the semi-empirical estimates stemming from the basic modeling, or to refine over the statistical grouping of levels. One may, for instance, replace one of the Layzer complex by a set of configurations and fine structure levels.

The tabulated data that we include is usually calculated using other numerical codes such as AVERROES \cite{Peyrusse99,Peyrusse00,Peyrusse01} or CFAC \cite{Gu08,StambulchikCFAC}, depending on the kind of approach we need. Some data can also be taken from databases such as the NIST Atomic Structure Database \cite{NISTASD,Ralchenko20}, which includes precise experimental measurements.

Of course, splitting Layzer complexes into smaller super-configurations, configurations or fine-structure levels can impact strongly on the rank of the CR matrix. This requires an insightful choice of the statistical objects to split, and at which level of detail they should be treated. This choice depend both on the plasmas conditions and on the application considered. One may, for instance, put the focus on a refined description of the main levels contributing to a particular spectral region, or choose to split a too-large supershell at plasma conditions where it is expected to be open. At some conditions, metastable levels may also have a strong impact on the atomic-state distribution and require a detailed accounting.

In order to solve for the CR equilibrium, we can use various direct or iterative solvers, our default being a Gauss-Jordan elimination algorithm of our own, optimized to take advantage of the particular structure of the CR matrix.

In the case of low density plasmas, we usually truncate the set of levels by fixing a limiting value for the principal quantum number and excitation number.

When there is a significant effect of density on the atomic structure, we resort to a quantum-ion-cell model (see Section~\ref{sec_quantum_ion_cell}) with subshell populations fixed to the non-LTE mean populations. This model is used to evaluate which subshells are delocalized. Many-electron energy levels having nonzero populations in these delocalized subshells are gradually removed, first using a degeneracy-reduction scheme, and ultimately removing the level from the CR model. Using such an approach, we manage to mimic equilibrium results from ion-cell approaches (see Figure~\ref{fig_DEDALE_comparison_INFERNO}). However, all the limitations discussed in Section~\ref{sec_bound_state_suppr} apply to the present approach as well and a relevant modeling of dense-plasmas out of equilibrium is an open problem (see Section~\ref{sec_collisional_processes}).

The DEDALE standalone code does not aim at being used within a numerical scheme solving the radiation transfer equation. For that reason, we can afford to have spectral features that are not strictly consistent with the atomic energy levels used in the CR modeling.

In order to get realistic spectral features, even when using our most basic CR model, radiative properties are estimated using quantum calculations based on the Pauli approximation. For levels taken from tabulated data, the tabulated data is used also in the calculation of radiative properties.

Optionally, when calculating the radiative properties, we can use additional tabulated data in order to split some features which are  accounted for in the CR matrix as a single statistical object. In this case, the population of the more-detailed levels are inferred from an LTE hypothesis within their parent statistical object. This enables refining further the spectra, without increasing the rank of the CR matrix.

For transition arrays among statistical objects, the statistical broadening is accounted for as a contribution to the Gaussian width. For all radiative transitions, the physical broadening of lines is accounted for as follows. Doppler broadening related to the ion temperature $T_\text{i}$ gives a contribution to the Gaussian width. Electron Stark broadening is estimated in the impact approximation, as described in \cite{Dimitrijevic80}, and gives a contribution to the Lorentzian width. Finite-lifetime broadening is obtained using the inverse of diagonal terms of the CR matrix, and yields a supplementary contribution to the Lorentzian width. Line profiles are then modeled as Voigt functions, using the Gaussian and Lorentzian widths. At present, we do not account for any ion Stark broadening.

Our treatment of the radiative properties allows us, from a relatively simple and fast model of the CRE, to get rather realistic emissivity and opacity spectra allowing comparisons with more sophisticated approaches. This kind of comparison allows us, in particular, to assess the relevance of the simplified CR modeling, which is used in our atomic physics package for radiation hydrodynamics (see next section).

\begin{figure}[h]
\begin{center}
\includegraphics[width=8cm] {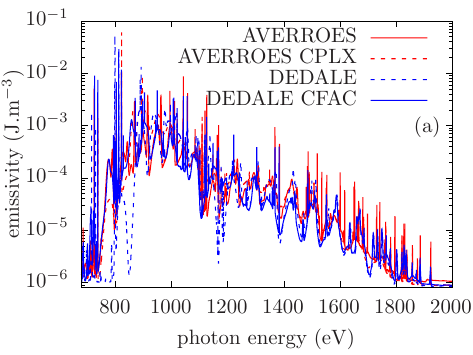} \hspace{1cm}
\includegraphics[width=8cm] {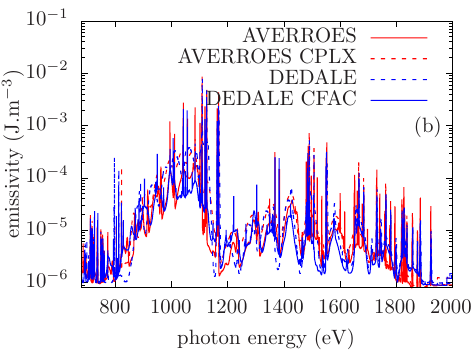} 
\end{center}
\caption{Comparison between results from the DEDALE and AVERROES codes on Iron emissivity at free-electron density of $10^{13}$\,e$^{-}$.cm$^{-3}$ and temperatures of $1$\,keV (a) and $1.6$\,keV (b). Calculations were performed with and without partial detailed-configuration accounting for the AVERROES code (AVERROES and AVERROES CPLX, respectively), and with and without using some tabulated configuration data from CFAC in the DEDALE code (DEDALE CFAC and DEDALE, respectively).
\label{fig_DEDALE_comparisons}}
\end{figure}

To illustrate this, figures~\ref{fig_DEDALE_comparisons} a and b presents comparisons between results from AVERROES with pure super-configuration accounting based on Layzer complexes (AVERROES CPLX), AVERROES with partial detailed-configuration accounting, DEDALE code using the basic CR model  and using also some detailed-configuration accounting from CFAC. The present cases of Iron are of interest for astrophysical applications and were suggested for the NLTE-12 workshop by Timothy Kalman (NASA).

The flexibility of modeling that we obtain by adding tabulated data with various level of detail allows us to cope with specific situations such as problems with metastable states and also to use DEDALE for detailed spectroscopic analysis that requires some fine-structure description (e.g., spectra involving lines from closed-shell configurations). 

\begin{figure}[t]
\includegraphics[width=8cm] {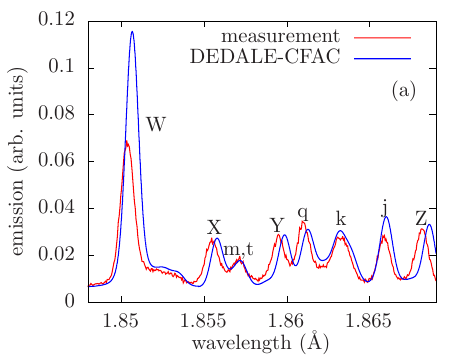} 
\hspace{1cm}
\includegraphics[width=8cm] {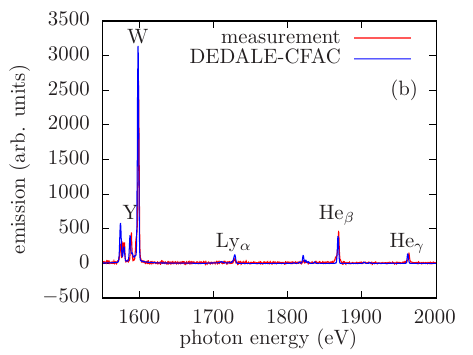} 
\caption{Interpretations of spectroscopy experiments using the DEDALE code. (a) K-shell features of He-like Iron, spectrum obtained on the TORE SUPRA tokamak \cite{MarchukPhD,Marchuk05}. Line notations are those of \cite{Gabriel72}. Assuming zero radiation field and electron density in the $10^{14}$ e$^-$.cm$^{-3}$ (coronal-limit case), the electron temperature inferred from the DEDALE computations was $1.9$ keV.
(b) K-shell features of He-like Aluminum. Preliminary interpretation of an emission spectroscopy experiment performed on the ELFIE facility (see \cite{MoranaPhD}).  Assuming zero radiation field, the conditions inferred from DEDALE computations are electron density $3.10^{20}$\,e$^{-}$.cm$^{-3}$ and electron temperature of $325$ eV.
\label{fig_DEDALE_interpretations_Kshell}}
\end{figure}

Figures~\ref{fig_DEDALE_interpretations_Kshell} a and b illustrate the use of DEDALE for the interpretation of K-shell emission spectroscopy experiments. For these calculations detailed level accounting data from the CFAC code was used. 

Figure~\ref{fig_DEDALE_interpretations_Kshell}a corresponds to an experiment performed on the TORE SUPRA tokamak, and was proposed as a case study by O. Marchuk \cite{MarchukPhD,Marchuk05} for the NLTE-9 workshop \cite{Piron17}. The plasma is at coronal conditions and results are thus not sensitive to electron density. In this spectrum one can observe He-$\alpha$ resonance line (W) with its Li-like satellites, and He-$\alpha$ intercombination line (Y). The detailed level accounting using CFAC, in particular for the fine-structure levels $1s\,2s\,^3S_0$, $1s\,2s\,^3S_1$ and $1s\,2p\,^3P_2$, enables a better description of the total $1s\,2s$ population and of the dipole-forbidden lines (X, Z).

Figure~\ref{fig_DEDALE_interpretations_Kshell}b present the interpretation of an experiment performed on the ELFIE laser facility during the Ph.D. thesis of Ambra Morana, supervised by Serena Bastiani \cite{MoranaPhD}. In the conditions we were expecting for this interpretation, our calculations were showing that the He$_\alpha$ resonance line was likely reabsorbed. In order to infer the plasma conditions from this emission spectrum, we have tried to reproduce the experimental intensity ratio of the Li-like satellites of the He$_\alpha$ resonance line to He$_\alpha$ intercombination line, as well as Ly$_\alpha$ to He$_\alpha$ intercombination line.

\begin{figure}[p]
\hspace{0.5cm}
\includegraphics[width=8cm] {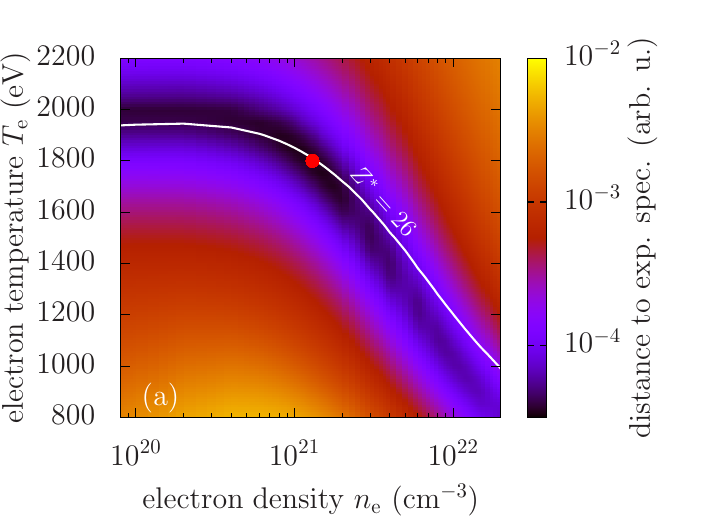} \hspace{1cm} 
\includegraphics[width=8cm] {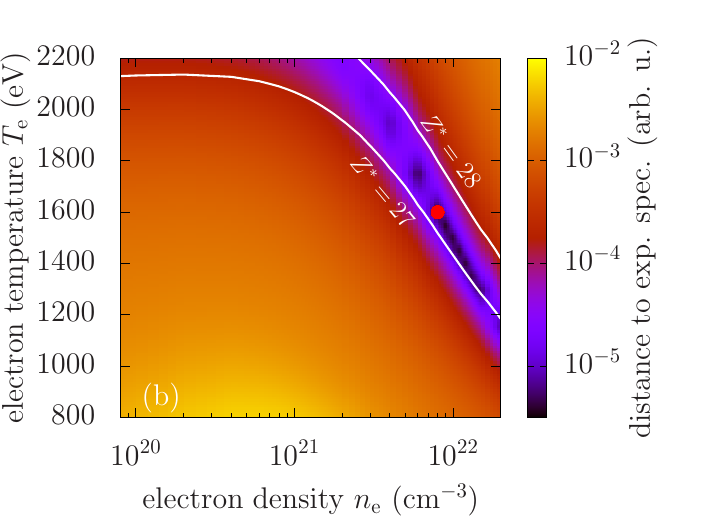} \\
\includegraphics[width=8cm] {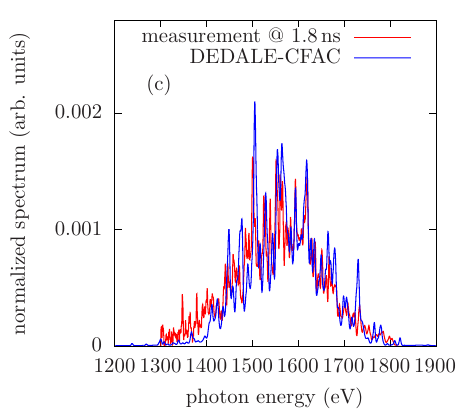} \hspace{1cm} 
\includegraphics[width=8cm] {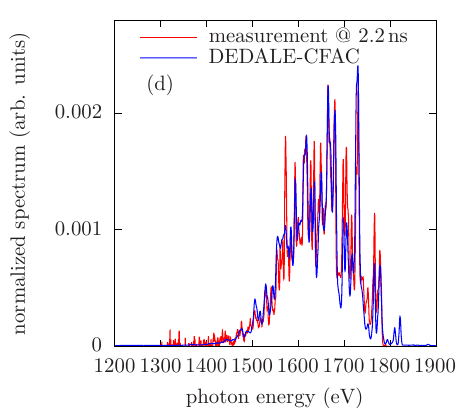} 
\caption{Interpretation of an L-shell spectroscopy experiments using the DEDALE code. The spectra correspond to measurements of L-shell features of open-L-shell Germanium, obtained on the OMEGA laser facility \cite{Bishel23}. (c) and (d) correspond to the gated spectra obtained with delays of $1.8$ ns and $2.2$ ns, respectively. For these calculations detailed configuration accounting data from the CFAC code was used. Interpretation was done assuming zero radiation field and making no hypothesis on the ion density.
In order to quantify the agreement among synthetic and experimental spectrum, we performed a principal-component analysis on a set of DEDALE synthetic spectra. Experimental spectrum was projected onto the 10 first principal components and this projection was used to define the distance to minimize. Map of distance to experimental spectrum (c) (respectively (d)) is shown in (a) (respectively (b)). The red dot corresponds to the DEDALE best fit respresented on (c) (respectively (d)).
\label{fig_DEDALE_interpretations_Lshell}}
\end{figure}

Figures~\ref{fig_DEDALE_interpretations_Lshell}\,a and b illustrate the use of DEDALE for the interpretation of an L-shell emission spectroscopy experiment. It presents two spectra obtained at different time, corresponding to L-shell features of open-L-shell Germanium, obtained on the OMEGA laser facility. The preliminary analysis of these spectra \cite{Bishel23} was a case study proposed by Edward Marley for the NLTE-11 workshop. It is part of a series of experiments performed on OMEGA and NIF \cite{Marley18,Bishel23,MarleyAPIP23} that was stimulated by the feedbacks of the NLTE workshops \cite{Piron17,Hansen20}. For our calculations, detailed configuration accounting data from the CFAC code was used. Interpretation was done assuming zero radiation field and making no hypothesis on the ion density. In order to quantify the agreement between synthetic and experimental spectra of such complexity, we performed a principal-component analysis on a set of DEDALE synthetic spectra covering the plausible range of temperatures and electron densities. We then projected the experimental spectrum onto the 10 first principal components and used this projection to define a distance to minimize. On figures \ref{fig_DEDALE_interpretations_Lshell}\,a,b, one can see how the minimal distance roughly follows the curves of same mean-ionization.

The analysis published more recently in \cite{Bishel23}, along with the experimental data, resorts to a joint measurement of Scandium K-shell emission spectra, and to an estimation of the ion density. The latter is inferred from the emitting volume observed using a side-on gated pinhole camera. This allows to better constrain the interpretation.% Adding such information would have of course changed the temperature of the DEDALE interpretation.

Work is in progress towards a robust accounting of line reabsorption in our DEDALE code using the escape-factor approach (see, in particular \cite{Drawin73,Irons79,RybickiKalkofenbook}). However, in the task of inferring plasma conditions from spectroscopic analysis, using lines that are significantly reabsorbed is always questionable. The results highly depends on the line profile, which is difficult to evaluate precisely, and is most often hidden by instrumental resolution. It also requires an estimate of the reabsorption length, which is usually not known and ends up as a parameter enabling to tweak the line intensity of reabsorbed features.

\section{Applications to Radiation-Hydrodynamics Simulations\label{sec_NLTE_rad_hydro}}
A way of producing X-rays is to heat a material of high atomic number up to temperatures at which significant part of its emission spectrum will lie in the X-ray range. Such heating can be achieved using absorption of infrared or optical light of a high-power laser (e.g., case of a hohlraum), or Joule effect in a pulsed-power device (e.g., case of a Z-pinch). However, at high frequencies, the opacity of the material drops. Non-thermal radiation (in the sense of $T_\text{e}$) escaping from the interior causes significant departure from the ideal black-body behavior and from the local thermodynamic equilibrium in the regions of moderate density.

Modeling the energy balance of X-ray sources is crucial for estimating their conversion efficiency. In this matter, a detailed description of the emission spectrum is often not required, but there is an interplay between hydrodynamics, heat conduction, radiation transfer and collisional-radiative dynamics. This requires radiation-hydrodynamics simulations of a non-LTE plasma flow.

In order to estimate the non-LTE plasma radiative properties, models based on a two-level CR approach were proposed and used in radiation-hydrodynamics simulations since many-years now \cite{Busquet93,Bowen03}. These models resort to the notion of an effective ionization temperature $T_Z$ at which the LTE (i.e.\ Saha) model yields the same mean ionization. Having $T_Z$, one estimates the radiative properties starting from the LTE properties at the temperature $T_Z$, and then applying some correction to get the free-free contribution corresponding to the electron temperature $T_\text{e}$. Such an extremely simplified approach enables very fast calculations, relying on tables of LTE radiative properties.

Average-atom approaches to CR modeling \cite{Lokkereport77,Rozsnyai97,Hansen23} are also used in the context of radiation-hydrodynamics simulations. In particular CR screened-hydrogenic average-atom models are used for this purpose. One may cite for instance the XSN \cite{Lokkereport77} and NOHEL (see in \cite{Bowen03}, Section 2.3) packages. The average-atom CR model is much more intensive computationally than the two-level approach. In the average-atom approach, the levels considered correspond to one-electron orbitals, which are grouped into shells in the case of the screened-hydrogenic approximation. This leads to a CR matrix of low rank, of the order of a few units to a few tens, while offering unbeatable completeness.  

However, average-atom CR models have two limitations. First, as in the equilibrium average-atom model, the orbitals, as well as the transition rates of the various processes, have to be calculated self-consistently with the orbital populations. This is required in order to have the CR matrix pertaining to the CRE. Such an implicit dependency makes the CR system strongly nonlinear, which notably impacts on the robustness of the solution algorithm. Second, dielectronic processes crucially depends on the two-occupation correlation functions \cite{Zhdanov78,MoreUCRL87,More88b,Albritton99}. Although an average atom CR approach including the description of such correlation functions was proposed \cite{Dallot98,Mirone98}, it seems beyond reach for a use within radiation-hydrodynamics simulations.

In order to perform integrated simulations of hohlraum and other X-ray sources, we developed a fast non-LTE-atomic-physics package called ICARE (Inline Collisional And Radiative Equilibrium).  This atomic-physics package can be run inline the CEA radiation hydrodynamics code, named TROLL. TROLL is used for the simulations of hohlraums and other types of X-ray sources (see \cite{Lefebvre19,Jacquet23}). For this project, we collaborate closely with the group in charge of developing the TROLL numerical code, as well as with the people in charge of designing X-ray sources. On the basis of their daily use of our ICARE package, they provide us with precious feedback.

ICARE is a lighter version of the DEDALE standalone code. In this version, we limit ourselves to the model based on the screened-hydrogenic atomic structure and semi-empirical formula for the rates and cross-sections. The considered set of levels is limited to a few tens of Layzer-complex-type super-configurations for each charge state. For the application to radiation hydrodynamics, the input is the ion density, so we resort to an iteration scheme to solve for $n_\text{e}$. In the course of the iterative procedure, we increase gradually the set of levels and perform a detection of LTE, in order to speed up the convergence steps. The approach remains even more intensive computationally than the CR average-atom approach. However, the solution algorithm seems somewhat more robust.

The spectral emissivities and opacities, which we subsequently calculate, are used for solving the radiation transfer problem. For that reason, there is a need for consistency in the treatment of radiative transitions between the CR model and the calculation of radiative properties.
Super-configuration populations, together with the same radiative transition energies, rates and cross-sections as in the CR model, are thus used to compute the radiative properties. In practice, contributions to the spectral opacities and emissivities are first averaged over the whole set of super-configurations and then accounted for as statistical features corresponding to each mono-electronic jump.

As a result of the simplified modeling and optimized implementation, the typical duration of a ICARE calculation lies in the 0.01 to 0.1 s range on an Intel Xeon E5630 @2.53 GHz. This enables significantly complex simulations to be performed within human-acceptable time. It nevertheless constitutes one of the most time-consuming part of the radiation-hydrodynamics simulation in virtually any case involving a non-LTE plasma flow. To give an idea, a typical 2-dimensional simulation of a megajoule-class-laser hohlraum may require a few $10^{7}$ to a few $10^{8}$ calls to the ICARE atomic-physics package. A 3-dimensional simulation requires typically a thousand times more.

\begin{figure}[t]
\includegraphics[width=8cm] {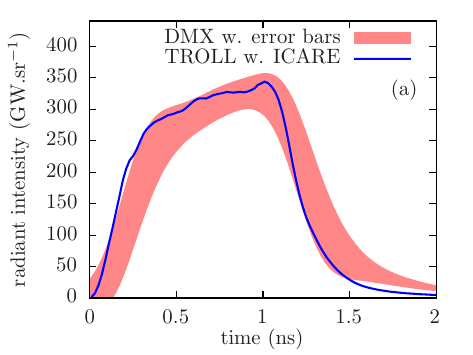} 
\hspace{1cm}
\includegraphics[width=8cm] {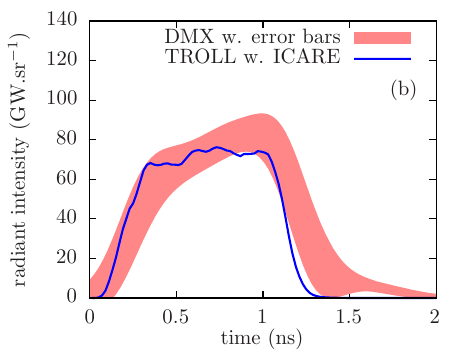} 
\caption{Radiant intensity as a function of time of a silver X-ray source driven by the OMEGA laser (shot \#95066). Detailed plots of radiant intensity integrated over the 0 - 2\,keV (a) and >2\,keV (b) spectral ranges, respectively, as measured by the DMX broadband spectrometer (red) compared to calculations from the TROLL radiation hydrodynamics code using the ICARE NLTE atomic physics package. Detailed descriptions of the experiment and simulation are given in \cite{Jacquet23}, as well as analysis of other shots.
\label{fig_OMEGA_X_ray}}
\end{figure}

A recent publication \cite{Jacquet23} gives an example of application of the ICARE package. It deals with the experimental validation of the design of an X-ray source. The experiment was performed on the OMEGA facility. In this experiment, a silver foil is heated by the beams of the OMEGA laser. Time-resolved, partially-frequency-integrated emission is recorded by the DMX broad-band spectrometer. Details about the experiment and interpretation using the TROLL and ICARE tools are given in \cite{Jacquet23}. Figure~\ref{fig_OMEGA_X_ray} shows the comparison of the simulated radiant intensity on the DMX axis, compared to the measurement, on a shot that was rather well reproduced by the TROLL/ICARE simulation. However, beside these encouraging results, there were also other shots showing larger discrepancies.

Radiative properties of non-LTE plasmas have a functional dependency on the radiative intensity. For that reason, it is difficult to use tables. An approach that was implemented by Howard Scott (see \cite{Ralchenkobook}, Chapter 4 by H. Scott) is to make tables of the whole linear response function of the radiative properties with respect to the input radiative intensity, for a given reference intensity. From the response function, one can correct the radiative properties using the difference between the considered radiative intensity and the reference. Encouraging results were reported for some cases of application, including hohlraum simulations. Of course such table is made for a chosen reference radiative intensity, which should be suited to the application, in order to remain within the linear response domain.

An alternative to making tables is to fit the results using regression analysis techniques.
In the course of an integrated simulation of an X-ray source, many calculations are performed with the ICARE package at closely similar conditions. Moreover, this kind of simulations are often likely to be repeated with small variations of many modeling parameters. In order to take advantage of this strong similarity among the calculations made with the ICARE package, an effort is now ongoing to mimic its results using specifically trained neural networks. In this view, we initiated a collaboration with Gilles Kluth, who started to work on this topic during a stay at the Lawrence Livermore National Laboratory \cite{Kluth20}.  A typical application of this method would be to first perform an approximate 2-dimensional simulation of an X-ray source of a given design using the ICARE package, then train a specific neural network and use it for a more realistic, and more demanding, 3-dimensional simulation \cite{Kluth24}.

Finally, let us remark that the effort of making tables of linear-response matrices, just as the training of a neural network, requires in itself a rather fast atomic-physics numerical code, based on a simplified model.

%Work is now in progress in order to gradually by a massive use of pre-calculated data, in order to refine over our simplified model.

\section{Collisional Processes in Dense Plasmas and Research\\ Prospects\label{sec_collisional_processes}}
The accounting for continuum electrons and pressure ionization in the modeling of non-LTE plasma is an open problem, and relevant research prospect. Dense plasmas are often associated to LTE conditions, because collisional processes tend to dominate at high density. However, there is a rather broad range of conditions for which both pressure ionization of orbitals and departure from LTE significantly matters. Moreover, some experiments using short radiation pulses can generate dense matter which strongly departs from equilibrium (short-pulse, high intensity lasers, X-ray free-electron lasers...).

As explained in Section~\ref{sec_CR_modeling}, the CR model of non-LTE plasmas is based on the physical picture of an ideal plasma of isolated ions. The problem of truncating the set of states, and of introducing non-ideality correction in the model is just as relevant as in the case of LTE. In the CR context, this problem also has some specific implications like the limitation of dielectronic channels. The neglect of resonances notably yields a discontinuity among the 1- and 2-electron collisional processes. There is thus an interest for a description of collisional processes in the framework of dense-plasma atomic models.

In the heuristic bridging to the Ziman formula (see Section~\ref{sec_reg_low_freq}), we obtained the collisional frequency by considering the net rate of electron-ion elastic-scattering out of momentum $\vec{k}$. In this context, we used the electron-ion elastic-scattering cross-section that stems from the limit of weak collisions. The latter only depends on the wave functions through the phase shifts (see, for instance,~\cite{SobelmanVainshteinYukov}):
\begin{align}
\sigma_{\text{scatter}}(\varepsilon_k)
=a_0^2\,\frac{4\pi}{k^2}
\sum_{\ell}(2\ell+1)\sin^2(\Delta_{k,\ell})
\end{align}
The elastic scattering of electrons by ions may be categorized as an elementary collisional atomic process, even if it does not change the ion electronic state.

\begin{figure}[t]
\includegraphics[width=8cm] {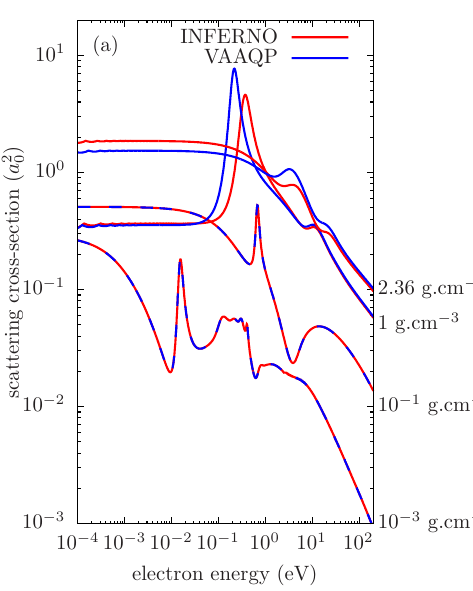} 
\hspace{1cm}
\includegraphics[width=8cm] {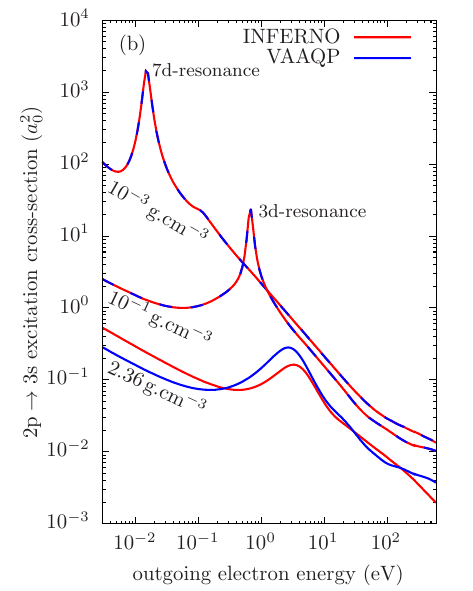} 
\caption{ Total electron-ion elastic-scattering cross-section ({a}) and 2p$\rightarrow$3s collisional-excitation cross-section ({b}) for a silicon plasma at a temperature of 5 eV, for various values of the matter density. Comparison between results from the INFERNO and VAAQP models, which are in close agreement up to matter density of 0.1 g.cm$^{-3}$.
\label{fig_electron_scatter}
\label{fig_coll_excitation}
}
\end{figure}

Figure~\ref{fig_electron_scatter}a displays the total electron-ion elastic-scattering cross-section for a silicon plasma at a temperature of 5 eV, for various values of the matter density. Contributions of some resonances are clearly visible in these cross-sections. Below matter density of 1 g.cm$^{-3}$, the results from the INFERNO and VAAQP models agree well.

For CR modeling, more relevant is the case of collisional excitation. The distorted-wave approximation (DWA; see, for instance,~\cite{SobelmanVainshteinYukov}) is widely used for the calculation of cross-sections of collisional processes for isolated ions (see, for instance \cite{Sampson09,Peyrusse99}). A straightforward approach to collisional ionization in dense plasmas consists of extending heuristically the DWA to models of pressure-ionized plasmas. One may calculate the collisional-excitation cross-section by generalizing the configuration-averaged collision strength~\cite{Peyrusse99} to the orbitals and fractional occupation numbers given by an average-atom model such as INFERNO or VAAQP.

As an illustration, in Figure~\ref{fig_coll_excitation}\,b we display the 2p$\rightarrow$3s collisional-excitation cross-section, obtained using the distorted-wave approach using the INFERNO and VAAQP models, respectively. The chosen case is a silicon plasma at 5 eV temperature, for matter densities of $10^{-3}$, $10^{-1}$, and 2.36 g.cm$^{-3}$. At high energies, the Born approximation behavior (plane-wave incoming and outgoing electrons) is recovered. At matter density of $10^{-3}$ g.cm$^{-3}$, the 7d bound orbital is delocalized, but still contributes a sharp resonance in the continuum.  At matter density of $10^{-1}$ g.cm$^{-3}$, the same occurs for the 3d bound orbital.

As a direct consequence of these resonances, we obtain sharp, quasi-discrete features in the corresponding cross-sections. These near-threshold sharp features may be seen as the remnants of the corresponding dielectronic recombination channels, namely, 2p$\rightarrow$3s,7d, and 2p$\rightarrow$3s,3d.

One thus can see how, prior to effectively removing them, screening of the potential may redispatch the collisional channels. Part of the dielectronic recombination channels then becomes collisional excitation.

However, the application of the usual distorted-wave method to dense plasma may be questionable in the context of dense plasmas. For instance, the effect of transient spatial localization of electrons was pointed as a potentially relevant effect, lacking in the usual distorted-wave approach~\cite{Zeng22}.

More generally, one may question the usual approach to collisional processes in the context of dense-plasma models. The usual CR approach is based on the ideal plasma picture. Collisional processes are introduced as a perturbative accounting for continuum electrons, seen as free, which supplements an isolated-ion Hamiltonian that completely disregards them (see, for instance,~\cite{SobelmanVainshteinYukov}). In this context, collisions are the only way continuum (free) electrons interact with the ion electronic structure. 

In principle, following strictly the ideal plasma model would even mean sticking with the Born approximation, because free electrons are supposed to be free, and thus not to interact at all with the ions. However, one usually considers this approximation not to be valid close to the ion, and the distorted-wave approximation is often preferred.

Models of pressure-ionized plasma account for the continuum electron as part of the ion electronic structure. They account for the static screening effect by continuum electrons. In this framework, one should probably adapt, re-interpret, or redefine the whole approach to collisional processes. To the best of the author's knowledge, such a rigorous approach to collisional processes in pressure-ionized plasma is an open question.

Finally, regarding the experimental side, let us mention that measures of a collisional-ionization cross-section in a dense plasma were recently performed and showed significant impact of density effects~\cite{vandenBerg18}.

{ 
\newpage
\thispagestyle{empty}~
}

%-------------------------
\chapter{Conclusions\label{ch_conclusions}}
%-------------------------
Atomic models of dense plasmas in themselves are still an active field of research, facing open questions as regards their theoretical foundations. The main shortfalls of models based on the notion of continuum lowering are rather well identified. However, models used in order to go beyond the continuum lowering picture are still mostly based on the picture of a Wigner-Seitz cavity. This picture may be seen as a practical, heuristic way of introducing the pressure ionization in the models but seems poorly motivated, especially for weakly and strongly coupled plasmas. 

Progress is still ongoing towards achieving a better modeling of pressure ionization, closer to the first principles. That means obtaining pressure ionization directly as a consequence of the plasma structure as an ion fluid. Recent research efforts were carried out in this direction with the models based on the QHNC approach and the VAMPIRES model. These models may be viewed as steps in the understanding of the problem but, for sure, they do not exhaust the theoretical challenge of the consistent modeling of nucleus-electron plasmas.

It is to be expected that defining a relevant notion of ion may not be possible at all plasma conditions. Defining the validity domain of atomic physics of plasma, if not of a particular model, remains among the most challenging issues.

Finding a satisfactory model of ions in dense plasmas is the essential first step towards applying many methods and notions from atomic physics. This notably includes the calculation of radiative properties or atomic processes typically involved in collisional-radiative modeling.

As regards radiative properties, most approaches are based on the independent-particle approximation. In these methods, the high number of excited states in high-temperature, mid-to-high-Z plasmas may still constitute an implementation challenge, requiring a tradeoff between completeness and level of detail.

A self-consistent linear-response approach was successfully applied as regards bound electrons and their related contribution to radiative properties since the 1980s~\cite{Zangwill80,Zangwill84,Doolen87}. However, despite a significant theoretical effort~\cite{Blenski06,Caizergues14,Caizergues16}, the consistent treatment of the continuum electron remains an issue in the quantum framework. Yet, this would constitute an important step towards a first-principles approach to the collective effects on radiative properties, neglected by the independent-particle approximation. 

Approaches to the effects of collisions on the radiative properties are known~\cite{Perrot96,Johnson06,Kuchiev08,Johnson09}. These approaches are in fact very similar and focus on the description of electron scattering by ions in the framework of the Boltzmann equation. However, in practice, they are limited to a heuristic bridging to Ziman's formula of static conductivity. A better-founded approach would be of great interest.

Regarding the modeling of non-LTE plasma, collisional-radiative modeling has been used for many years and still remains challenging in itself. Some long-standing issues are well-identified, notably the truncation of the set of levels and the description of dielectronic channels. These issues are in fact related to the modeling of collisional processes in dense-plasmas.

Addressing collisional processes is required for the collisional-radiative modeling of dense plasmas. Work is in progress to study the collisional processes in the framework of fully quantum models of screening in dense plasmas. From a direct application of the distorted-wave approach, it appears that screening of the potential can result in a different distribution of transitions among excitation, ionization, and dielectronic channels. However, the theoretical justification for applying the distorted-wave approach to models that account for continuum electrons deserves a deeper investigation. 

Some equilibrium dense-plasma models, such as INFERNO, have now been studied and used for many years. However, experimental checks of their validity are scarce and do not really allow one to discriminate among the various models. Equation-of-state measurements often have large experimental uncertainties and rarely access the temperature of the plasma in regimes properly relevant to hot plasmas. On the other hand, measurements of radiative properties are most frequently performed on diluted plasmas (see, for instance,~\cite{Davidson88,Bruneau90,DaSilva92,Bailey07,Blenski11,Bailey15}), addressing regimes in which differences among models are not pronounced. Direct-current conductivity measurements~\cite{Renaudin02,Korobenko07} or X-ray Thomson scattering may address relevant regimes but usually require one to take a further step in the modeling in order to interpret the measurements~\cite{Gregori04,Sperling17}.

Efforts to improve atomic models of dense plasmas are timely, in view of the growing concern for understanding the warm-dense matter, with applications to stellar astrophysics and planetology in mind. These efforts are also in sync with the recent advances in experiments on warm and hot dense plasmas, enabled by the advent of new facilities and experimental platforms. One may cite, for instance, the recent convergent-spherical shockwave experiments at NIF~\cite{Kraus16,Swift18}, which give access to equation-of-state and X-ray Thomson scattering data at Gbar pressures; the opacity measurements of compressed plasma at OMEGA~\cite{Hu22}; the measurement of spectral emission of dense, near-equilibrium plasma using buried layers at ORION~\cite{Hoarty13,Beiersdorfer19}; or the experiments on the photoionization of metals using tunable X-ray free-electron laser at LCLS~\cite{Ciricosta12}. These recent improvements in experimental techniques may allow one to better investigate the models' limitations.

Regarding the validation of collisional radiative models of non-LTE plasmas, benchmarks experiments remain hard to set up. Characterizing precisely the state of the non-LTE plasma that is probed is difficult. It is also difficult to conceive an experiment that would selectively address the collisional-radiative part of the modeling, leaving aside any issue of radiation transfer. However, progress is occurring rapidly and recent experiments using multiple diagnostics may provide stringent tests of non-LTE-plasma models. One may cite, for instance, experiments coupling L- and K-shell spectroscopy \cite{Marley18,Bishel23}, experiments measuring absorption and emission simultaneously \cite{Loisel17}, or measuring emission spectra along two lines of sight with distinct reabsorption lengths \cite{PerezCallejo19}.

%Efforts to improve the modeling of non-LTE plasma 
%On the NLTE side, experiments by Ed Marley, S. Bastiani, photoionized, face-on/side-on exp

%{ 
%\newpage
%\thispagestyle{empty}~
%}

{
%\phantom{\chapter{References}}
%\titleformat{\chapter}[display]{\huge}{\vspace{-3cm}}{12pt}{\huge\bf}
\titleformat{\chapter}[display]{\huge}{\vspace{-3cm}}{12pt}{\huge\bf \thechapter.~}
\renewcommand{\bibname}{References}
\bibliographystyle{unsrturl}
\bibliography{biblio-utf8.bib}
}

{ 
\newpage
\thispagestyle{empty}~
}

%-------------------------
\chapter{Symbols and Notations\label{list_symbols}}
%-------------------------
This table gives systematic rules of notation rather than an exhaustive list of symbols. Symbols may play a slightly different role in the various chapter, in particular when the same symbol is used in the description of different models (e.g. $n_\text{e}$). Tildas are sometimes used in order to make a distinction among two functions that are closely related to each other.

\noindent
\begin{minipage}{8cm}
\begin{flushleft}
\begin{tabular}{c l}
%\hline
&Physical constants\\
\hline
$\hbar$ & Planck constant\\
$q_\text{e}$ & Elementary charge\\
$\epsilon_0$ & Electric constant\\
$\ee^2$& $q_\text{e}^2/(4\pi\epsilon_0)$\\
$m_\text{e}$ & Electron mass\\
$k_\text{B}$ & Boltzmann constant\\
$\alpha$ & Fine-structure constant\\
$c$ & Speed of light \\
$a_0$ & Bohr radius \\
$\mathcal{N}$ & Avrogadro's number\\
\hline
\smallskip\\
&Mathematical notations\\
\hline
$\underline{\bullet}$ & Functional dependency\\
$\bar{\bar{\bullet}}$ & Matrix\\
$\bar{\bullet}$ & Vector of arbitrary space\\&(vectors of 3D space in bold)\\
$\theta$ & Heaviside function\\
$\delta$ & Dirac distribution\\
$\mathcal{PP}$ & Cauchy principal part\\
$I_{1/2}$, $I_{3/2}$ & Fermi-Dirac integrals\\&(order 1/2, 3/2)\\
$j_\ell$, $y_\ell$ & Regular and irregular\\&spherical Bessel functions\\
$F_\ell^\text{C}$, $G_\ell^\text{C}$ & Regular and irregular\\&Coulomb wavefunctions\\
Re, Im & Real and Imaginary parts\\
Tr & Trace\\
$[\bullet,\bullet]$ & Commutator\\
\hline
\end{tabular}
\end{flushleft}
\end{minipage}
\hfill
\begin{minipage}{9cm}
\begin{flushright}
\begin{tabular}{c l}
&General plasma parameters\\
\hline
$\bullet_\text{e}$ & Free-electron quantity\\
$\bullet_\text{i}$ & Ion quantity\\
$n_\bullet$ & Number density\\&(homogeneous system)\\
$T$ & Temperature \\
$Z$ & Bare charge\\
$Z^*_\bullet$ & Effective charge\\
$Z^*$ & Mean ionization\\
$M_\text{mol}$ & Molar mass\\
$\Lambda_\bullet$ & Thermal length \\
$R_\text{WS}$ & Wigner-Seitz radius \\
$\lambda_\text{D}$ & Debye length \\
$\omega_\text{P}$ & Plasma frequency (angular) \\
\hline
\smallskip\\
&General mechanical quantities\\
\hline
$\vec{r}$, $\vec{R}$ & Position vector\\
$\vec{p}$, $\vec{P}$ & Momentum vector\\
$u(r)$ & Interaction potential \\
\hline
\smallskip\\
&Classical mechanics\\
\hline
$\mathcal{H}$ & Hamiltonian \\
$\mathcal{W}$ & Coulomb interaction\\
\hline
\end{tabular}
\end{flushright}
\end{minipage}

\noindent
\begin{minipage}{9cm}
\begin{flushleft}
\begin{tabular}{c l}
&Quantum mechanics\\
\hline
$\hat{\bullet}$ & Quantum many-electron operator\\
$\tilde{\bullet}$ & Quantum 1-electron operator\\
$\bullet^\dagger$ & Hermitian conjugate\\
$\ket{\bullet}$, $\bra{\bullet}$ & State vector\\&(ket, bra of Dirac's notation)\\
$\ket{1:\bullet\,2:\bullet}$ & Tensorial-product state vector\\ 
$H$ & Hamiltonian\\
$K$ & Kinetic energy\\
$V$, $v$ & External or self-consistent potential\\
$W$ & Coulomb interaction\\
$a_\bullet$ & Fermion annihilation operator\\
$n_\bullet$ & number operator\\
$\rho$ & Density operator \\
$\chi^\bullet$ & Susceptibility assoc. to operator\\
$\mathcal{D}^\text{R}$ & Retarded density susceptibililty\\
$\Psi$ & Many-electron state or energy level\\
$E$ & Many-electron energy\\
$\xi$, $\zeta$ & orbital labels\\
$\varphi_\bullet$ & 1-electron state or wavefunction\\
$P(r)$ & 1-electron radial wavefunction\\
$\varepsilon$ &  1-electron energy or eigenvalue\\
$g_\bullet$ & Degeneracy\\
$n$ & Principal quantum number\\
$\ell$ & Orbital quantum number\\
$m$ & Magnetic quantum number\\
$\Delta_\bullet$ & Scattering phase-shift\\
$\bullet_\text{xc}$ & exchange-correlation contribution\\
\hline
\smallskip\\
&Radiative Properties\\
\hline
$\vec{k}$ & Wave vector\\
$\nu$ & Frequency\\
$\omega$ & Angular frequency\\
$\chi$ & Electric susceptibility\\
$\epsilon$ & Dielectric function\\
$\sigma$ & Conductivity\\
$k_\text{abs}$ & absorption coefficient\\
$\kappa$ & opacity (per unit mass)\\
$n^\text{ref}$ & Refraction index\\
$\vec{A}$, $\Phi$ & Vector potential, scalar potential\\
$\vec{E}$ & Electric field\\
\hline
\end{tabular}
\end{flushleft}
\end{minipage}
\hfill
\begin{minipage}{9cm}
\begin{flushright}
\vspace{1cm}

\begin{tabular}{c l}
&Statistical Physics \& thermodynamics\\
\hline
$g(r)$ & Pair distribution function\\
$h(r)$ & Pair correlation function\\
$n(r)$ & Electron density\\&(inhomogeneous system)\\
$q(r)$ & Displaced electron density\\
$\dot{\bullet}$ & Quantity per ion \\
$\bullet_\text{eq}$ & Equilibrium quantity \\
$\bullet_\text{ex}$ & Excess quantity \\
$\bullet_\text{id}$ & Ideal-gas quantity \\
$\bullet^\text{F}$ & Fermion-gas quantity\\
$\bullet^\text{DH}$ & Debye-Hückel quantity\\
$\bullet^\text{HNC}$ & Hypernetted chain quantity\\
$\bullet^\text{IS}$ & Ion-sphere quantity\\
$N_\bullet$ & Particle number\\
$\Omega$ & Grand potential \\
$F$ & Free energy \\
$U$ & Internal energy \\
$f$ & Free energy per unit volume \\
$u$ & Internal energy per unit volume \\
$s$ & Entropy per unit volume \\
$P$ & Pressure \\
$V$ & Volume \\
$\mu$ & Chemical potential\\
$p_\text{F}$ & Fermi-Dirac distribution\\
$\beta$&$(k_\text{B}T)^{-1}$ \\
\hline
\smallskip\\
&Atomic processes\\
\hline
$\tau_{\bullet\to\bullet}$ & Transition rate\\
\begin{tabular}{@{}p{1.cm}@{}}
$\sigma^\bullet_{\bullet\to\bullet}$\\ $F^\bullet_{\bullet\to\bullet}$\\
$G^\bullet_{\bullet\to\bullet}$
\end{tabular}
 & Transition cross-section\\
PE, LE & Photoexcitation, line emission\\
PI, RR & Photoionization,\\&radiative recombination\\ 
CE, CD & Collisional excitation, de-excitation\\
CI, 3R & Collisional ionization,\\&3-body recombination\\
AI, DC & Autoionization, dielectronic capture\\
\hline
\end{tabular}
\end{flushright}
\end{minipage}

\newpage
\thispagestyle{empty}
~

%\newpage
%\thispagestyle{empty}
%~

\newpage
\thispagestyle{empty}
%-------------------------
%\chapter*{Outline}
\par\noindent\textbf{\Huge Outline}
\vspace{0.5cm}\par
%-------------------------
{
\small
Modeling plasmas in terms of atoms or ions is theoretically appealing for several reasons. When it is relevant, the notion of atom or ion in a plasma provides us with an interpretation scheme of the plasma's microscopic structure. From the standpoint of quantitative estimation of plasma properties, atomic models of plasma allow extending many theoretical tools of atomic physics to plasmas. This notably includes the statistical approaches to the detailed accounting for excited states, or the collisional-radiative modeling of non-equilibrium plasmas, which is based on the notion of atomic processes.

This habilitation manuscript is mostly focused on the studies to which the author has contributed in the field of atomic modeling of dense, non-ideal plasmas. 

First we introduce the problem of atomic physics of plasma by reviewing a selection of atomic models, from ideal plasmas to non-ideal and pressure-ionized plasmas. We discuss the limitations of these models, closing this selected review with the Variational Average-Atom in Quantum Plasma (VAAQP) model. 

We then address the applications of the VAAQP model to the calculation of radiative properties. This includes an extension of the model for detailed configuration accounting,  and a study of the self-consistent, dynamic linear response in the framework of the VAAQP model.

We then address the extension of atomic models to the accounting of ion-ion correlations in plasmas. We discuss the problem, review some studies on the generalized free-energy functionals for classical fluids. Finally, we outline our recent work on a variational atomic model of plasma that account for both the electron structure of ions and the ion structure of the plasma, seen as a classical fluid of ions.

At the end of the manuscript, we  briefly address the studies to which the author contributed in the field of collisional-radiative modeling of non-LTE plasmas and sketch the prospect of bridging with dense-plasma models.
}

%-------------------------
\vspace{2cm}
\par\noindent\textbf{\Huge Résumé}
\vspace{0.5cm}\par
%-------------------------
{
\selectlanguage{french}
\small
Modéliser les plasmas en utilisant la notion d'atome présente un grand intérêt théorique. Lorsqu'elle est pertinente, la notion d'atome fournit un schème interprétatif de la structure microscopique du plasma. Du point de vue de l'estimation quantitative des propriétés des plasmas, les modèles atomiques de plasma permettent d'étendre de nombreux outils théoriques de la physique atomique aux plasmas. Ceci inclut notamment les approches statistiques de la description des états atomiques excités, ou l'approche collisionnelle-radiative des plasmas hors d'équilibre, qui se fonde sur la notion de processus atomique.

Ce manuscrit d'habilitation se concentre sur les études auxquelles l'auteur a contribué sur la modélisation atomique des plasmas denses, non-idéaux.

Nous introduisons d'abord le sujet en passant en revue une sélection de modèles atomiques de plasma, allant des plasmas idéaux, aux plasmas non-idéaux et ionisés par la pression. Nous discutons des limitations de ces modèles, terminant cette revue partielle par le modèle variationnel d'atome dans un plasma VAAQP.

Nous abordons ensuite les applications du modèle VAAQP au calcul des propriétés radiatives. Ceci inclut une extension du modèle VAAQP aux calculs détaillés en configurations, et une étude de la réponse linéaire dynamique autocohérente dans le cadre du modèle VAAQP.

Nous nous penchons ensuite sur l'extension des modèles atomiques pour prendre en compte les corrélations ion-ion dans le plasma. Nous discutons du problème, passons en revue nos travaux sur les fonctionnelles d'énergie libre généralisées pour les fluides classiques. Pour finir, nous décrivons succinctement nos travaux récents sur un modèle atomique variationnel de plasma qui décrit à la fois la structure électronique des ions, incluant les électrons du continuum, et la structure ionique du plasma, vu comme un fluide classique d'ions.

Enfin, nous décrivons brièvement les travaux auxquels l'auteur a contribué dans le domaine de la modélisation collisionnelle-radiative des plasmas hors d'équilibre, esquissant une perspective de faire le lien avec les modèles de plasmas denses.
}

\end{document}